\documentclass{article}
\usepackage{amssymb}
\usepackage{amsmath}
\usepackage{comment}
\newtheorem{example}{Example}
\newtheorem{theorem}{Theorem}
\newtheorem{definition}{Definition}
\newtheorem{lemma}{Lemma}
\newtheorem{corollary}{Corollary}
\newtheorem{proposition}{Proposition}
\newtheorem{claim}{Claim}
\newtheorem{remark}{Remark}
\title{Regular resolution for CNFs with almost bounded one-sided treewidth}
\author{Andrea Cal{\`{\i}} and Igor Razgon
\thanks{The second author would like to thank Stefan Szeider and Friedrich Slivovsky 
for very useful personal communication.}}
\begin{document}
\maketitle
\begin{abstract}
We introduce a one-sided incidence tree decomposition of a CNF $\varphi$.
This is a tree decomposition of the incidence graph of $\varphi$
where the underlying tree is rooted and the set of bags containing
each clause induces a directed path in the tree.
The one-sided treewidth is the smallest width of a one-sided incidence tree decomposition.

We consider a class of unsatisfiable CNF $\varphi$ that can be turned into one of one sided
treewidth at most $k$ by removal of at most $p$ clauses.
We show that the size of regular resolution for this class of CNFs is FPT parameterized
by $k$ and $p$. The results contributes to understanding the complexity of resolution
for CNFs of bounded incidence treewidth, an open problem well known in the areas of
proof complexity and knowledge compilation. In particular, the result significantly generalizes
all the restricted classes of CNFs of bounded incidence treewidth that are known to admit
an FPT sized resolution. 

The proof includes an auxiliary result and several new notions that may be of an independent interest. 
\end{abstract}
\section{Introduction}
It is well known that an unsatisfiable CNF $\varphi$ has a resolution proof of unsatisfiability
of size FPT in the \emph{primal} treewidth of $\varphi$. On the other hand, it is a well known open problem
whether there is an FPT sized resolution parameterized by the \emph{incidence} treewidth of $\varphi$ \cite{ResIncOpen}.
The only progress related to this problem we are aware of (besides the FPT upper bound parameterized
by the primal treewidth) is establishing that resolution is FPT if parameterized by incidence \emph{pathwidth} \cite{Imanishi}. 
In this paper we establish an FPT upper bound significantly generalizing the parameterization
by both primal treewidth and incidence pathwidth. 

First of all, we consider the \emph{regular} resolution rather than the full-fledged one. 
Second, we introduce the notion of a \emph{one-sided treewidth} of a CNF $\varphi$. 
This is the smallest width of a \emph{one-sided tree decomposition} of the incidence graph of $\varphi$
meaning that the underlying tree is rooted and the set of bags containing
each clause induces a directed path in the tree. Suppose that $\varphi$ has $p$ clauses such that after their
removal, the resulting CNF has one-sided treewidth at most $k$.    
We prove that in this case, unsatisfiability of $\varphi$ can be proved by regular resolution
of size $n\cdot 2^{O(k^2+kp+p\log p)}+n^2 \cdot 2^{O(k^2+kp)}$.
Intuitively speaking the regular resolution is FPT for CNFs whose one-sided treewisth is \emph{almost bounded}.  As we said above, the considered class of CNFs significantly generalizes all the
restricted classes of CNFs of bounded incidence treewidth for which an FPT upper bound on resolution
is known. The result also carries an important message for researchers attempting to prove an \emph{XP lower bound
for the general case}: the target class of CNFs must be \emph{involved} in the sense that in any \emph{rooted} tree decomposition
of the incidence graph of a CNF of this class there must be many clauses each of them appearing in bags of more than
one branch of the underlying tree. 

An important part of the proof is an auxiliary result stating that \emph{transitional resolution}
is FPT for CNFs of bounded one-sided treewidth. The transitional resolution is a 
novel generalization of regular resolution that may be of an independent interest so let us consider
it in more detail. First of all, throughout this paper, we regard regular resolution as a read-once branching
program $R$ with clauses associated with the sinks \cite{Kraj95}. The semantic of this representation is the 
following. Let $u$ be a sink of $R$ and $C$ be the clause associated with $u$. Let $A$ be an assignment 
'carried' by a path from the source of $R$ to $u$. Then $A$ must falsify $C$. In the transitional resolution,
we set aside a subset ${\bf TR}$ of clauses of $\varphi$ and call them \emph{transitional clauses}.
In a transitional resolution, the sinks may be \emph{non-transitional} and \emph{transitional} ones.
A non-transitional sink is associated with a non-transitional clause $C$ with the same semantic as described
above. A transitional sink $u$ is associated with a subset ${\bf C}$ of ${\bf TR}$ and the semantic is as follows. 
Let $A$ be an assignment carried by a path from the source of $R$ to $u$.  Then $A$ \emph{falsifies} all the
clauses of ${\bf C}$ and there is an extension $A^*$ of $A$ that satisfies the remaining clauses of $\varphi$.

To understand the motivation, let $V$ be a proper subset of $Var(\varphi)$ the set of variables of $\varphi$.
Consider $\varphi_V$, a CNF obtained by projection to $V$ of all the clauses having at least one
occurrence of a variable of $V$.  Suppose $\varphi_V$ is unsatisfiable. Can we use a regular resolution
proof for $\varphi_V$ as a building block for the regular resolution proof for $\varphi$? 
Yes if it is transitional resolution! The transitional clauses are those that  have occurrences of both 
$V$ and $Var(\varphi) \setminus V$. Indeed, a non-transitional clause $C$ of $\varphi_V$ is also a clause of $\varphi$.
Therefore, a sink labelled with $C$ can also be a sink for a regular resolution for $\varphi$. However, each 
transitional clause of $\varphi_V$ is a \emph{proper subset} of a clause of $\varphi$. Therefore, if such a clause
is falsified, further reasoning is needed to derive a falsified clause for the whole $\varphi$. 
The transitional sinks provide all the required information for this reasoning. They serve as 'interface' or 
'connection ports' that allow to 'plug in\ a resolution for $\varphi_D$ into a resolution for $\varphi$. 
This approach is critical for our main result and we believe it will be useful in other contexts. 

Let us now state formally our result regarding the transitional resolution. 
Let $\varphi$ be an unsatisfiable CNF of one-sided treewidth $k$ and with $p$ 
transitional clauses. Then there is a transitional resolution proof for $\varphi$ 
of size $n \cdot 2^{O(kp+\log^2 p)}$. Put it differently, the resolution size remains FPT in $k$
only if $p=O(\log n/\log \log n)$. 

Both results of this paper are obtained constructively, that is we demonstrate an algorithm for 
constructing the required resolution subject to the size upper bound.
The transitional resolution is constructed using \emph{top-down dynamic programming}.
This means that we define \emph{local DAGs} associated with the nodes $t$ of the tree of the
underlying tree decomposition, the sinks are associated with sources of local DAGs of children of $t$.
This approach is quite standard in knowledge compilation \cite{DarwicheJACM,DesDNNF,SDD}.  
The difficulty of handling transitional clauses is that we need to implement the conjunction in the FBDD style \cite{BeameDNNF}.
This may potentially lead to exponential explosion. The most non-trivial aspect of the proof is to demonstrate 
that the explosion is tamed and the required upper bound is indeed achieved.

The regular resolution witnessing the main result is constructed using bottom up dynamic programming: 
the local DAGs are associated with prefixes of the postorder traversal of the underlying tree.
To obtain the desired upper bound, we use the following nice fact. If for each non-leaf node of a rooted binary
tree $T$, the number of nodes in the left subtree is greater than or equal to the number of nodes in the right subtree
then the subgraph induced by each prefix of the postorder traversal  is the union of at
most $\lfloor \log n \rfloor+1$ vertex-disjoint subtrees of $T$.
Bottom-up dynamic programming has been used, example in \cite{Florentbeta} and \cite{FlorentLICS}.
However, we are not aware of other results utilizing the above combinatorial statement about
postorder traversals.

Let us now overview related results that we have not considered so far.
Arguably, the best known parameter in the area of proof complexity 
is the width of a resolution proof. A classical result \cite{reswidthexp}
demonstrates that the size of a resolution proof is exponential in its width. 
It was open for some time whether the resolution size is FPT in its width 
until it was resolved negatively in \cite{reswidthnonfpt}

The equivalent definition of regular resolution as read-once branching program
connects it to the area of \emph{knowledge compilation}.
On the surface, regular resolution is very similar to free binary decision diagram (FBDD):
the difference only in the labelling of sinks and the corresponding semantics of source-sink
paths. However, regular resolution is in fact much closer to decision decomposable negation
normal forms (Decision DNNF) \cite{DesDNNF}: they both can easily handle conjunction of two variable
disjoint CNFs. Decision DNNF simply has a decomposable decision gate at its disposal.
Regular resolution does not have luxury of using such a gate but instead it has a not less
impressive power of \emph{forgetting}. Indeed if $\varphi=\varphi_1 \wedge \varphi_2$ is unsatisfiable
and $\varphi_1$ is variable-disjoint with $\varphi_2$ then one of them is unsatisfiable.
If, say $\varphi_1$ is unsatisfiable then a regular resolution proof for $\varphi_1$ is in fact such a
proof for the whole $\varphi$, so that $\varphi_2$ can simply be discarded! 
On the contrast, FBDDs do not possess the gift of forgetting and have no conjunction gate at
their disposal. As a result they cannot even efficiently represent CNFs of bounded \emph{primal} treewidth \cite{RazgonAlgo}.
It is also interesting to note that it is open whether CNFs of bounded incidence treewidth can be represeted
by FPT-sized Decision DNNFs. 

We conclude the literature overview by saying 
that investigation of new graph parameters is currently a bustling research direction.
Notable examples include twin-width \cite{twinwidth1}  and several variants of maximum matching width
\cite{mimwidth1}, the latter set of parameters is known to be of a significant relevance for knowledge compilation 
\cite{RazgonIPEC21}.

We conclude the introduction by overviewing the structure of the paper.
There are five sections in the main body of the paper and three sections in the appendix.
Section 2 is preliminaries. The auxiliary result (Theorem \ref{transres}) is proved in Section 3.
The main result (Theorem \ref{mainres}) is proved in Section 4. The conclusion is provided in Section 5.  
The proofs of both theorems consist of establishing correctness of the dynamic programming
constructions and proving upper bounds on their sizes. While the latter is neat and compact,
the former is rather tedious. Therefore, in the main body of the paper we provide only
sketches of proofs of Theorems \ref{transres} and \ref{mainres} postponing detailed proofs
to the appendix that is structured as follows.  Section A provides theorems essentially reducing
the correctness proofs to proving that the local DAGs satisfy certain properties.
A detailed proof of Theorem \ref{transres} is provided in Section B and a detailed proof of Theorem \ref{mainres}
is provided in Section C of the appendix.

\section{Preliminaries}

\subsection {Set of literals, CNFs, and DAGs.}
In this paper when we consider a set $S$ of literals of Boolean variables,
we mean that $S$ is \emph{well formed} in the sense that it does not contain 
both positive and negative literals of the same variable. A variable $x$ \emph{occurs}
in $S$ if either $x \in S$ or $\neg x \in S$. In the former case, we say that $x$
\emph{occurs positively} or has a \emph{positive assignment} in $S$.
In the latter case, we say that $x$ \emph{occurs negatively} or has 
a \emph{negative assignment} in $S$.

We use $Var({\bf A})$ to denote the set of variables of an object ${\bf A}$
which may be a set of literals, a CNF, a graph with sets of variables associated with
its vertices (e.g. a tree decomposition) or a branching program.

We consider a CNF $\varphi$ as a set of clauses and
each clause is just a set of literals. 
Let $C$ be a clause and let $S$ be a set of literals.
We say that $S$ \emph{satisfies} $C$ if $C \cap S \neq \emptyset$.
If $S$  does not satisfy $C$ 
then $C|_S=C \setminus \neg S$ where $\neg S=\{\neg x|x \in S\}$.
We denote by $\varphi|_S$ the set of all $C|_S$ such that $C$ is a clause
of $\varphi$ that is not satisfied by $S$.
A CNF $\varphi'$ is a \emph{subCNF} of $\varphi$ if $\varphi' \subseteq \varphi$
and the subCNF is \emph{modular} if $Var(\varphi) \cap Var(\varphi \setminus \varphi')=\emptyset$. 

For a set $S$ of literals and a set $V$ of variables,
we denote by $Proj(S,V)$  the \emph{projection} of $S$ to $V$
that is $S' \subseteq S$ such that $Var(S')=Var(S) \cap V$.  

The \emph{primal} graph of a CNF has vertices corresponding to
$Var(\varphi)$ and two variables are adjacent if they occur in the same clause.
The \emph{incidence} graph of a CNF has vertices corresponding to the variables and
clauses of $\varphi$ and there is an edge between $x \in Var(\varphi)$ and $C \in \varphi$
if $x$ occurs in $C$.

In this paper we often consider directed acyclyc graphs (DAGs)
(including rooted binary trees). For a DAG $Z$ and $u \in V(Z)$,
we denote by $Z_u$ the subgraph of $Z$ including $u$ and all the vertices
reachable from $u$.

\subsection{One-sided tree decomposition}
A \emph{tree decomposition} $(T,{\bf B})$ of a graph $G$ is a pair where $T$ is a tree,
${\bf B}$ is a set of bags $B(t)$ associated with each node $t \in V(T)$.
The bags must obey the rules of (i) union ($\bigcup_{t \in V(T)} B(t)=V(G)$)
(ii) containment ($\forall e \in E(G) \exists t \in V(t) e \subseteq B(t)$) and
(iii) connectivity (for each vertex $v \in V(G)$, the set of nodes whose bags contain
$v$ induces a connected subgraph of $T$). The width of the tree decomposition
is the largest size of a bag minus one. The treewidth of a graph is the smallest
width of its tree decomposition.

In this paper we will consider tree decompositions for the incidence graph
of a CNF $\varphi$ (\emph{incidence tree decompositions} of $\varphi$). 
In this case, for each node $t$ of the underlying tree,
$B(t)$ is partitioned into $Var(t)$ and $CL(t)$ respectively corresponding 
to variables and clauses. 


\begin{definition}
Let $(T,{\bf B})$ be a \emph{rooted} incidence tree decomposition
of $\varphi$. For each clause $C$ of $\varphi$, let $V_C$ denotes
the set of nodes of $T$ whose bags contain $C$.
Then $(T,{\bf B})$ is \emph{one-sided} if for each $C \in \varphi$,
$T[V_C]$ is a directed path.

The \emph{one-sided (incidence) treewidth} of $\varphi$
is the smallest width of a one-sided incidence tree decomposition of
$\varphi$.
\end{definition}

It is not hard to see that the one-sided treewidth of $\varphi$
'dominates'  both the incidence pathwidth and the primal treewidth of $\varphi$
in the following sense.

\begin{proposition} \label{domoneside}
The onde-sided treewidth of $\varphi$ does not exceed
the incidence pathwdith of $\varphi$ and is at most the
primal treewidth of $\varphi$ plus one.  
\end{proposition}

{\bf Proof.}
An incidence path decomposition can be turned
into a one-sided one by making an end vertex
of the underlying path the root.

Let $(T,{\bf B})$ be a primal tree decomposition of $\varphi$.
Apply the following standard transformation. 
For each non-empty clause $C$, identify a node $t$ of $T$
such that $B(t)$ includes all the variables of $C$.
Introduce a new node $t'$ connected to $t$
and make $Var(C) \cup \{C\}$ the bag of $t'$.

Clearly we obtained an incidence tree decomposition of $\varphi$
of width at most the width $(T,{\bf B})$ plus one where each clause $C$ 
is present in at most one node. Therefore, whatever node of the underlying
tree is made the root, the resulting decomposition will be one-sided.
$\blacksquare$

\begin{proposition} \label{onesidebinary}
Let $(T,{\bf B})$ be a one-sided incidence tree decomposition 
of $\varphi$. Then $(T,{\bf B})$ can be replaced by a one sided
incidence tree decomposition $(T^*,{\bf B}^*)$ of $\varphi$ 
with the number of bags linear in $|\varphi|+|Var(\varphi)|$,
whose width is not larger than that of $(T,{\bf B})$ and 
$T^*$ being a binary tree.
\end{proposition}

{\bf Proof.}
First we produce a tree decomposition of a linear size while keeping one-sidedness.
This is done is the following two stages.

\begin{enumerate}
\item Remove each leaf whose bag is a subset of its parent.
\item For each non-root node $u$ having one child such that $B(u) \subseteq B(v)$
where $v$ is the parent of $u$, contract $u$. 
\end{enumerate}

Note that the result of the first step is that each leaf is associated with a unique
element of $\varphi \cup Var(\varphi)$. This ensures that the number of leaves is
at most $|\varphi|+|Var(\varphi)|$. The same argument bounds the number of 
non-root nodes with one child. Taking into account that the number of non-root nodes with
two or more children is not greater than the number of leaves and, adding the non-root nodes
implies that the total number of nodes is at most $3*(|\varphi|+|Var(\varphi)|)$.
Clearly, these transformations do not violate the read-onceness. 

If $T$ is binary we are done.
Otherwise, let $u$ be a node of $T$ with children
$v_1, \dots, v_q$ such that $q \geq 3$.
We obtain a one-side incidence tree decomposition
$(T',{\bf B'})$ of $\varphi$ of width not larger than that
of $(T,{\bf B})$ by applying the following transformation.

\begin{enumerate}
\item Introduce a new node $u'$.
\item Let $u',v_q$ be the children of $u$ and
$v_1, \dots, v_{q-1}$ the children of $u'$.
\end{enumerate}

Clearly by repeated application of this transformation
we will eventually obtain an underlying tree which is binary.
It remains to specify the content of the bags of the nodes
involved. The bags of $u,v_1, \dots, v_q$ are exactly the same
as for $(T,{\bf B})$. The bag for $u'$ is constructed as follows.
\begin{enumerate}
\item $Var(u')=Var(u)$.
\item $CL(u')=CL(u) \setminus CL(v_q)$.
\end{enumerate}

Let $C \in CL(V_q) \cap CL(u)$.
By the one-sidedess property, $C$ does not occur n bags of
any $v_1, \dots, v_{q-1}$. Therefore, the connectedness property
regarding $C$ is preserved. The rest of the required properties 
of a one-sided incidence tree decomposition are easy to verify
by direct inspection.

Note that this transformation does change the number of leaves nor the number of
nodes having one child. Hence the counting argument made before
the transformation applies and we still have the upper bound
of $3*(|\varphi|+Var(\varphi))$. 
$\blacksquare$

The notions of $Var(t)$ and $CL(t)$ naturally extend 
to structures containing several nodes of $T$.
For example, if $T_1$ is a subtree $T$ then $Var(T_1)=\bigcup_{t \in V(T_1)} Var(t)$,
If $x \in Var(T_1)$, we may also say that $x$ is \emph{contained} in $T_1$.
Also, if ${\bf T}$ is a subset of subtrees of $T$ then $Var({\bf T})=\bigcup_{T' \in {\bf T}} Var(T')$.

\subsection{Postorder traversals} \label{sec:post}
Let $T$ be a rooted binary tree.
For each non-leaf node $u$ we identify its left and right
children as follows. If $u$ has only one child $v$ then $v$
is considered the left child of $u$. 
If $u$ has two children $v_1$ and $v_2$
and, say, $|V(T_{v_1})|>|V(T_{v_2})|$ then $v_1$ is the left child of $u$
and $v_2$ is the right child of $u$. Finally, 
$|V(T_{v_1})|=|V(T_{v_2})|$ then the left and the right children are assigned
in an arbitrary (but fixed) way. 

Having defined the left and right children for the 
non-leaf nodes of $T$, we can define the permutation $\pi_T$
of the nodes of $T$ explored according to the \emph{postorder}
traversal (left subtree of $T$, if any, is recursively traversed then the right
subtree, if any, is recursively traversed, then the root is traversed).  

Let $\pi$ be a prefix of $\pi_T$.
It is not hard to see that $T[\pi]$ \footnote{Here and in several other places we slightly
abuse notation by identifying a sequence with its underlying set. The correct interpretation
will always be clear from the context.} is the union of vertex disjoint
trees $T_u$ (meaning subtrees of $T$ rooted by some vertices $u \in V(T)$).
We denote the set of these trees by $Trees_{\pi}$.

\begin{proposition}  \label{ordertrees1}
Let $T_1,T_2$ be two distinct elements of $Trees_{\pi}$
Then one of them, say, $T_1$ occurs before the other in the following
sense: all the nodes of $T_1$ precede in $\pi$ all the nodes of $T_2$.
\end{proposition}
 
{\bf Proof.}
By induction on $|V(T)|$
The statement is clearly true for $|V(T)|=1$
so assume that $|V(T)|>1$. 

Let $rt$ be the root of $T$.
If $T$ has only one child, te statement is easily seen
to hold by the induction assumption. 
So, assume that $rt$ has two children $t_1$ (the left child)
and $t_2$ (the right child). 
Denote the respective postorder traversals for $T_{t_1}$
and $T_{t_2}$ by $\pi_1$ and $\pi_2$. Clearly $\pi_T=\pi_1+\pi_2+rt$.
Assume that $rt \in \pi$. Then $T_{rt}$ is the only elements of
$Trees_{\pi}$ and hence the statement is vacuously true.
Next, assume that $t_1 \notin \pi$.
It follows that $\pi \subseteq \pi_1$ and hence the statement holds
by the induction assumption. 

Assume now that $t_1 \in \pi$.
This means that $\pi_1$  is a prefix of $\pi$.
Let $\pi=\pi_1+\pi'$. Clearly $T_{t_1} \in Trees(\pi)$ and it
is the largest element of $Trees_{\pi}$ according to the
order specified in the statement of the proposition. 
Moreover, $Trees_{\pi} \setminus \{T_{t_1}\}$ are all in $\pi'$.
The order between them exists by the induction assumption.
It remains to say that the order between the trees in $\pi'$ is the
same as in $\pi$.
$\blacksquare$

Proposition \ref{ordertrees1} naturally defines a linear order 
on $Trees_{\pi}$. In what follows, if we say that for $T_1,T_2 \in Trees_(\pi)$
$T_1<T_2$, we mean that $T_1$ occurs before $T_2$ in $\pi$ as specified
in Proposition \ref{ordertrees1}.

\begin{definition}
Let $\pi$ be a proper prefix of $\pi_T$ and let $t$ be its 
immediate successor (that is, the node immediately following $\pi$ in $\pi_T$). 
If $t$ is a leaf of $T$, we call $t$ an
\emph{expanding} node. Otherwise $t$ is a \emph{contracting}
node. 
\end{definition}  

\begin{proposition} \label{logtrees}
Let $\pi$ be a prefix of $\pi_T$.
Then $|Trees_{\pi}| \leq \lfloor \log |V(T)| \rfloor +1$.
\end{proposition}

{\bf Proof.}
By induction on $|V(T)|$
The statement is clearly true for $|V(T)|=1$
so assume that $|V(T)|>1$. 

Let $rt$ be the root of $T$.
If $T$ has only one child, the statement is easily seen
to hold by the induction assumption. 
So, assume that $rt$ has two children $t_1$ (the left child)
and $t_2$ (the right child).  Then the reasoning is done
through the following case analysis.

\begin{enumerate}
\item $rt \in \pi$. This is possible only if $\pi=\pi_T$
But in this case $|Trees_{\pi}|=1$.
\item $t_1 \notin \pi$.
It follows that $\pi \subseteq V(T_{t_1})$ and hence 
the statement holds regarding $V(T_{t_1})$ by the induction assumption
and hence holds regarding $V(T)$.
\item Assume that $t_1 \in \pi$. 
Let $\pi'$ be the subsequence of $\pi$ consisting of all elements of $T_{t_2}$.
Clearly $Trees(\pi)=\{T_{t_1}\ \cup Trees_{T_{t_2},\pi'}$
(the subscript $T_{t_2}$ si used to emphasize that the set is considered
w.r.t. $T_{t_2}$ rather than $T$. By the induction assumption,
$|Trees(\pi) \leq 2+\lfloor  \log |V(T_{t_2})| \rfloor$.
By selection of the left and right children $|V(T_{t_2})|<|V(T)|/2|$ and hence
$\lfloor  \log |V(T_{t_2})| \rfloor \leq \lfloor  \log |V(T)| \rfloor-1$.
\end{enumerate}
$\blacksquare$


Finally, we introduce two more notations.
For a set ${\bf T}$ of subtrees of $T$, we denote 
the set of roots of ${\bf T}$ by $Roots({\bf T})$.
Also, the last tree of $Trees_{\pi}$ according to the above
order is denoted by $last(Trees_{\pi})$.

\subsection{Regular resolution and $\varphi$-based functions}

The compressed decision tree defined below
is essentially a read-once branching program where the sinks are not
labelled and no related semantics is provided. 
In terms of programming languages, this notion can be thought of
as an abstract class. We find this generic notion very handy as we can
then easily define several related models by simply specifying
labelling of links and providing constraints on source-sink paths. 

\begin{definition} [{\bf Compressed Decision Tree}]
A \emph{Compressed Decision Tree} (CDT)  $H$ is 
a DAG with a single source with the following additional properties.
\begin{enumerate}
\item Each non-terminal node is associated with a variable. 
\item Each non-terminal node $u$ has exactly two out-neighbours, one of them is labelled
with $\neg x$, the other is labelled with $x$, where $x$ is the variable labelling $u$. 
\item \emph{Read-onceness} property: on each directed path $P$ of $H$, no variable
occurs twice as a literal labelling an edge of $P$.  
\end{enumerate}

For a directed path $P$ of $H$, we denote by $A(P)$ the set of literals labelling
the edges of $P$. We sometimes refer to $A(P)$ as the assignment \emph{carried}
by $P$. 

If the underlying graph of $H$ is a tree then $H$ is called a \emph{decision tree}. 
If ${\bf A}$ is the set of all assignments carried by root-leaf paths of $H$. we sometimes
say that $H$ is a decision tree \emph{over} ${\bf A}$.
\end{definition}

\begin{definition} [{\bf Regular resolution}]
A regular resolution (RR) of a CNF $\varphi$ is a CDT $H$
with clauses associated with the sinks such that
the following holds.

Let $u$ be a sink of $H$, let $C$ be a clause associated with $u$
and let $P$ be a path from the root to $u$.
Then $A(P)$ falsifies $C$.  
\end{definition}

\begin{definition} [{\bf $\varphi$-based functions}]
Let $\varphi$ be a CNF. A $\varphi$-based function $f$
is a function with the domain $dom(f) \subseteq \varphi$
and, for each $C \in dom(f)$, $f(C) \subseteq C$.
We denote by $range(f)$ the set of all clauses $C'$ such that
there is a clause $C \in dom(f)$ such that $f(C)=C'$.

We identify two special $\varphi$-based functions.
The first is the $C \rightarrow ()$ function for each $C \in \varphi$.
where $dom(C \rightarrow ())=\{C\}$ and $C$ is mapped to $()$.
The second is the \emph{identity} function ${\bf 1}={\bf 1}_{\varphi}$ with 
$dom({\bf 1})=\varphi$ and for each $C \in \varphi$, ${\bf 1}(C)=C$. 

We also say that a $\varphi$-based function $f$ is \emph{unsatisfiable}
if $range(f)$ is unsatisfiable. 
\end{definition}

\begin{definition}
\begin{enumerate}
\item Let $f_1$ and $f_2$ be two $\varphi$-based functions
such that for each $C \in dom(f_1) \cap dom(f_2)$,
$f_1(C)=f_2(C)$. 
The union $f=f_1 \cup f_2$ is a function with $dom(f)=dom(f_1) \cup 
dom(f_2)$, for each $C \in dom(f_1)$, $f(C)=f_1(C)$ and for 
each $C \in dom(f_2)$, $f(C)=f_2(C)$. 
Functions $f_1 \cap f_2$ and $f_1 \setminus f_2$ are defined accordingly.

\item Let $f$ be a $\varphi$-based function and let $A$ be a set
of literals.
Let ${\bf C} \subseteq dom(f)$ be the subset of $dom(f)$
consisting of all clauses $C$ such that $f(C)$ is satisfied by $A$.
Them $f|_A$ is a function whose domain is 
$dom(f) \setminus {\bf C}$ and for each $C \in dom(f|_A)$,
$f|_A(C)=f(C)|_A$.
\item Let $f$ be a $\varphi$-based function
${\bf C} \subseteq dom(f)$. Then the \emph{restriction} of $f$
to ${\bf C}$ is a function $f'$ with $dom(f')={\bf C}$ and for
each $C \in {\bf C}$, $f'(C)=f(C)$. We also sometimes refer to $f'$
as a \emph{subfunction} of $f$. 
Another way to define a restriction is to specify the set of clauses ${\bf C^*}$
to be removed from the domain. The resulting function is denoted by $f \setminus {\bf C^*}$. 
\end{enumerate} 
\end{definition}

{\bf Remark.}
Let $f$ be a $\varphi$-based function and let ${\bf C} \subseteq dom(f)$.
Then $f({\bf C})=\{f(C)|C \in {\bf C}\{$. 

\begin{definition}  [{\bf Functional regular resolution}]
A Functional Regular Resolution (FRR) is a CDT $H$
with $\varphi$-based functions associated with (labelling) its root and its 
sinks so that the following holds. Let $f^*$ be the function associated 
with the root of $H$. Let $u$ be a sink of $H$ and let $f$ be the function
associated with $u$. Then $dom(f) \subseteq dom(f^*)$.
Moreover, let $P$ be a root-$u$ path. Then for each $C \in dom(f)$,
$A(P)$ does not satisfy $C$ and $f(C)=f^*(C)|_{A(P)}$. 
$H$ is a \emph{falsifying} FRR is each sink of $H$ is labelled with $C \rightarrow()$
for some $C \in dom(f^*)$.
\end{definition}

FRR is a more general structure than RR due to the following easy to
establish argument.

\begin{proposition} \label{transfer1}
Let $Z$ be a falsifying FRR for ${\bf 1}_{\varphi}$.
Then a RR for $\varphi$ can be obtained from $Z$ by changing
each label $C \rightarrow ()$ of a sink just to $C$.
\end{proposition}

In light of the above proposition, we will construct FRRs instead
of RRs in the subsequent sections.  The benefit of this increased
generality is that a composition of FRRs is easier to describe formally
that a composition of RRs because in the former case we \emph{do not need
to change the labels of sinks}. Indeed, suppose that $R$ is an FRR for a
$\varphi$-based function $F$ and let $u$ be a non-root node. 
Let $A$ be an assignment carried by a path from the source of $R$ to 
$u$. Then it can be shown that $R_u$ is an FRR for $F|_A$.
On the other hand, $R$ is also an RR for $range(F)$ (subject to changing 
all the sink labels $C \rightarrow ()$ to $C$. However, in this case 
$R_u$ is \emph{not} a RR for $range(F)|_A$ because the labels of the sinks
are the clauses of $range(F)$ not of $range(F)|_A$!


\section{Regular resolution with transitional clauses}
Throughout this section $\varphi$ is an \emph{unsatisfiable} CNF and $(T,{\bf B})$
is a one-sided tree decomposition of the incidence graph of
$\varphi$ having width at most $k$. In light of Proposition \ref{onesidebinary},
we assume that $T$ is a binary tree.   
To define a transitional resolution we need to identify for
the given CNF $\varphi$ a subset ${\bf TR}$ of clauses that we call
\emph{transitional}. We need two auxiliary definitions.

\begin{definition} [{\bf Falsifier}]
Let $f$ be a $\varphi$-based function and let 
${\bf C} \subseteq dom(f) \cap {\bf TR}$.
An assignment $A$ is a ${\bf C}$-falsifier (for $f$) if the following
conditions hold.
\begin{enumerate}
\item for each $C \in {\bf C}$, $f(C)$ is falsified by $A$.
\item There is a \emph{witnessing} extension $A'$ of $A$ such that
for each $C \in dom(f) \setminus {\bf C}$, $A \cup A'$
satisfies $f(C)$.
\end{enumerate}
\end{definition}

\begin{definition} [{\bf Transition function}]
A $\psi$-based function $f$ is called \emph{transitional}
if $dom(f) \subseteq {\bf TR}$ and $range(f)=\{()\}$.
We refer to a transition function $f$ with ${\bf C}=dom(f)$
by $trans_{\bf C}$. 
\end{definition}

\begin{definition} [{\bf Transitional resolution}]

A \emph{transitional resolution} (TRes) is a CDT
$R$ whose source is associated 
with a $\varphi$-based function $f$ and the sinks are 
partitioned into two types of nodes: 
\emph{terminal} and \emph{transitional} ones. 

Each terminal node $u$ is associated with a function
$C \rightarrow ()$ for some $C \in dom(f) \setminus {\bf TR}$ 
and for each path $P$ from the source to $u$, $A(P)$ falsifies $f(C)$.

Each transitional node $u$ is associated with 
a transitional function $trans_{\bf C}$  with ${\bf C} \subseteq dom(f) \cap {\bf TR}$
such that for each path $P$ from the source to $u$,
$A(P)$ is a ${\bf C}$-falsifier.

$R$ is a TRes for $\varphi$ if $f={\bf 1}_{\varphi}$. 
\end{definition}

\begin{theorem} \label{transres}
Let $n=|Var(\varphi)|+|\varphi|$.
Then there is a transitional resolution for $\varphi$ of size
$n \cdot 2^{O(k \cdot |{\bf TR}|+ \log^2(|{\bf TR}|))}$ in case ${\bf TR} \neq \emptyset$
and $n \cdot 4^k$ in case ${\bf TR}=\emptyset$.
In other words, if $|{\bf TR}|=O(\log n/ \log \log n)$, the size if FPT in $k$ only.  
\end{theorem}

In the next section, we use a TRes as a building block for construction
of the FRR witnessing the main result (Theorem \ref{mainres}).  Theorem \ref{transres}
provides an upper bound on the size of the building blocks.
In fact, the TRes occurring in the proof of Theorem \ref{mainres} has
its sets of transitional clauses of size at most $k$, so the full
power of Theorem \ref{transres} as expressed in its last sentence will not be 
needed.  The idea of the use of a TRes for construction of FRRs may be of an independent
interest, therefore, we briefly describe the underlying intuition. 

Let $V_1,V_2$ be a partition of $Var(\varphi)$.
The idea is to construct first a resolution for a $\varphi$-based function
mapping clauses to their projections to $V_1$ (of course the only clauses for
which these projections are non-empty) and then to somehow extend the 
resolution to one for the whole ${\bf 1}_{\varphi}$.

$\varphi$ can be seen as the disjoint union of sets ${\bf C_0}, {\bf C_1}, {\bf C_2}$,
where ${\bf C_0}$ consists of clauses whose sets of variables have non-empty
intersections  with both $V_1$ and $V_2$ and for each $i \in \{1,2\}$, 
${\bf C_i}=\{C|C \in \varphi, Var(C) \subseteq V_i\}$. 
Let $F_1$ be a $\varphi$-based function with $dom(F_1)={\bf C_0} \cup {\bf C_1}$
and for each $C \in dom(F_1)$, $F_1(C)=Proj(C,V_1)$. 
What is a possible use of $F_1$ for the construction of an FRR for ${\bf 1}_{\varphi}$?
First, if $F_1$ is satsfiable, then ${\bf C_2}$ is an unsatisfiable set of clauses,
so we can construct ${\bf 1}_{\bf C_2}$ instead of ${\bf 1}_{\varphi}$.
If $F_1$ is unsatisfiable then the situation is more complicated.

Let $Z_1$ be an falsifying FRR for $F_1$. Can it be extended to a falsifying FRR for 
${\bf 1}_{\varphi}$?  No, not quite. A sink labelled with $C \rightarrow ()$ for $C \in {\bf C_1}$ 
can indeed be a sink for a falsifying FRR for ${\bf 1}_{\varphi}$. But if $C \in {\bf C_0}$,
this sink needs to be further expanded.  However, the information providced by the
sink (falsification of $Proj(C,V_1)$ is nit sufficient for such an expansion.  
The issue of 'information transfer'  is completely resolved by the use of transitional resolution. 
Indeed, suppose that $Z_1$ is a transitional resolution for $F_1$  with ${\bf C_0}$  being
the set of transitional clauses. Then the clauses of ${\bf C_0}$ only take part in the labels
for transitional sinks and these labels provide sufficient information for their further 
expansion. Indeed, let $u$ be such a sink and assume that it is labelled with $trans_{\bf C}$
for some ${\bf C} \subseteq {\bf C_0}$. This means that each assignment carried by a source-$u$
path of $Z_1$ can be extended to an assignment over the whole $V_1$ that satisfies
all the clauses of ${\bf C_0} \cup {\bf C_1} \setminus {\bf C}$ but falsifies the projections
to $V_1$ of the clauses of ${\bf C}$.  Let now $F_2$ be a $\varphi$-based function with domain
${\bf C_2} \cup {\bf C}$ mapping each clause of the domain to its projection to $V_2$. 
From $trans_{\bf C}$ being a label of $u$, we conclude that $F_2$ is unsatisfiable.
Let $Z_2$ be a falsifying FRR for $F_2$. Identify the source of $Z_2$ with $u$ and perform
the same operation for the rest of transitional sinks of $Z_1$ (of course, w.r.t. their corresponding
labels). It can be shown that the resulting construction is a falsifying FRR for ${\bf 1}_{\varphi}$. 
In the actual construction in the next section, the expansions of different transitional sinks are
not necessarily vertex disjoint, thus the construction is made more compact.  


In fact what we need in the next section is the following corollary
of Theorem \ref{transres}.

\begin{corollary} \label{col:transres}
Let $f$ be a $\varphi$-based unsatisfiable function.
Then there is a transitional resolution of $f$ of size as specified
in Theorem \ref{transres}..
Moreover, the set of variables of this transitional resolution is
a subset of $Var(range(f))$
\end{corollary}

{\bf Proof.}
Let $\psi=range(f)$.
First we show that the one-sided trewidth of $\psi$ cannot be large than $k$.
Indeed, $dom(f) \subseteq \varphi$, therefore the one-sided treewidth of
$dom(f)$ s at most $k$. Now, replace each $C$ with $f(C)$ (removing double occurrences 
in case the mapping is not one to one). Clearly,  the tree decomposition for $dom(f)$
 (after removal of the set of variables not occurring in $\psi$) applies to $\psi$.

We are going to consider a transitional decomposition for ${\bf 1}_{\psi}$.
However, we need to define the set ${\bf TR}_{\psi}$ of transitional 
clauses for $\psi$. In order to do this, we need to look at the set
$f^{-1}(C)$ for each $C \in \psi$.
${\bf TR}_{\psi}$ consists precisely of clauses $C$ for which
$f^{-1}(C) \subseteq {\bf TR}$. 
As clauses of ${\bf TR}_{\psi}$ correspond to a partition 
of a subset of ${\bf TR}$, $|{\bf TR}_{\psi}| \leq |{\bf TR}|$.
Therefore by Theorem \ref{transres}, there is a transitional resolution $R$
for ${\bf 1}_{\psi}$ with ${\bf TR}_{\psi}$ being the set of transitional 
clauses and whose size is upper bounded as specified in Theorem \ref{transres} by $k$ and $|{\bf TR}|$

We transform $R$ into a transitional resolution $R^*$ for $\varphi$ w.r.t ${\bf TR}$
by simply changing the labelling of the sinks in the following way. 
\begin{itemize}
\item A label $C \rightarrow ()$ is replaced by $w(C) \rightarrow ()$ where
$w(C)$ is an arbitrary but fixed clause in $f^{-1}(C) \setminus {\bf TR}$
(note that since $C$ is non-transitional for $\psi$, such a clause must exist). 
\item A label $trans_{\bf C}$is replaced with $trans_{\bf C^*}$ where
$C^*=\bigcup_{C \in {\bf C}} f^{-1}(C)$.
\end{itemize}

We now need to verify that $R^*$ is indeed a transitional resolution 
Let $u$ be a sink of $R^*$ and let $P$ be a path from the source to $u$.
Assume first that, in $R$, $u$ is associated with $C \rightarrow()$.
This means that $A(P)$ falsifies $C$. In $R^*$, $u$ is associated 
with $w(C)$, hence $A(P)$ falsifies $f(w(C))=C$.
Next we assume that, in $R$, $u$ is associated with $trans_{\bf C}$.
We observe that that $A(P)$ falsifies $f(C)$ for each $C \in {\bf C^*}$
simply because $f(C) \in {\bf C}$ by definition of ${\bf C^*}$ and 
$A(P)$ falsifies all the clauses of ${\bf C}$ by definition of $R$.
Next, there is an extension $A$ of $A(P)$ satisfies $f(C')$ for
each $C' \in \psi \setminus {\bf C}$. Now, we claim that $A(P)$
satisfies $f(C)$ for each $C \in \varphi \setminus {\bf C^*}$ simply
because $f(C) \notin {\bf C}$ by definition of ${\bf C^*}$
as a 'saturated' set. We have thus verified that $R^*$ is indeed
a transitional resolution. 

Finally, we note that the proof of Theorem \ref{transres}
uses only variables of ${\bf 1}_{\psi}$ simply by construction
of local graphs for the resolution, This means that,
by construction $Var(R^*) \subseteq Var({\bf 1}_{\psi})=Var(range(f))$.
$\blacksquare$

\subsection{Proof of Theorem \ref{transres}: Informal overview}
It is instructive to first assume that the number of transitional
clauses is zero. 
In this case, we associate with each node $t \in V(T)$ 
an FPT number of $\varphi$-based functions.
Each such a function is 'induced' by an assignment to the variables 
lying in the bags of the ancestors of $t$. 
Then we construct a DAG $R$  where, 
for each such a function $f$,  there is a node $u$
such that  $R_u$ is an FRR of $f$.
One of these functions is ${\bf 1}_{\varphi}$ hence $R_u$ , where $u$ is the node corresponding
to ${\bf 1}_{\varphi}$ is the required FRR. 
 
In order to construct $R$, we order the nodes
of $T$ so that children occur before their parents. 
Then we order the functions in an arbitrary order \emph{adhering} 
to the order of their respective nodes. This means that if $t_1<t_2$ 
then all the functions associated with $t_1$ are ordered before all 
th functions associated with $t_2$.  Let $f_1,f_2, \dots, f_q$ be the resulting sequence of functions,
the nodes of $R$ corresponding to them as above are denoted by $u_1, \dots, u_q$.
For each $f_i$, we construct an
FPT sized \emph{local subgraph} $L_i$ whose source is $u_i$ 
and each sink is either a sink of the whole $R$ or is 
identified with some $u_j$ for $j<i$.  

In order to see how the local subgraphs $L_i$ are created
let us see in a greater detail how the functions are defined.
In fact what we associate with each node $t \in V(T)$ are not
functions but \emph{types}, that is tuples  of the form ${\bf tp}=(t,CL,S)$
where $C \subseteq CL(t)$ and $S$ is an assignment to
$Var(t) \cap Var(p(t))$ where $p(t)$ is the parent of $t$
(of course $S=\emptyset$ if $t$ is the root).
Intuitively, $CL$ is the set of clauses satisfied by the assignment 
to the variables in the bags of the ancestors of $t$ and $S$ is the projection
of this assignment to $Var(t) \cap Var(p(t))$. 
The functions are associated with the types, in particular, the function
associated with ${\bf tp}$ is denoted by $f_{\bf tp}$ and is defined
as follows. The domain $dom(f_{\bf tp})$ is $CL(T_t)$ but without $CL$ and without
the clauses satisfied by $S$. 
For each $C \in dom(f_{\bf tp})$, $f_{\bf tp}(C)=Proj(C, Var(T_t) \setminus S)$. 
Intuitively, $f_{\bf tp}$ registers the effect on ${\bf 1}_{\varphi}$ of the assignment to the ancestors
of $t$ on $CL(T_t)$. 

As the number of types per node is at most $2^k$,
so the total number of types and hence the number of functions is
at most $2^k \cdot |V(T)|$. We consider only \emph{unsatisfiable} types ${\bf tp}$,
that is those for which $f_{\bf tp}$ is unsatisfiable. 
A detailed consideration is provided
in the next subsection. For now, we consider only a single case that conveys the idea
of the design of the local graphs. 

So, let $t \in V(T)$ be a non-leaf having two children $t_1$ and $t_2$ 
and let ${\bf tp}=(t,CL,S)$ be an unsatisfiable type such that all clauses in the range of $f_{\bf tp}$
are non-empty and, in addition, $Var(t) \setminus Var(S) \neq \emptyset$. 
In this case, the local subgraph $L_{\bf tp}$ corresponding to $tp$ is nothing else
then a decision tree over all possible assignments $A$ to $Var(t) \setminus Var(S)$
with the root corresponding to $tp$ and each leaf corresponding to a type of one of the children
of $t$. Let us consider more in detail how we choose this type. 

Each $i \in \{1,2\}$, ${\bf tp}$ and $A$ naturally \emph{induces}
a type ${\bf tp_i}=(t_i,CL_i,S_i)$, where $S_i=Proj(S \cup A, Var(t) \cap Var(t_i))$
and $CL_i$ consists of $C \cap CL(t_i)$ and, in addition, those clauses of
$CL(t_i)$ that are satisfied by $S \cup A$. It is not hard to see that
one of ${\bf tp_1},{\bf tp_2}$ must be unsatisfiable. The algorithm arbitrarily chooses
an unsatisfiable type ${\bf tp_i}$  and associates the vertex corresponding to $A$
with ${\bf tp_i}$. The other type is simply discarded even if it is also unsatisfiable.

Why do we discard the other type? The intuition is the following.
Imagine an unsatisfiable CNF $\psi$ consisting of two variable disjoint subCNFs $\psi_1,\psi_2$.
Clearly, one of these subCNFs must be unsatisfiable, let it be say $\psi_1$.
Then, in order to prove the unsatisfiability of $\psi$ using regular resolution, it is enough
to do so for $\psi_1$. The subCNF $\psi_2$ can be discarded even if it is unsatisfiable as well. 

\begin{example}
Let $\varphi=(x_1 \vee x_2) \wedge (\neg x_1 \vee x_3) \wedge (\neg x_1 \vee x_4) \wedge
(\neg x_3 \vee \neg x_4) \wedge (\neg x_2 \vee x_5) \wedge (\neg x_2 \vee x_6) \wedge (\neg x_5 \vee \neg x_6)$
Consider the incidence tree decomposition $(T,{\bf B})$ where $V(T)=\{t,t_1,t_2,t_3,t_4\}$.
We set $t$ to be the root with children $t_1$ and $t_2$, $t_3$ being the child of $t_1$ and $t_4$
being the child of $t_2$. 
Further on, we let $Var(t)=\{x_1,x_2)\}$, $CL(t)=\{(x_1 \vee x_2\}$,
$Var(t_1)=\{x_1,x_3,x_4\}$, $CL(t_1)=\{(\neg x_1 \vee x_3),(\neg x_1 \vee x_4)\}$,
$Var(t_2)=\{x_2,x_5,x_6\}$, $CL(t_2)=\{(\neg x_2 \vee x_5), (\neg x_2 \vee x_6)\}$,
$Var(t_3)=\{x_3,x_4\}$, $CL(t_3)=\{(\neg x_3 \vee \neg x_4)\}$,
$Var(t_4)=\{x_5,x_6\}$, $CL(t_4)=\{(\neg x_5 \vee \neg x_6)\}$.

Consider the type ${\bf tp}=(t, \emptyset, \emptyset)$.
It is not hard to see that $range(f_{\bf tp})$ is just $\varphi$.
Now let $A=\{\neg x_1,x_2\}$ and consider 
the types of the children of $t$ induced by ${\bf tp}$ and $A$.
These types are ${\bf tp_1}=(t_1,S_1,C_1)$ and ${\bf tp_2}=(t_2,S_2,C_2)$
where $S_1=\{\neg x_1\}$ and $C_1=\{(\neg x_1 \vee x_3), (\neg x_1 \vee x_4)\}$,
$S_2=\{x_2\}$, $C_2=\emptyset$. 

It follows that $dom(f_{\bf tp_1})=\{(\neg x_3 \vee \neg x_4)\}$ and the clause is
mapped to itself. In other words, $tp_1$ is a satisfiable type.
On the other hand, $dom(f_{\bf tp_2)}=\{(\neg x_2 \vee x_5), (\neg x_2 \vee x_6), (\neg x_5 \vee \neg x_6)\}$
and $f_{\bf tp_2}(\neg x_2 \vee x_5)=(x_5)$ \footnote{Note a slight abuse of not using the double brackets
for the sake of a better readability.}, 
$f_{\bf tp_2}(\neg x_2 \vee x_6)=(x_6)$, $f_{tp_2}(\neg x_5 \vee \neg x_6)=(\neg x_5 \vee \neg x_6)$. 
Thus $range(f_{\bf tp_2})=\{(x_5),(x_6),(\neg x_5 \vee \neg x_6)\}$ and clearly it forms an unsatisfiable CNF
thus ${\bf tp_2}$ is an unsatisfiable type. The branch of the local subgraph of ${\bf tp}$ corresponding to $S$
corresponds to ${\bf tp_2}$.

Note, however that the assignment $\{x_1,x_2\}$ induces two unsatisfiable types, one 
for each child and hence an arbitrary one of them can be chosen as the sink of the local
sugraph of ${\bf tp}$ corresponding to $\{x_1,x_2\}$. 
 \end{example}

Let us now discuss the general case where the set ${\bf TR}$ of transitional clauses
is not empty. In this case the types are of the form ${\bf tp}=(t,CL,S,ET)$ where
$ET \subset {\bf TR}$. The idea is that $ET$ is a subset of transitional clauses
that are falsified by assignment to variables located in the bags of the ancestors of $t$.
The function corresponding to ${\bf tp}$ is $f_{\bf tp} \cup trans_{ET}$ where $f_{\bf tp}$ (with a slight
abuse of notation) is the function corresponding to the shortened type $(t,CL,S)$ as described
above. Like in the case with ${\bf TR}=\emptyset$, the types are ordered and DAGs are gradually
built so that for each type ${\bf tp}$, the resulting DAG contains a subgraph which is a 
transitional resolution of $f_{\bf tp} \cup trans_{ET}$. Like in the case with ${\bf TR}=\emptyset$,
this DAG is constructed by adding a local subgraph for each type in the considered order.
Most of the cases of construction of local graphs are easy extensions of the case of ${\bf TR}=\emptyset$.
However, there is one case that is conceptually different. To understand it, consider the situation
below.

Suppose that a CNF $\varphi$ is a variable disjoint union of two unsatisfiable CNFs $\varphi_1$
and $\varphi_2$. Suppose next that both ${\bf TR} \cap \varphi_1$ and ${\bf TR} \cap \varphi_2$ are
non-empty. Because of the last assumption we cannot use just one of $\varphi_1$, $\varphi_2$ and
to discard the other one. Indeed, suppose we consider only $\varphi_1$ and discard $\varphi_2$.
If all the sinks of the resulting resolution are terminal ones, good and well. But what if we have
a transitional sink $u$? Consider an assignment $A$ carried by a path ending at $u$. Then $A$ is a
falsifier of some ${\bf C} \subseteq {\bf TR} \cap \varphi_1$ w.r.t. ${\bf 1}_{\varphi_1}$. However, $A$ is
\emph{not} such a falsifier w.r.t. the whole ${\bf 1}_{\varphi}$ as, by assumption, $A$ cannot be extended to an 
assignment satisfying all the clauses of $\varphi_2$! Put it differently, a transitional node needs 
information about \emph{all} the clauses of ${\bf TR}$ not just part of them.

Thus in the considered situation the resolution must be applied to a conjunction of $\varphi_1$ and
of $\varphi_2$. Obviously,  the resolution does not have conjunction gates, so we use the standard
trick of putting the resolution for say ${\bf 1}_{\varphi_1}$ 'on top' of  the one for ${\bf 1}_{\varphi_2}$. 
We postpone the nuances
of this arrangement to the stage of formal description and discuss here only the aspect of combinatorial
explosion. If the DAGs of resolutions for ${\bf 1}_{\varphi_1}$ and ${\bf 1}_{\varphi_2}$ have already been defined then
the 'on top' configuration requires creating a \emph{separate copy} of the resolution of ${\bf 1}_{\varphi_1}$.
Thus the local graph corresponding to ${\bf 1}_{\varphi}$ is of size comparable to the union of all the DAGs that
have been created before. We refer to this situation as \emph{doubling}. To mitigate its effect we, essentially
compare the quantities $|\varphi_1 \cap {\bf TR}|$ and $|\varphi_2 \cap {\bf TR}|$ and put on top 
the one whose respective quantity is smaller (ties can be broken arbitrarily). This way we ensure that
the sequence of nested doublings is of size at most $\log |{\bf TR}|$.  

\subsection{Proof of Theorem \ref{transres}: formal description}
\begin{definition} [{\bf Bag-related type}]
A \emph{(bag-related) type} ${\bf tp}$ 
is a quadruple $(t,S,CL,ET)$ 
where $t \in V(T)$, 
$S$ is an assignment with $Var(S) = Var(t) \cap Vat(t^*)$
where $t^*$ is the parent of $t$ ($Var(S)=\emptyset$ if $t$ is the root),
$CL \subseteq CL(t)$, and $ET \subseteq {\bf TR}$.
We may refer to the respective components
of ${\bf tp}$ by $t({\bf tp}), S({\bf tp}), CL({\bf tp}), ET({\bf tp})$. 
\end{definition}

\begin{definition} [{\bf Associated functions}]
The function $f_{\bf tp}$ \emph{associated} with ${\bf tp}$
is a $\varphi$-based function with the domain consisting of the clauses
of $CL(T_t)$ that do not belong to $CL \cup ET$ and are not satisfied by $S$. 
For each $C \in dom(f_{\bf tp})$,
$f_{\bf tp}(C)=Proj(C,Var(T_t) \setminus Var(S))$. 

We also denote the function $f_{tp} \cup trans_{ET}$ by $h_{tp}$
\end{definition}

\begin{definition} [{\bf Unsatisfiable and basic types}]
A type ${\bf tp}$ is \emph{unsatisfiable}
if $h_{\bf tp}$ is unsatisfiable. 
An unsatisfiable type ${\bf tp}$ is \emph{basic}
if either (i)  there is a non-transitional clause $C \in dom(f_{\bf tp})$ such 
that $f_{\bf tp}(C)=()$ or
(ii) $f_{\bf tp}$ is satisfiable. 
In case (i), ${\bf tp}$ is a \emph{non-transitional} basic type
with $C$ being a \emph{witnessing} clause 
and in case (ii) it is a \emph{transitional} one. 
\end{definition}


\begin{definition} [{\bf Extension for a type}]
Let ${\bf tp}$ be a type with $t=t({\bf tp})$. An extension $S'$ 
of ${\bf tp}$ is an assignment to $Var(t) \setminus Var(S)$.
We denote the set of all extensions of ${\bf tp}$ by 
${\bf Ext}({\bf tp})$. In particular, if $Var(t) \setminus Var(S)=\emptyset$
then ${\bf Ext}({\bf tp})=\{\emptyset\}$.  
\end{definition}

\begin{definition} \label{defsucc1} [{\bf Successor of a type}]
Let ${\bf tp}$ be a non-basic unsatisfiable type such that $t=t({\bf tp})$
is not a leaf of $T$. Let $A \in {\bf Ext}({\bf tp})$
and let $t_1$ be a child of $t$.
The $(t_1,A)$-successor of ${\bf tp}$ is a
type ${\bf tp_1}$ such that 
$t({\bf tp})=t_1$, 
$S({\bf tp_1})=Proj(S \cup A,Var(t_1))$,
$CL({\bf tp_1})$ consists of $CL({\bf tp}) \cap CL(t_1)$
and all $C \in CL(t_1)$  such that $C$ is satisfied by $S({\bf tp}) \cup A$,
$ET({\bf tp_1}) $ is the union of $ET({\bf tp})$ and 
the set clauses $C \in dom(f_{\bf tp}) \cap {\bf TR}$
such that $f_{\bf tp}(C)$ is falsified by $A$. 
\end{definition}

If $t$ has two children $t_1$ and $t_2$ then each extension $A$ of
${\bf tp}$ identifies the $(t_1,A)$ and $(t_2,A)$-successors of ${\bf tp}$. 
In most cases, during the construction of the transitional resolution, 
one of them will be chosen as the preferred successor.
making the resulting construction similar to the case with ${\bf TR}=\emptyset$.
However, there is one particular case where the preferred successor cannot be chosen. 
We define this case through the notion of a \emph{special} pair. 

\begin{definition} [{\bf Special pair}]
Let ${\bf tp}$ be a non-basic unsatisfiable type such that $t=t({\bf tp})$
has two children $t_1$ and $t_2$.
Let $A \in {\bf Ext}({\bf tp})$. 
Then $({\bf tp},A)$ is a \emph{special pair}
if (i) both $(t_1,A)$ and $(t_2,A)$-successors of ${\bf tp}$ are unsatisfiable
and nonbasic and (ii) both $CL(T_{t_1}) \cap {\bf TR}$ and $CL(T_{t_2}) \cap {\bf TR}$ are
non-empty. 
\end{definition}

We also need two concepts related to special pairs: those of the main child and the main 
successor. 

\begin{definition} [{\bf The main child}]
Let $t$ be a node of $T$ having two children and let $t_1$ and $t_2$
be the children of $t$. Assume that
$|CL(T_{t_1}) \cap {\bf TRANS}|<|CL(T_{t_1}) \cap {\bf TRANS}|$.
Then we say that $t_1$ is the \emph{main child} of $t$.
If both intersections are of the same size then we arbitrarily
fix a child of $t$ and name it the main child of $t$.
\end{definition}

\begin{definition} [{\bf The main and secondary sucessors.}]
Let $({\bf tp},A)$ be a special pair and let $t=t({\bf tp})$.
Then clearly, $t$ has two children $t_1$ and $t_2$. 
Let ${\bf tp_1}$ and ${\bf tp_2}$ be the respective $(t_1,A)$ and $(t_2,A)$ successors
of ${\bf tp}$. Assume w.l.o.g that 
$t_1$ is the main child of $t$.
Then we refer to ${\bf tp_1}$ and ${\bf tp_2}$ as, respectively,
the \emph{main} and \emph{secondary} successors of $({\bf tp},A)$. 
\end{definition}

Now, we establish a linear order on the union of the set of unsatisfiable non-basic types
and the set of special pairs. 

\begin{definition} \label{deford}
Let $\mathcal{NS}$ be the set of all unsatisfiable types and let $\mathcal{SP}$
be the set of all special pairs. We fix a linear order $\mathcal{ORD}$ satisfying 
the constraints as specified below.

For $r \in \mathcal{NS} \cup \mathcal{SP}$, the type ${\bf tp}(r)$ is defined as follows.
If $r={\bf tp}$ then ${\bf tp}(r)={\bf tp}$ and if $r=({\bf tp},A)$ then ${\bf tp}(r)={\bf tp}$.
Let $r_1,r_2 \in \mathcal{NS} \cup \mathcal{SP}$, let ${\bf tp_1}={\bf tp}(r_1)$,
${\bf tp_2}={\bf tp}(r_2)$, and let $t_i=t({\bf tp_i})$ for $i \in \{1,2\}$.
Then $r_1<r_2$ is mandatory in the following two cases
\begin{enumerate}
\item $t_1$ is a descendant of $t_2$.
\item $t_1=t_2$ but $r_1$ is a special pair while $r_2$ is a type. 
\end{enumerate}
\end{definition}

The above definition is somewhat tedious because of the need to consider
types and special pairs within the same set. The idea, however, is simple.
We first look at the nodes corresponding to $r_1$ and $r_2$ and if $t_1$ is an ancestor
of $t_2$ in $T$ then $r_1$ is ahead of $r_2$.
In the other case, $r_1$ and $r_2$ are associated with the same type. In this case $r_1$ must precede
$r_2$ if $r_1$ is a special pair while $r_2$ is not. Otherwise, the ordering is arbitrary. 

\begin{definition}  [{\bf Local subgraphs for elements of $\mathcal{ORD}$}] \label{defconstr}
We define a graph $R$ as the union of \emph{local graphs} $L_r$ for each $r \in \mathcal{ORD}$.
The graphs $L_r$ are defined inductively along $\mathcal{ORD}$. In particular, when we define
$L_r$ for some $r \in \mathcal{ORD}$, we assume that for all $r'<r$, $L_{r'}$ have been defined. 
The set of vertices of each $L_r$ is disjoint with $\bigcup_{r'<r} V(L_{r'})$ except some sinks 
that may be identified with the sources of previous $L_{r'}$. 

Assume first that $r={\bf tp}$.
$L_{\bf tp}$ is a single source DAG whose source is $u_{\bf tp}$. 
If $Var(t) \setminus Var(S)=\emptyset$
Then $u_{\bf tp}$ is the only node of $L_{\bf tp}$
and the assignment $A=A(u_{\bf tp})$ corresponding to it is $\emptyset$.
Otherwise, $L_{\bf tp}$ is a decision tree (with the root $u_{\bf tp}$).
Let $u$ be a sink of the $L_{\bf tp}$. The assignment $A=A(u)$ corresponding to $u$
is $A(P)$ where $P$ is the $u_{\bf tp}$ to $u$ path of $L_{\bf tp}$. 

Now we make a decision regarding the sinks $u$ of $L_{\bf tp}$.
The decision is specified in the list below. Each item in the list specifies
a condition and the decision made in case this condition holds. For each item
but the first one, we assume that the conditions of the previous items
do not hold. 

We denote $A(u)$ by $A$.
The children of $t$ (if any) are denoted by $t_1$ and $t_2$, if $t$ has only
one child then it is $t_1$. Their respective successors of ${\bf tp}$ through
$A$ are denoted by ${\bf tp_1}$ and ${\bf tp_2}$. 
 
\begin{enumerate}
\item $A$ falsifies $f_{\bf tp}(C)$ for some $C \in dom(f_{\bf tp}) \setminus {\bf TR}$. 
In this case $u$ is labelled with $C \rightarrow ()$. If there are several such $C$ choose
one arbitrarily.
\item $A$ falsifies $f_{\bf tp}({\bf C})$ for ${\bf C} \subseteq dom(f_{\bf tp}) \cap {\bf TR}$
while $(f_{\bf tp}|_A) \setminus {\bf C}$ is satisfiable. In this case, $u$
is labelled with $trans_{ET({\bf tp}) \cup {\bf C}}$.
\item $({\bf tp},A)$ is a special pair. Then $u$ is identified with the source of $L_{({\bf tp},A)}$.
\item $t$ has only one child. Then $u$ is identified with the source of $L_{\bf tp_1}$. 
\item Satisfiable sibling case: $t$ has two children and, say $f_{\bf tp_1}$ is satisfiable.
Then $u$ is identified with the source of $L_{\bf tp_2}$.
\item Non-transitional child case:  $t$ has two children and, say $dom(f_{\bf tp_1}) \cap {\bf TR}=\emptyset$.
Then $u$ is identified with the source of $L_{\bf tp_1}$. 
\end{enumerate}

In the last three cases, $u$ is identified with the source of the local graph of one
successor of ${\bf tp}$ through $A$. We refer to this successor as the {\bf preferred successor}
of ${\bf tp}$ through $A$. 

Now, we assume that $r=({\bf tp},A)$, where
$({\bf tp},A)$ is a special pair. 
Let $t=t({\bf tp})$, $t_1,t_2$  be the children of $t$,
${\bf tp_1}$, ${\bf tp_2}$ be the respective successors through $A$
with ${\bf tp_1}$ being the main successor.
Let $Q$ be a transitional resolution for $h_{\bf tp_1}$
such that $Var(Q) \subseteq Var(range(h_{\bf tp_1}))$.
Then $L_{({\bf tp},A)}$ is obtained from $Q$ by the follwing
modification of the transitional sinks.
Let $u$ be a sink of $Q$ labelled with $trans_{\bf C}$.
Then, in $L_{({\bf tp},A)}$, $u$ is identified with the source of 
$L_{\bf tp^*}$ where $ET({\bf tp^*})=ET({\bf tp_2}) \cup {\bf C}$ 
and the rest of the components remain the same. 

\end{definition}

\begin{remark}
Note that if $t=t({\bf tp})$ is a leaf then any extension of ${\bf tp}$
assigns all the variables of $range(f_{\bf tp})$, therefore, the first or the second
case must occur.

Note also that for the definition to be valid, whenever a sink of a local graph is identified
with the source of local graph of another type, it needs to be established 
that this type is an element of $\mathcal{ORD}$. In the correctness proof in the appendix, we show that this is 
indeed the case.

\end{remark}

{\bf Proof sketch of Theorem \ref{transres}.}
The proof consists of two stages. 
First we establish correctness of the construction as per Definition \ref{defconstr}
then we establish an upper bound on its size. 

For the correctness, we let $R=\bigcup_{r \in \mathcal{ORD}} L_r$. 
Then we show that for each ${\bf tp} \in \mathcal{NS}$, $R_{u_{\bf tp}}$ is a transitional
resolution for $h_{\bf tp}$ and for each $({\bf tp},A) \in \mathcal{SP}$,
$R_{({\bf tp},A)}$ is a transitional resolution for $h_{\bf tp}|_A$. 
In particular, this is so for the \emph{starting type}
${\bf st}=(rt,\emptyset,\emptyset,\emptyset)$, where $rt$ is the root of $T$. 
We observe that $h_{\bf st}={\bf 1}_{\varphi}$. 
Hence, we conclude that $R_{\bf st}$ is the required transitional resolution for $\varphi$.
The proof uses the machinery developed in Section A of the appendix enabling a 'piecewise'
construction of a transitional resolution. In particular, given a sequence of 'local' CDTs satisfying
certain criteria, it is shown that their union is a transitional resolution for a certain function. 
Because of this framework, in the actual proof of Theorem \ref{transres} (Section B of the Appendix), 
we only need to show that the graphs $L_{\bf tp}$ and $L_{({\bf tp},A)}$ satisfy 
the required criteria. The proof is mostly straightforward checking of conditions, the most
interesting aspect is the use the properties of tree decompositions and one-sidedness.

To upper-bound the size of $R$, we essentially, upper bound the sum of sizes of local graphs. 
We observe
an immediate difficulty: there is uncertainty in the definition of the local graph $L_{({\bf tp},A)}$.
Definition \ref{defconstr} says that is is a transitional resolution for $h_{\bf tp_1}$ (where ${\bf tp_1}$
is the main successor of ${\bf tp}$ through $A$) 
subject to a certain variable constraint, but, beyond this, no specification is provided and hence
a size upper bound is not clear. We know, however that $R_{u_{\bf tp_1}}$ is a transitional
resolution for $h_{\bf tp_1}$. So we can simply take a copy of $R_{u_{\bf tp_1}}$ 
as $L_{({\bf tp},A)}$. This, however, immediately creates another problem: we take a 'whole'
graph $R_{u_{\bf tp_1}}$ to serve as a local graph, the total of the sizes essentially 'doubles'
and this may lead to exponential explosion. Demonstration that this exponential explosion
is controlled by parameters requires a more careful view.

First of all, we observe that the number of nodes $t$ such that $t=t({\bf tp})$ and 
$({\bf tp},A)$ is a special pair is at most $|{\bf TR}|$. This means that the number $sp$
of special pairs is upper bounded by a function of $k$ and $|{\bf TR}|$.
Next, we define a rank of an element of $\mathcal{ORD}$.
For this, we first say that if a sink $L_{r_1}$ is identified with $u_{r_2}$ then $r_2$
is a child of $r_1$. The notion of a child naturally leads to the notion of a descendant.
With this in mind the rank is defined as follows. The rank of ${\bf tp}$ is the maximum
over ranks of its descendants (if ${\bf tp}$ has no descendants then the rank is zero).
The rank of a special pair $({\bf tp},A)$ is the rank of ${\bf tp_1}$ plus one where ${\bf tp_1}$
is the main successor of ${\bf tp}$ through $A$. 

We prove that the rank of an element of $\mathcal{ORD}$ is at most $max(0,\log(|{\bf TR}|))$.
Then we show that the local graph size for a special pair of rank $i$
is upper bounded by $n\cdot sp^i \cdot 2^{O(k+|{\bf TR}|)}$ where $n=|Var(\varphi)|+|\varphi|$.
We conclude that, since the rank is upper bounded by $\log(|{\bf TR}|)$ and the number of elements of $\mathcal{ORD}$
is upper bounded by $2^{O(k+|{\bf TR}|)}$, the required upper bound follows.
$\blacksquare$

\section{The main result}

\begin{theorem} \label{mainres}
Let $\varphi$ be a CNF, ${\bf LONG} \subseteq \varphi$ to
which we refer as \emph{long clauses} and $k>0$ be an integer.
Assume that $\varphi \setminus {\bf LONG}$ has a one-sisded 
tree decomposition $(T,{\bf B})$ of the incidence graph of width
at most $k$ (due to Proposition \ref{onesidebinary}, we assume that 
$T$ is a binary tree).  Then there is a resolution of $\varphi$ 
of size $(n+|{\bf LONG}|^{|{\bf LONG}|}) \cdot n 
\cdot 2^{O(k^2+k \cdot |{\bf LONG}|)}$. 
\end{theorem}

In the rest of this section, we introduce definitions towards the proof of Theorem \ref{mainres}
and provide a sketch of the proof. The complete proof is available in Section C of the Appendix.
Formal statements in this section alternate with fragments of informal discussion.
For a reader interested in the formal reasoning only, the informal discussion can be safely omitted.

Similarly to the proof of Theorem \ref{transres}, the resolution is constructed in a piecewise
manner as a union of local graphs of types. However, the types are defined in a different manner.
In particular, while for the proof of Theorem \ref{transres}, the types were associated with nodes 
of the underlying search tree, in the considered case, the types are associated with prefixes of $\pi_T$.
Before we continue the discussion, let us define the types and the corresponding 
classification of variables.

\begin{definition} [{\bf Type}]
A type ${\bf tp}$ is a quadruple 
$(\pi,map,CN,RA)$ 
where $\pi$ is a (possibly empty) prefix of $\pi_T$ and 
$map$ is function from ${\bf LONG}$ to $Trees_{\pi} \cup \{none\}$ 
To define the remaining two components, we need to have a more
detailed look at the domain and range of $map$. 
So, let ${\bf LS}$  be the subset of ${\bf LONG}$ that are \emph{not} mapped
to $none$  and let ${\bf MT}=map({\bf LS})$.
Then $CN$ is a subset of $CL(Roots({\bf MT}))$ and 
$RA$ is an assignment to $Var(Roots({\bf MT}))$. 

When it may be not clear from the context which type the above structures
are related to, we prove the name of the type in the brackets, for example,
$\pi({\bf tp})$, ${\bf MT}({\bf tp})$ and so on. 
\end{definition}

\begin{definition} [{\bf Inner, fixed, and outer variables}] \label{innfixout}
Let ${\bf tp}$ be a type.
We call $Var({\bf MT})$ the \emph{inner variables} of ${\bf tp}$ and denote the set
by $Var_{in}$.
Let $x \in Var(\pi) \setminus Var_{in}$.
We say that $x$ is \emph{fixed} if there is a long clause $C$ such that $x \in Var(C)$
and one of the following holds:
(i) $map(C)=none$ or $map(C) \in Trees_{\pi}$ and $MT_{\pi}(x)<map(C)$
where $MT_{\pi}(x)$ is the smallest tree $T'$ of $Trees_{\pi}$ such that
$x \in Var(T')$. 
We call $C$ a \emph{witnessing clause} of $x$. If there are several
clauses matching the definition, pick an abitrary one. 
We denote by $Var_{fix}$ the set of all fixed variables.
For each $x \in Var_{fix}$, the \emph{fixed assignment} of $x$ is the
literal opposite to its occurrence in the wintessing clause $C$.
We denote by $FA$ the set of all the fixed literals.  
Finally the variables of $Var(\varphi) \setminus (Var_{fix} \cup Var_{in})$
are called \emph{outer} variables and their set is denoted by $Var_{out}$.
\end{definition}

In order to interpret the above definitions, we need to note that,
in the resulting resolution, each type ${\bf tp}$ (reachable from the source)
is associated with a node $u_{\bf tp}$.  The types describe invariant properties of the 
assignments $A$ carried by the paths from the source to $u_{\bf tp}$. 
For example, the long clauses satisfied by $A$ are ${\bf LS}({\bf tp})$. 
In this case, $map(C)$ is the smallest tree $T'$  of $Trees(\pi)$ 
such that $Var(T') \cap Var(A \cap C) \neq \emptyset$ or, to put it differently, 
the minimal tree containing a variable whose occurrence in $A$ satisfies $C$. 
Then ${\bf MT}$ is the set of such minimal trees. To continue the discussion,
let us first define the function of the type. 

\begin{definition} [{\bf The type function}]
For a type ${\bf tp}$,
$F_{\bf tp}$ is a 
$\varphi$-based function.
For each $C \in dom(F_{\bf tp})$ 
$F_{\bf tp}(C)=Proj(C,V_{out})$. 

The domain of $F_{\bf tp}$ 
$C \in \varphi$ such that one of the following three conditions hold:
(i) $C \in {\bf LONG} \setminus {\bf LS}$; (ii) $C \in CN$; (iii) 
$C \notin CL(Roots({\bf MT}))$ and 
$Var(C) \cap Var_{out} \neq \emptyset$
and $C$ is not satisfied by $RA \cup FA$.
\end{definition}

Thus the assignment $A$ as above does not assign $Var_{out}({\bf tp})$. 
$Var_{fix}({\bf tp})$ occur in $A$ with assignment $FA$,
$Var(Roots({\bf MT}))$ occur with assignment $RA$ but what about the 
rest of the variables of$Var_{in}$?  To understand this, note that variables
of $Var_{in}$ 'communicate' with the rest of the variables through the vaiables
of the roots of ${\bf MT}$ and through the clauses having variables both
inside $Var_{in}$ and outside $Var_{in}$ (let us call them 'connecting' clauses
for the sake of the argument). In particular let $B_1$ and $B_2$ be two assignments
to $Var_{in}$ that do not falsify any clause, have the same projection to
$Var(Roots({bf MT}))$, and satisfy the same set of connecting clauses.
Then it can be shown that $\varphi|_{B_1}=\varphi|_{B_2}$. 
The meaning of the component $CN$ of ${\bf tp}$ is exactly the set of connecting clauses 
that are not satisfied by $A$. So, the assignment to the roots of ${\bf MT}$ plus $CN$
do provide a complete description and, subject to this invariant, the assignment
to the rest of the variables of $Var_{in}$ may be arbitrary!

\begin{definition} [{\bf Basic, final, and unsatisfiable types.}]
A type ${\bf tp}$ is \emph{unsatisfiable} if
$F_{\bf tp}$ is unsatisfiable. 
An type ${\bf tp}$ is \emph{basic}
if there is $C \in dom(F_{\bf tp})$ such that
$F_{\bf tp}(C)=()$. 
A type ${\bf tp}$ is \emph{final} if $\pi({\bf tp})=\pi_T$
\end{definition}

Now we start defining local subgraphs.
We will consider only non-basic types. 
One useful observation is that for a long clause $C \in dom(F_{\bf tp})$,
$Var(F_{\bf tp}(C))$ does not intersect with $Var(\pi)$. 
Indeed, by definition, $map(C)=none$. Therefore, any $x \in (Var(\pi) \setminus Var_{in}) \cap Var(C)$
must belong to $Var_{fix}$ and cannot be in $Var_{out}$. 
This immediately simplifies construcrtion of local graphs for the final types.
Indeed, if ${\bf tp}$ is final then $Var_{out}({\bf tp}) \subseteq Var(\pi)$. Since we assumed
that ${\bf tp}$ is non-basic $dom(F_{\bf tp})$ does not contain long clauses.  
In other words, $F_{\bf tp}$ is a $\varphi \setminus {\bf LONG}$-based function
and hence there is an FPT size falsifying FRR for $F_{\bf tp}$ by Corollary \ref{col:transres}.

For a non-final type ${\bf tp}$, the construction of the local subgraph depends on whether
the immediate successor $t$ of $\pi({\bf tp})$ is an expanding or an contracting node. 
If $t$ is expanding then the local graph is effectively a decision tree with some nuances
related to definition of sinks. The case when $t$ is contracting is more involved.
In order to informally discuss it, we need to introduce several additional definitions.

\begin{definition}
A linear order $\mathcal{ORD}$ is an arbitrary but fixed
ordering over unsatisfiable types obeying the following constraint.
Suppose that ${\bf tp_1}$ and ${\bf tp_2}$ are unsatisfiable types
such that $\pi({\bf tp_1})>\pi({\bf tp_2})$. Then ${\bf tp_1}<{\bf tp_2}$ in $\mathcal{ORD}$.  
\end{definition}

\begin{definition} \label{def:establ}  
Let ${\bf tp}$ be a non-basic non-final unsatisfiable type and let $t$ be 
the immediate successor of $\pi({\bf tp})$. 
We say that $t$ is \emph{established} if $t$ is a contracting node
and at least one element of ${\bf MT}({\bf tp})$ is a subtree of
$last(Trees_{\pi({\bf tp})+t})$.  
\end{definition}

\begin{definition}
With the data as in Definition \ref{def:establ},
a variable $x \in Var(t) \cap Var_{out}$ is \emph{determining}
if there is a clause $C \in {\bf LONG} \setminus {\bf LS}$ 
such that $x \in Var(C)$. We denote by $Det({\bf tp})$
the set of all determining variables, omitting the brackets if the
type is clear from the description. 
\end{definition}

\begin{definition}
Let ${\bf tp}$ be a non-basic non-final unsatisfiable type and $t$ be the immediate successor of $\pi({\bf tp})$.
The \emph{dome} $D_{\bf tp}$ of ${\bf tp}$ is defined as follows.
\begin{itemize}
\item If $t$ is established then $D_{\bf tp}$ is a single node if $Var(t) \cap Var_{out}=\emptyset$.
Otherwise, $D_{\bf tp}$ is a decision tree over all the assignments to $Var(t)  \cap Var_{out}$.
\item If $t$ is non-established then $D_{\bf tp}$ is a single node if $Det=\emptyset$. 
Otherwise, $D_{\bf tp}$ is constructed
in two stages. On the first stage, we define $D^*$ as a decision tree over all assignments
of $Det$. If $Var(t) \cap Var_{out}=Det$ then $D_{\bf tp}=D^*$. Otherwise, for each sink $u$ of $D^*$
such that the assignment $A$ carried by the path from the root of $D^*$ to $u$ satisfies at least one $C \in {\bf LONG} \setminus {\bf LS}$,
make $u$ the root of a decision tree over all the assignments to $(Var(t) \cap Var_{out}) \setminus Det$. 
\end{itemize}
\end{definition}

\begin{definition}
Let ${\bf tp}$ be a non-basic non-final unsatisfiable type and $t$ be the immediate successor of $\pi({\bf tp})$.
Assume that $t$ is contracting and let $A$ be an assignment to $Var(t) \cap Var_{out}$
that does not falsify any clause in $range(F_{\bf tp})$. 
Then we say that $A$ is \emph{potentially complex} if one of the following two conditions
holds. 
\begin{itemize}
\item $t$ is established.
\item $A$ satisfies a clause $C \in {\bf LONG} \setminus {\bf LS}$.
\end{itemize}
\end{definition}

The local subgraph of any type ${\bf tp}$ of $\mathcal{ORD}$ includes the dome
$D_{\bf tp}$ plus subgraphs whose sources are sinks of $D_{\bf tp}$ 
Why is not a a sink $u$ of $D({\bf tp})$ not necessarily a sink of the whole
local subgraph $L_{\bf tp}$ of ${\bf tp}$? This may happen when the assignment $A$ carried by the
path from the source of $D_{\bf tp}$ to $u$ is potentially complex. In this case
$T_t$ is going to be a minimal tree in any successor ${\bf tp'}$ of ${\bf tp}$ identified with a sink
of $L_{\bf tp}$ reachable from $u$, meaning that all the variables of $T_t$ will be inner for ${\bf tp'}$.
Yet, $T_t$ may include variables of $Var_{out}({\bf tp})$ that are not yet assigned 
after assgning the variables of $A$. These are precisely the variables taken care by $[L_{\bf tp}]_u$.
To see how this is done, we need one more definition. 

\begin{definition} \label{def:cavity} [{\bf Filling function and its transitional resolution}]
Let ${\bf tp}$ be a non-basic non-final unsatisfiable type and $t$ be the immediate successor of $\pi({\bf tp})$.
Let $A$ be a potentially complex assignment.
For the sake of brevity, let us denote $last(Trees_{\pi({\bf tp})+t})$ by $T^*$.  
Denote the set $Var_{out} \cap Var(T^*) \setminus Var(t)$ by $Var_{free}$ and refer
to them as \emph{free variables}. 

The \emph{filling function} $F^*$ (${\bf tp}$ and $A$ may be added in brackets if
not clear from the context) is a $\varphi$-based function such that for each 
$C \in dom(F^*)$, $F^*(C)=Proj(C,Var_{free})$.
The domain of $F^*$ consists of all clauses $C$ of $dom(F_{\bf tp})$ such that
$C$ is not satisfied by $A$ and $Var(F(C)) \cap Var_{free} \neq \emptyset$. 

If $range(F^*)$ is unsatisfiable, let $R^*$ be a transitional resolution
of $F^*$ with the transitional clauses being $dom(F^*) \cap CL(t)$
and with the extra
constraint that $Var(R^*) \subseteq Var_{free}({\bf tp})$.
\end{definition}

Continuing the informal discussion, we introduce a subgraph of $L_{\bf tp}$
with $u$ being the source if the corresponding filling function $F^*$ is unsatisfiable.
This subgraph is nothing but the transitional resolution $R^*$ for $F^*$
as defined above. Earlier types of $\mathcal{ORD}$ are identified with the transitional sinks of $R^*$.
The precise construction is provided in the definition below. 
It is worth noting that due to the same reason as for the final types, $dom(F^*)$ does not
include long clauses, hence $R^*$ can be of FPT size in $k$.

\begin{definition} \label{defloc}
The \emph{local} graph $L_{\bf tp}$ for each unsatisfiable type ${\bf tp}$
is constructed recursively along $\mathcal{ORD}$.
All the nodes of $L_{\bf tp}$, except sinks, are unique for ${\bf tp}$ in the sense that they
do not occur in the earlier types. The sinks may be identified with sources of local
graphs for earlier types.

If ${\bf tp}$ is final then $L_{\bf tp}$ is a falsifying FRR for $F_{\bf tp}$. 
In the rest of the definition, we assume that ${\bf tp}$ is non-basic and non-final.

The dome $D_{\bf tp}$ is a subgraph of $L_{\bf tp}$
and the source of $D_{\bf tp}$ is the source of $L_{\bf tp}$. 
$L_{\bf tp}$ is obtained from $D_{\bf tp}$ by processing of each sink and making one
of the following three decisions (i) associating the sink with some $C \rightarrow ()$
or (ii) identifying the sink with the source of the local graph of an earlier type or
(iii) deciding that the considered sink $u$ of $D_{\bf tp}$ is not a sink of $L_{\bf tp}$ and constructing
the graph $[L_{\bf tp}]_u$. The detailed construction is specified below.

Let $t$ be the immediate successor of $\pi({\bf tp})$. Let $u$ be a sink of $D_{\bf tp}$.
Then $A(u)$, the \emph{assignment corresponding} to $u$ is $\emptyset$ if $u$ is the only node 
of $D_{\bf tp}$. Otherwise, $D_{\bf tp}$ is a decision tree and $u$ is its leaf. In this case,
$A(u)$ is $A(P)$ where $P$ is the root-$u$ path of $D_{\bf tp}$.

So, let $u$ be the considered sink of $D_{\bf tp}$.
Suppose that $A(u)$ falsifies $F_{\bf tp}(C)$ for some
$C \in dom(F_{\bf tp})$. Then $u$ is associated with $C \rightarrow()$.

Otherwise, we consider separately the cases where $t$ is an expanding node
and where $t$ is a contracting node. 
Suppose first that $t$ is an expanding node. 
If $A(u)$ does not satisfy any clause of ${\bf LONG} \setminus {\bf LS}$
then $u$ is identified with source of $L_{\bf tp'}$ where  ${\bf tp'}=(\pi({\bf tp})+t,map,CN',RA)$
where $CN'$ is obtained from $CN$ by removal of clauses that are satisfied by $A(u)$.
Otherwise, $u$ is identified with the source of $L_{\bf tp'}$ where ${\bf tp'}=(\pi({\bf tp})+t,map',CN^*,RA')$
where $map'$ is obtained from $map$ by replacing $C \rightarrow none$ with $C \rightarrow last(Trees_{\pi({\bf tp}+t)})$
for each $C \in {\bf LONG} \setminus {\bf LS}$ that is satisfied by $A(u)$.
$CN^*$  s obtained from $CN$ by removal of all clauses that are satisfied by $A(u)$
and adding those clauses of $dom(F_{\bf tp}) \cap CL(t)$ that are not satisfied by $A(u)$.
Finally $RA'=RA \cup A(u) \cup Proj(FA({\bf tp}), Var(t))$.

Assume now that $t$ is a contracting node.
If $A(u)$ is not potentially complex then $u$ is 
associated with the type $(\pi({\bf tp})+t,map,CN',RA)$
where $CN'$ is obtained from $CN$ by removal of clauses that are satisfied by $A(u)$.

The most interesting case though is where $A(u)$ is potentially complex.
For this case we need to introduce additional notations.

Let $CN_1$ be the subset of $CN$ that are not satisfied by $A(u)$.
Let $Trees^*=({\bf MT}({\bf tp}) \setminus Trees'_{\pi}) \cup \{T_t\}$.
Let $CN_2$ be the subset of $dom(F_{\bf tp}) \setminus ({\bf LONG} \cup CL(Roots({\bf MT})))$
consisting of all clauses $C$ such that $C$ is not satisfied by $A(u)$ and 
$C \in CL(t)$ and $Var(F_{\bf tp}(C)) \cap Var_{free}({\bf tp})=\emptyset$. 

Let $map'$ be a function from ${\bf LONG}$ to $Trees_{\pi+t}$ obtained from $map$
as follows.
\begin{itemize}
\item For each $C \in {\bf LONG}$ such that $map(C)=none$ and $C$ is satisfied by $A(u)$,
$map'(C)=T_t$.
\item For each $C \in {\bf LONG}$ such that $map(C) \in Trees'_{\pi}$,
           $map'(C)=T_t$.
\end{itemize}
Let $RA'=Proj(RA,Var(Roots(Trees^*))) \cup Proj(FA,Var(t)) \cup A(u)$.



Assume first that $F^*$ is satisfiable.
Then $u$ is identified with the source of $L_{\bf tp'}$ where ${\bf tp'}=(\pi+t,map',CN_1 \cup CN_2,RA')$.


Finally assume that $F^*$ is not satisfiable. Let
$R^*$ be the transitional resolution for $F^*$ as per Definition \ref{def:cavity}.
Identify $u$ with the source of $R^*$. Then for each transitional sink $v$ of $R^*$
do the following. Let $trans_{\bf C}$ be the function associated with $v$.
Then, identify $v$ with the source of $L_{\bf tp'}$, where ${\bf tp''}=(\pi+t,map',CN_1 \cup CN_2 \cup {\bf C},RA')$
\end{definition}

{\bf Proof sketch of Theorem \ref{mainres}.}
We start the proof from defining a well-formed type.
Let $x$ be a fixed variable for ${\bf tp}$. Recall 
that for $x$ we define the witnessing clause $C$ for $x$
and choose $C$ arbitrarily if there are several candidates. 
The type ${\bf tp}$ is well formed if $x$ has the same occurrence
in all the candidates, hence it does not matter which candidate
we choose. We did not introduce the definition in the main body 
of the paper to improve readability. The proof of Theorem \ref{mainres}
is applied not to $\mathcal{ORD}$ but rather to its subsequence $\mathcal{ORD^*}$
consisting of all the well-formed types. 

Let $R$ be the union of all the local subgraphs of the elements of $\mathcal{ORD^*}$.
For each ${\bf tp} \in \mathcal{ORD^*}$, let $u_{\bf tp}$ be the source of $L_{\bf tp}$.
We prove that each $R_{u_{\bf tp}}$ is a falsfying FRR for $F_{\bf tp}$.
The proof uses Theorem \ref{assemble1} provided in Appendix A. 
In order to apply the theorem, we need to demonstrate for each ${\bf tp} \in \mathcal{ORD^*}$
that $L_{\bf tp}$ is valid in the following sense. First, whenever a sink $u$ of $L_{\bf tp}$ is labelled
with $C \rightarrow ()$ for each path $P$ from $u_{\bf tp}$ to $u$, $A(P)$ falsifies $F_{\bf tp}(C)$.
Second, whenever a sink $u$ of $L_{\bf tp}$ is identified with an earlier type ${\bf tp'}$ 
then for each path $P$ of $L_{\bf tp}$ from $u_{\bf tp}$ to $u$, $F_{\bf tp'}$ is a subfunction of $F_{\bf tp}|_{A(P)}$
and ${\bf tp'} \in \mathcal{ORD^*}$ (that is ${\bf tp'}$ is well-formed, non-basic, an unsatisfiable).
Next, we observe that there is one particular type ${\bf stp}=(\emptyset,\emptyset,\emptyset,\emptyset)$ 
(the first $\emptyset$ denotes the empty prefix) such that $F_{\bf tp}={\bf 1}_{\varphi}$.
(That is, $R_{u_{\bf st}}$ is, in fact a regular resolution for $\varphi$ by Proposition \ref{transfer1}.) 
This completes the correctness proof and it remains to establish the upper bund on
the construction size.

The upper bound can be obtained by multiplying an upper bound 
on the size of a local subgraph by the number
of types.  The most critical part in assessing the local subgraph
size is that it may include a falsifying FRR. However, as we explained 
before, the resolution is always an FPT size in $k$ since the domain of the 
corresponding function does not contain long clauses. 
For the number of types, the number of possible first components 
is $O(n)$. With the first component fixed, by Proposition \ref{logtrees} the number of possible second
components is $O(\log n^{|{\bf LONG}|})$, which is well known to be FPT. 
With the first two components fixed, the number of possible third and forth
components is at most $2^{k \cdot |{\bf LONG}|}$. 
$\blacksquare$

\section{Conclusion}
We have proved that regular resolution is
FPT for CNFs whose one-sided treewidth is almost bounded. 
We have also demonstrated how a resolution under a more restricted
parameterization can serve as a building block for the construction
towards the main result, increasing by one the degree of $n$.
We believe that using this approach FPT algorithms can be defined
for more and more general parameters at the price of higher and higher 
degree of the polynomial dependence on $n$. The main question however
is: can we reach this way the general case of bounded incidence treewidth?
We believe that the answer is no and that an XP lower bound needs to be
sought at least in the case of regular resolution.

We believe that the first step in the design of hard instances should be
understanding the properties of the underlying hypergraphs that make the
instances hard. Such properties would be most convenient to investigate
if the CNFs were monotone. In the context of resolution, this is clearly
impossible since the CNFs must be unsatisfiable (we may, of course,
allow empty clauses but the resulting class would hardly be interesting). 
We think that the right approach is to consider a closely related model of Decision DNNFs
representing monotone CNFs of bounded incidence treewidth.
As mentioned in the introduction, this model has a lot in common with the
regular resolution. Therefore, we believe that understanding the complexity of the former
on CNFs of bounded incidence treewidth will provide an important insight
for the latter.

\appendix

\section{Assembling a resolution from small fragments}

\begin{definition} \label{def:falsesequence}
Let $f_1, \dots, f_m$ be $\varphi$-based functions.
For each $1 \leq  i \leq m$ we define a CDT $L_i$ with source $u_i$
satisfying the following conditions.
\begin{itemize}
\item Each sink of $L_i$ is labelled with $C \rightarrow ()$
or, otherwise, is identified with the source of $L_j$ for some $j<i$,
(Of course the latter condition is possible only if $i>1$).
\item For each $i$ and each $v \in V(L_i)$,  except those that
are identified with roots of earlier $L_j$, $v \notin \bigcup_{j=1}^{i-1} V(L_j)$.
\item Let $u$ be a sink of $L_i$ and let $P$ be a path from $u_i$ to $u$.
Then if $u$ is associated with $C \rightarrow ()$ then $C \in dom(f_i)$ and
$f_i(C)$ is falsified by $A(P)$. Otherwise, $u$ is associated with 
the source of $L_j$ for some $j<i$. In this case $f_j$ is a subfunction of
$f_i|_{A(P)}$. 
\end{itemize}
We call $L_1, \dots L_m$ a falsifying sequence. 
\end{definition}

\begin{definition}
With data as in the previous definition
we say that $f_j$ is a \emph{child} of $f_i$
is $L_i$ has a sink identified with the root of $L_j$
(clearly $j<i$).
Let $i_1, \dots, i_r$ be a sequence such that
for each $1 \leq k<r-1$ $f_{i_k}$ is a child of $f_{i_{k+1}}$.
We then say that $f_{i_1}$ is a \emph{descendant} of
$f_{i_r}$.
The falsifying sequence is \emph{read-once}
if for each $i,j$ such that $F_j$ is a descendant of $F_i$,
$Var( L_i) \cap Var(L_j)=\emptyset$
\end{definition}

\begin{theorem} \label{assemble1}
Let $L_1, \dots L_m$ be a falsifying read-once sequence 
for the respective $\varphi$-based functions 
$f_1, \dots f_m$. 
Let $R=\bigcup_{i=1}^m L_i$. 
Let $u_1, \dots u_m$ be the respective sources of
$L_1, \dots L_m$.  
Then each $R_{u_i}$ is a falsifying FRR for $f_i$. 
\end{theorem}

{\bf Proof.}

\begin{claim} \label{clm:twodags}
Let $D_1$  and $D_2$ be DAGs wth labels on their edges.
Suppose that $V(D_1) \cap V(D_2) \subseteq sinks(D_2)$.
Then the following statements hold.
\begin{enumerate}
\item $D=D_1 \cup D_2$ is a DAG. 
\item $D[V(D_1)]=D_1$.
\item $sinks(D)=sinks(D_1) \cup (sinks(D_2) \setminus V(D_1))$.
\end{enumerate}
\end{claim}

{\bf Proof.}
The second statement is immediate by definition
as $D$ does not add nor removes edges between
vertices of $D_1$ (note that there are no edges
between sinks).
That is $D[V(D_1)]$ is a DAG and also 
$D[V(D_2) \setminus V(D_1)]$ but there is no
path from the former to the latter. This proves 
the first statement.  
Finally, for the third statement, $sinks(D_1)$ are sinks of 
$D$ because of the second statement and $sinks(D_2) \setminus V(D_1)$
are sinks of $D$ simply by construction.  The rest of the
vertices are not sinks either in $D_1$ or in $D_2$ so, obviously,
they remain non-sinks in $D$.
$\square$

For $1 \leq i \leq m$,
let $R^i$ be $\bigcup_{1 \leq j \leq i} L_i$.
Using inductively the above claim we make the following 
list of observations that will be referred by their numbers in the rest of the proof.

\begin{enumerate}
\item Each $R^i$ is a DAG all sinks of which
are labelled with $C \rightarrow ()$. Apply inductively the first and the third statements of the above claim.
\item For each $v \in D(R)$ that is not a sink, $v$ has two outgoing edges one labelled with the positive and
one labelled with the negative literal of the same variable. 
Let $i$ be the smallest index such that $v \in V(R^i)$. Note that if $v$ is a sink in $R^i$ then, by inductive
application of the second statement of the claim, starting from $R^i$ and up to $R^m=R$, we observe 
tat $v$ remains a sink in $R$. Hence $v$ is not a sink in $R^i$. By definition $v \in V(L_i)$. Moreover, if 
$i>1$ then $v$ is not a vertex of $R^{i-1}$ by the minimality assumption. Hence, as $v$ is not a sink of 
$R^i$, it follows from the first statement of the claim that $v$ is not a sink of $L_i$. But then $v$ has
two outgoing edges as specified in the stement of the current item. Again applying part 2 of the claim
inductively until $R=R^m$, we observe that this situation with two outgoing edges is preserved in $R$. 
\item For each $1 \leq i \leq m$, $R^i_{u_i}=R^j_{u_i}$ for all $j \geq i$. Again,   starting from $R^i_{u_i}$,
apply inductively the second part of the claim 
\end{enumerate}

Now, we need to prove that each $R_{u_i}$ is read-once. 
For this purpose, we prove that  for each path $P$ in $R$ from $u_i$ to a sink
$P=P_1+ \dots+P_r$ where each $P_j$ is a source sink path in one of $L_1, \dots, L_m$. 
The proof is by induction on $i$.  
By the last item in the above list, it is enough to consider
$R^i_{u_i}$. $R^1_{u_1}=L_1$, so we are done by construction.
Assume now that $i>1$.  Then $P$ has a prefix $P'$ which is a source-sink
path of $L_i$. If $P=P'$, we are done. 
Otherwise, $P=P'+P''$, where $P''$ starts from $u_j$ for some $j<i$.
By the induction assumption, $j$ satisfies the requirement. 
Hence, the desired concatenation of subpaths for $P''$ plus $P'$
at the beginning provides the desired concatenation of paths for $P$.
Now, consider the concatenation $P=P_1+ \dots +P_r$ as above.
Each $P_j$ is read-once by definition so, if there is a repetition,
there are some $P_j$ and $P_k$, $j<k$ such that $Var(A(P_j)) \cap Var(A(P_k))$
intersect. But $P_j$ is a path of some $L_{j'}$ and $P_k$ is a path of some 
$L_{k'}$ such that $k'<j'$ and $f_{k'}$ is a descendant of $f_{j'}$.
By assumption, the variables of $L_{j'}$ and $L_{k'}$ must be disjoint, 
a contradiction.

It follows from the combination of the first three items the above list
and the read-onceness claim that each $R_{u_i}$ is a falsifying FRR.
It remains to prove that each $R_{u_i}$ has this capacity for $f_i$.
We proceed by induction on $1, \dots, m$. For $i=1$, as we have seen
above, $R_{u_1}=L_1$, so the statement holds by construction.
Assume now that $i>1$. Let $P$ be a source-sink path of $R_{u_i}$.
Let $C \rightarrow ()$ be the label of the final node of $P$
As  verified above $P$ has a prefix $P'$ that is a source-sink path of $L_i$.
Assume first that $u=u_j$ for some $j<i$.  Then $P=P'+P''$ and, by the induction
assumption $C \in dom(f_j)$ and $A(P'')$ falsifies $f_j(C)$.
By construction, $f_j$ is a subfunction of $f_i|_{A(P')}$. We conclude that $C \in dom(f_i)$
and $A(P)$ falsifies $f_i(C)$. If $u$ is not an earlier $u_j$ then the same conclusion follows
by construction. $\blacksquare$

In the rest of this section, ${\bf TR}$ is the set of transitional clauses of $\varphi$.

\begin{definition}
Let $f$ be a $\varphi$-based function.
We denote by $empty(f)$ the subset of $dom(f)$  consisting of all
clauses $C$ such that $f(C)=()$. 
We say that $f$ is \emph{interesting} if the following two conditions hold.
\begin{itemize}
\item $empty(f) \subseteq {\bf TR}$.
\item $f \setminus empty(f)$ is unsatisfiable 
\end{itemize}
\end{definition}

\begin{definition}  \label{def:goodfun}
Let $f$ be an interesting $\varphi$-based function.
Let $A$ be an assignment and suppose that $f|_A$ is also
an interesting $\varphi$-based function.
Let $g$ be an interesting subfunction of $f|_A$. We say that
$g$ is a \emph{good} function for $f,A$ if one of the following conditions
is true.
\begin{enumerate}
\item $dom(g) \cap {\bf TR} \subseteq empty(g)$. 
\item $(f|_A) \setminus g$ is satisfiable and $Var(range(g)) \cap Var(range(f|_A \setminus g))=\emptyset$. 
\end{enumerate}
\end{definition}

\begin{proposition} \label{prop:goodfun}
$f|_A$ is a good function for $f,A$. 
\end{proposition}

\begin{lemma} \label{falsecombine}
Suppose that $g$ is a good function for $f,A$.
Let $S$ be a ${\bf C}$ falsifier for $g$ for some ${\bf C} \subseteq {\bf TR}$ 
such that $Var(S)$ is disjoint with both $Var(A)$ and $Var(range(f|_A \setminus g))$.
Then $S \cup A$ is a ${\bf C}$-falsifier for $f$.
\end{lemma}

{\bf Proof.}
If the first condition of Definition \ref{def:goodfun} holds then the statement holds
in a vacuous way because no such a falsifier exists.
Indeed by assumption ${\bf C} \subseteq empty(g)$,
hence an extension of $S$ must satisfy
$g(dom(g) \setminus empty(g))$ in contradiction to the
definition of an interesting function. So, we assume 
that the second case holds. 

Let $S_1$ be an extension of $S$
satisfying $g(dom(g) \setminus {\bf C})$
and let $S_2$ be a satisfying assignment to 
$range(f|_A \setminus g)$.
We may assume w.l.o.g. that $Var(S \setminus S_1) \subseteq Var(range(g))$ 
and that $Var(S_2) \subseteq Var(range(f|_A \setminus g))$.
By assumption $Var(S_1)$, $Var(S_2)$, and $Var(A)$ are mutually
disjoint. Therefore $S^*=S_1 \cup S_2 \cup A$ is a well formed set of literals.
We claim that $S^*$ satisfies $f(C)$ for all $C \in dom(f) \setminus {\bf C}$. 
Indeed, if $C \in dom(f) \setminus dom(f|_A)$
then $f(C)$ is satisfied by $A$.
If $C \in dom(g)$ then $g(C)$ and hence $f(C)$ is satisfied by $S_1$.
It remains to assume that $C \in dom(f|_A \setminus g)$.
But then $f(C)$ is satisfied by $S_2$.
$\blacksquare$

\begin{definition} 
Let $f_1, \dots, f_m$ be interesting $\varphi$-based functions.
For each $1 \leq  i \leq m$ we define a CDT $L_i$ with source $u_i$
satisfying the following conditions.
\begin{itemize}
\item Each sink of $L_i$ is labelled with $C \rightarrow ()$ or $trans_{\bf C}$
or, otherwise, is identified with the source of $L_j$ for some $j<i$,
(Of course the latter condition is possible only if $i>1$).
\item For each $i$ and each $v \in V(L_i)$,  except those that
are identified with roots of earlier $L_j$, $v \notin \bigcup_{j=1}^{i-1} V(L_j)$.
\item Let $u$ be a sink of $L_i$ and let $P$ be a path from $u_i$ to $u$.
Then we consider the following subcases. 
    \begin{itemize}
    \item If $u$ is associated with $C \rightarrow ()$ then $C \in dom(f_i)$ and
           $f_i(C)$ is falsified by $A(P)$. 
    \item if $u$ is associated with $trans_{\bf C}$ 
    then $A(P)$ is a ${\bf C}$-falsifier for $f_i$ 
    \item Otherwise, $u$ is associated with 
the source of $L_j$ for some $j<i$. In this case $f_j$ is a good function for $f_i,A(P)$. 
     \end{itemize}
\end{itemize}
We call $L_1, \dots L_m$ a \emph{transitional falsifying sequence}. 
\end{definition}

We define children and descendants analogously to the non-transisitional
case. With this in mind, we are now in a position to state a version 
of Theorem \ref{assemble1} for the transitional case. 

\begin{theorem} \label{assemble2}
Let $L_1, \dots L_m$ be a transitional falsifying sequence 
for the respective $\varphi$-based functions 
$f_1, \dots f_m$. 
Let $R=\bigcup_{i=1}^m L_i$. 
Let $u_1, \dots u_m$ be the respective sources of
$L_1, \dots L_m$.  
Then each $R_{u_i}$ is a DAG with $Var(R_{u_i})$
being the union of $Var(L_i)$ and all $Var(L_j)$ 
such that $f_j$ is a descendant of $f_i$
and sinks labelled by either $C \rightarrow ()$ or $trans_{\bf C}$. 
Also, let $R^i=L_1 \cup \dots \cup L_i$
Then, for each $j \geq i$, $R^j_{u_i}=R^i_{u_i}$. 

Suppose  that two \emph{additional conditions} take place for each $R_{u_i}$
\begin{enumerate}
\item Each path of $R_{u_i}$ is read-once.
\item Let $P$ be a source-sink path of $L_i$ and suppose that
the final vertex of $P$ is $u_j$ for some $j<i$. 
Then $Var(R_{u_j})$ is disjoint with $Var(range(f_i|_{A(P)} \setminus f_j))$.
\end{enumerate}
Then $R_{u_i}$ is a transitional resolution for $f_i$. 
\end{theorem}

{\bf Proof.}
Claim \ref{clm:twodags} applies for this proof.
The three observations in the proof of Theorem \ref{assemble1}
also hold but with the  statement for the first one is slightly modified as specified below.
\begin{enumerate}
\item Each $R^i$ is a DAG all sinks of which
are labelled either with $C \rightarrow ()$ or with $trans_{\bf C}$. 
Apply inductively the first and the third statements of Claim \ref{clm:twodags}.
\item For each $v \in D(R)$ that is not a sink, $v$ has two outgoing edges one labelled with the positive and
one labelled with the negative literal of the same variable. 
 
\item For each $1 \leq i \leq m$, $R^i_{u_i}=R^j_{u_i}$ for all $j \geq i$. 
\end{enumerate}

These observations already prove the part of the theorem that does not take into
account the additional conditions. With these conditions in mind, let $P$ be a source-sink path of $R_{u_i}$. 
Due to read-onceness condition, $A(P)$ is a well formed set of literals. 
We need to prove that (i)  if the final vertex of $P$ is labelled with $C \rightarrow ()$ then $C \in dom(f_i)$ 
and $A(P)$ falsifies $f_i(C)$ and (ii) if the  final vertex of $P$ is associated with $trans_{\bf C}$
then ${\bf C} \subseteq dom(f_i)$ and $A(P)$ is a ${\bf C}$-falsifier for $f_i$.

The proof of (i) is analogous to the corresponding proof in Theorem \ref{assemble1},
so we concentrate on proving (ii). If $P$ is a source-sink path of $L_i$ the statement follows by
construction. Let $P'$ be the prefix of $P$ that is a source-sink path of $L_i$. 
Then the final vertex of $P'$ is $u_j$ for some $j<i$. Let $P''$ be the suffix of $P$ starting
from $u_j$
By the indudction assumption (the induction basis is correct by construction),
${\bf C} \subseteq dom(f_j)$ and $A(P'')$ is a ${\bf C}$-falsifier for $f_j$.
Recall that by construction, $f_j$ is a good finction for $f_i,A(P')$ and that
$Var(P'')$ is disjoint with $Var(P')$ (due to read-onceness of $P$) and with 
$Var(range(f_i|_{A(P')} \setminus f_j))$ (by construction). 
Therefore, all the premises of Lemma \ref{falsecombine} are satisfied
and we conclude that $A(P)$ is a ${\bf C}$-falsifier for $f_i$.
$\blacksquare$

\section{Proof of Theorem \ref{transres}}

\begin{lemma} \label{lem:ltp1}
Let ${\bf tp} \in \mathcal{ORD}$, $A=A(P)$ where $P$ is a source-sink
path of $L_{\bf tp}$. 
Assume that $f_{\bf tp}|_A$ is an interesting function.
Further on, we assume that $t=t({\bf tp})$ is not a leaf 
and ${\bf tp_1}$ and ${\bf tp_2}$ be successors of ${\bf tp}$ through $A$
(in case $t$ has only one child, let ${\bf tp_1}$ be the only successor of 
${\bf tp_1}$ through $A$). 
Then the following statements hold. 
\begin{enumerate}
\item $f_{\bf tp}|_A \setminus empty(f_{\bf tp}|_A)$ is the disjoint union of 
$f_{\bf tp_1}$, $f_{\bf tp_2}$. In case $t$ has only one child, 
$f_{\bf tp}|_A \setminus empty(f_{\bf tp}|_A)=f_{\bf tp_1}$
\item At least one of $f_{\bf tp_1}$, $f_{\bf tp_2}$ is unsatisfiable. 
\item  $h_{\bf tp}|_A$ is the disjoint union of
$f_{\bf tp_1}$, $f_{\bf tp_2}$, and $trans_{ET({\bf tp} \cup empty(f|_A)}$. 
\item For each $i \in \{1,2\}$, $h_{\bf tp_i}$ is the disjoint union 
of $f_{\bf tp_i}$ and $trans_{ET({\bf tp}) \cup empty(f|_A)}$. 
\end{enumerate}
The last two cases naturally adapted to the situation where $t$ has only one child.
In particular, $f_{\bf tp_2}$ is not part of the union in the third case and
the definition holds only for $i=1$ in the fourth case. 
\end{lemma}

{\bf Proof.}
First of all observe that by construction,
$empty(f_{\bf tp}|_A)$ is a subset of $ET({\bf tp_i})$ for each $i \in \{1,2\}$.
Therefore, to establish that  $f_{\bf tp_i}$ is a subfunction of
$f_{\bf tp}|_A \setminus empty(f)$, it is enough to show that
$f_{\bf tp_i}$ is a subfunction of $f_{\bf tp}|_A$.

First of all, we need to show that 
for each $C \in dom(f_{\bf tp_i}) \cap dom(f_{\bf tp}|_A)$,
$f_{\bf tp_i}(C)=f_{\bf tp}|_A(C)$.




Let $C \in dom(f_{\bf tp}|_A) \cap dom(f_{\bf tp_i})$. 
By definition, $C \in CL(T_{t_i})$. Hence, by one-sidedness of 
$(T,{\bf B})$, $C \notin CL(T_{t_{3-i}})$. In particular,
this implies that $f_{\bf tp}(C)=Proj(C,Var(T_t) \setminus Var(S) )=Proj(C,(Var(T_{t_i}) \cup Var(t))\setminus Var(S))$
(the rest of the variables are present only in $T_{t_{3-i}}$
so $C$  would need to be present in one of the bags of the subtree
due to the containement property). 
Then $f_{\bf tp}|_A(C)=Proj(C,(Var(T_{t_i}) \cup Var(t))\setminus (Var(S) \cup Var(A))$.
As $Var(S) \cap Var(A)=Var(t)$, We conclude that 
$Var(T_{t_i}) \cup Var(t))\setminus (Var(S) \cup Var(A)=Var(T_{t_i})\setminus Var(t)$
and hence $f_{\bf tp}|_A(C)=f_{\bf tp_i}(C)$.


Now, let $C \in dom(f_{\bf tp_i})$.
Then $C \in CL(T_{t_i}) \subseteq CL(T_t)$.
There are three reasons why $C$ may not belong to
$dom(f_{\bf tp})$ we will observe that none of these reasons holds.
The first reason is that $C \in CL({\bf tp})$.
This means that $C \in CL(t)$. 
Since $C \in CL(T_{t_i})$ by  the connectivity condition of tree decompositions,
$C \in CL(t_i)$. That is $C \in CL({\bf tp}) \cap CL(t_i) \subseteq CL({\bf tp_i})$,
a contradiction. 

The next reason is that $C \in ET({\bf tp})$. However, in this case, 
$C \in ET({\bf tp_i})$ in contradiction to $C \in dom(f_{\bf tp_i})$.

Finally, it may be that $C$ is satisfied by $S$. 
Then $C$ is satisfied by $S \cup A$. 
If $C \in CL(t_i)$ then, by definition, $C \in CL({\bf tp_i})$,
a contradiction. Otherwise, let $x \in Var(S \cup A)$ be a variable
whose occurrence in $S \cup A$ satisfies $C$. 
Since $C \notin CL(t_i)$, $C$ cannot occur in any bag outside $T_{t_i}$
by the treewidth connectivity condition. Then the containment condition implies 
that $x$ must also occur in a bag inside $T_{t_i}$. Since $x$ is occurring in a bag
outside $T_{t_i}$ (namely in the bag of $t$), $x \in Var(t_i)$ by the treewidth connectivity
condition. In particular, $x$ occurs in $Proj(S \cup A, Var(t_i))=S({\bf tp_i})$ in 
contradiction to $C \in dom(f_{\bf tp_i})$.

Thus we conclude that $C \in dom(f_{\bf tp})$. It remains to be seen that $C$
is not satisfied by $A$. But in this case $C$ is satisfied by $S \cup A$ and
we apply the reasoning of the previous paragraph. 

Now, we assume that $C \in dom(f_{\bf tp}|_A) \setminus empty(f)$
and that $C \in CL(T_{t_i})$, We claim that $C \in dom(f_{\bf tp_i})$.
Indeed, there may be two reasons why this is not so.
One such a reason is that $C \in CL({\bf tp_i})$.
Recall that $CL({\bf tp_i})$ consists of two subsets.
The first is $CL({\bf tp}) \cap CL({\bf tp_i})$ but in this case 
$C \notin f_{\bf tp}$, a contradiction.
The second set are those clauses that are satisfied by $S \cup A$.
But if $C$ is satisfied by $S$ it is not contained in $dom(f_{\bf tp})$
and if $C$ is satisfied by $A$ then it is not contained in $dom(f_{\bf tp}|_A)$,
a contradiction in both cases. 
The other reason that may cause $C$ to not be in $dom(f_{\bf tp_i})$
is that $C$ is satisfied by $S({\bf tp_i})$. But, by definition, $S({\bf tp_i})$
is a subset of $S \cup A$, so the last argument applies. 

Next, we observe that if $t$ has two children then $dom(f_{\bf tp_1})$
and $dom(f_{\bf tp_2})$ are disjoint. Indeed, for a clause $C$ to be in both
domains, it must be present in both $CL(T_{t_1})$ and $CL(T_{t_2})$
which is impossible due to one-sidedness. 

Finally, we observe that $f_{\bf tp}|_A \setminus empty(f_{\bf tp}|_A)$
does not have any clauses but $dom(f_{\bf tp_1}) \cup dom(f_{\bf tp_2})$.
Indeed, suppose that such a clause $C$ exists. By the proven above
$C \notin CL(T_{t_1}) \cup CL(T_{t_2})$. By the containement condition,
$Var(C)$ cannot intersect with $Var(T_t) \setminus Var(t)$. 
It follows that all the variables of $f_{\bf tp}(C)$ are assigned by $S \cup A$, in fact by $A$ by definition og $f_{\bf tp}$.
Then $f_{\bf tp}(C)$ is either satisfied by $A$ (contradiction to $C \in dom(f_{\bf tp}|_A)$) 
or falsified by $A$ (contradiction to $C \notin empty(f_{\bf tp}|_A)$). 
 So we have proved the first statement for the case of two children.
The proof applies for the case of one child simply by observing that
$dom(f_{\bf tp}|_A) \cap CL(T_{t_2})=\emptyset$. 

The second statement follows immediately if $t$ has only one child.
For the case of two children, we observe that $Var(range(f_{\bf tp_1})) \cap Var(range(f_{\bf tp_2}))=\emptyset$.
Indeed, by definition of the type functions
$Var(range(f_{\bf tp_i})) \subseteq Var(T_{t_i}) \setminus Var(t)$ and it is a well known
property of tree decompositions 
that $Var(T_{t_1}) \cap Var(T_{t_2}) \subseteq Var(t)$ (for otherwise the connectivity condition does not hold).
Assume that the second statement does not hold.
For each $i \in \{1,2\}$, let $S_i$ be an assignment satisfying $range(f_{\bf tp_i})$.
Clearly, we may assume that $Var(S_i) \subseteq Var(range(f_{\bf tp_i}))$ for each $i \in \{1,2\}$.
But then $Var(S_1) \cap Var(S_2)=\emptyset$ and hence $S=S_1 \cup S_2$ is a well formed set of literals. 
By the first statement $S$ satisfies $f_{\bf tp}|_A \setminus empty(f_{\bf tp}|_A)$ in contradiction to
our assumption about unsatisfiability of the latter. This proves the second statement. 

The third statement follows from the combination of the first statement and the observation 
that $h_{\bf tp}|_A=f_{\bf tp}|_A \cup trans_{ET({\bf tp})}$. 
For the fourth statement, recall that, by definition each $h_{\bf tp_i}$ is the union of $f_{\bf tp_i}$
and $trans_{ET({\bf tp_i})}$ and that $ET({\bf tp_i})=ET({\bf tp}) \cup (empty(f_{\bf tp}|_A) \cap {\bf TR})$.
However, by assumption $empty(f_{\bf tp}|_A) \subseteq {\bf TR}$, so the fourth statement follows.
$\blacksquare$

\begin{lemma} \label{validtype}
Let ${\bf tp} \in \mathcal{ORD}$
and let $A=A(P)$ where $P$ is a source-sink
path of $L_{\bf tp}$. 
Assume that for each non-transitional clause $C \in dom(h_{\bf tp})$,
$h_{\bf tp}(C)$ is not falsified by $A$ 
and that $h_{\bf tp}|_A \setminus empty(h_{\bf tp}|_A)$ is not satisfiable.
(In other words, we assume that the conditions of the first two items 
in the list in Definition \ref{defconstr} do not hold.) 

Then the following statements hold.
\begin{enumerate}
\item If $({\bf tp},A)$  is a special pair then $h_{\bf tp}|_A$ is a good function for $h_{\bf tp},A$.
\item Otherwise, let ${\bf tp'}$ be the preferred successor of ${\bf tp}$ through $A$. 
Then ${\bf tp'} \in \mathcal{ORD}$ (meaning that the definition of $L_{\bf tp}$ is valid)
and $h_{\bf tp'}$ is a good function for $h_{\bf tp},A$. 
\end{enumerate}

\end{lemma}

{\bf Proof.}
It is not hard to see that for each ${\bf tp} \in \mathcal{ORD}$,
$h_{\bf tp}$ is an interesting function.
Also, $h_{\bf tp}|_A$ is an interesting function by assumption of the lemma.
This means that the premises of the definition of a good function are satisfied.

The first statement of the lemma is immediate from Proposition \ref{prop:goodfun}. 
For the second statement, observe that $empty(f_{\bf tp}|_A) \subseteq empty(h_{\bf tp}|_A)$
and that $f_{\bf tp}|_A \setminus empty(f_{\bf tp}|_A)=
                 h_{\bf tp}|_A \setminus empty(h_{\bf tp}|_A)$.
It follows that $f_{\bf tp}|_A$ is an interesting function and,
in particular, that all the premises of Lemma \ref{lem:ltp1}
have been met. 

It follows from the first statement of Lemma \ref{lem:ltp1}
that $f_{\bf tp'}$ does not map any clause in its domain to $()$.
It also follows from the definition of $L_{\bf tp}$ and from the second 
statement of Lemma \ref{lem:ltp1} that $f_{\bf tp'}$ is unsatisfiable.
Thus we conclude that ${\bf tp'}$ is a non-basic unsatisfiable type
and hence belongs to $\mathcal{ORD}$. 
As $f_{\bf tp'}=h_{\bf tp'} \setminus empty(h_{\bf tp'})$, we also conclude
that $h_{\bf tp'}$ is an interesting function. To observe that remaining part of the definition of a good
function is satisfied, we consider the last three items of definition of $L_{\bf tp}$
separately.  

In the satisfiable sibling case, let ${\bf tp''}$ be the successor
of ${\bf tp}$ through $A$ other than ${\bf tp'}$. 
By combination of the last two statements of Lemma \ref{lem:ltp1},
$f_{\bf tp''}=h_{\bf tp}|_A \setminus h_{\bf tp'}$ and, by the assumption
of the case $f_{\bf tp''}$ is satisfiable thus $h_{\bf tp'}$ is a good
function by the second item of Definition \ref{def:goodfun}.
In the non-transitional child case 
$dom(f_{\bf tp'}=dom(h_{\bf tp'}) \setminus empty(h_{\bf tp'})$
does not intersect with ${\bf TR}$ thus $h_{\bf tp'}$ is a good
function by the first item of Definition \ref{def:goodfun}. 
Finally, in the single child case, the last two cases of Lemma \ref{lem:ltp1}
imply that $h_{\bf tp'}=h_{\bf tp}|_A$ thus, so the lemma follows from
Proposition \ref{prop:goodfun}.

$\blacksquare$

\begin{lemma}  \label{validspair}
Let $({\bf tp},A) \in \mathcal{ORD}$.
Let $P$ be a source-sink path $L_{({\bf tp},A)}$
and let  $u$ be the final vertex of $P$. 
Then the following statements hold.

\begin{itemize}
\item Suppose that $u$ is associated with $C \rightarrow ()$. 
Then $C \in dom(h_{\bf tp}|_A)$ and $A(P)$ falsifies
$h_{\bf tp}|_A(C)$.
\item Suppose that $u$ is associated with a source of $L_{\bf tp^*}$. 
Then ${\bf tp^*}$ is non-basic unsatisfiable type (implying 
validity of definition of $L_{({\bf tp},A)}$ in Definition \ref{defconstr}) 
and $h_{\bf tp^*}$ is a good subfunction for $h_{\bf tp}|_A, A(P)$. 
\end{itemize}
\end{lemma}

{\bf Proof.}

Assume first that $u$ is associated with $C \rightarrow ()$. 
Combining the third and fourth statements of Lemma \ref{lem:ltp1},
we observe that $h_{\bf tp_1}$ is a subfunction of $h_{\bf tp}|_A$.
By definition of a transitional resolution, 
$C \in dom(h_{\bf tp_1})$ and $A(P)$ falsifies 
$h_{\bf tp_1}(C)$. Hence the same holds if 
$h_{\bf tp_1}$ is replaced by $h_{\bf tp}|_A$. 

Let us now assume that $u$ is identified with the source of some 
$L_{\bf tp^*}$. By construction and the fourth statement of Lemma \ref{lem:ltp1},
$A(P)$ is a ${\bf C}$-falsifier for $h_{\bf tp_1}$ where ${\bf C}$ is the union of 
$ET({\bf tp_1})$ and some clauses of $dom(f_{\bf tp_1})$.
Applying again the fourth statement of Lemma \ref{lem:ltp1}, we observe
that $ET({\bf tp_1})=ET({\bf tp_2})$, hence by construction, 
$ET({\bf tp^*})={\bf C}$. By the third statement of Lemma \ref{lem:ltp1},
$dom(f_{\bf tp_1}) \cap dom(f_{\bf tp_2})=\emptyset$, hence by construction
$f_{\bf tp^*}=f_{\bf tp_2}$. In particular, it follows from ${\bf tp_2} \in \mathcal {ORD}$ 
that ${\bf tp^*} \in \mathcal{ORD}$ and hence $h_{\bf tp^*}$ is an interesting function. 
Thus it remains to be shown that $h_{\bf tp^*}$ is a subfunction of
$h_{\bf tp}|_{A \cup A(P)}$ and that the range of 
$h_{\bf tp}|_{A \cup A(P)} \setminus h_{\bf tp^*}$ is satisfiable.

Let $C \in dom(h_{\bf tp^*})$.
Assume first that $C \in dom(f_{\bf tp_2})$.
Note that, 
$Var(range(f_{\bf tp_2}) \cap Var(range(f_{\bf tp_1})=\emptyset$
(see the proof of the second statement of Lemma \ref{lem:ltp1} for an explanation).
On the other hand, by construction, $Var(A(P)) \subseteq Var(range(f_{\bf tp_1}))$.
It follows that $C \in dom(h_{\bf tp}|_{A \cup A(P)}$ and, moreover, that 
$h_{\bf tp}|_{A \cup A(P)}(C)=h_{\bf tp^*}(C)$. 
Next, if $C \in {\bf C}$, then $A(P)$ falsifies $h_{\bf tp}|_A(C)$.
Hence $C \in dom(h_{\bf tp}|_{A \cup  A(P)})$ and
$()=h_{\bf tp^*}(C)=h_{\bf tp}|_{A \cup A(P)}(C)$. 
Thus we have shown that $h_{\bf tp^*}$ is a subfunction of $h_{\bf tp}|_{A \cup A(P)}$. 

Let ${\bf C^*}=dom(h_{\bf tp}|_{A \cup A(P)}) \setminus dom(h_{\bf tp^*})$. 
We need to show that $h_{\bf tp}|_{A \cup A(P)}({\bf C^*})=
h_{\bf tp}|_A({\bf C^*})|_{A(P)}$ is satisfiable.
It follows from the previous paragraph that ${\bf C^*}$ is the subset
of $dom(h_{\bf tp_1}) \setminus {\bf C}$ that are not satisfied by $A(P)$.
As by Lemma \ref{lem:ltp1}, $h_{\bf tp_1}$ is a subfunciton of $h_{\bf tp}|_A$,
we in fact need to show that $h_{\bf tp_1}({\bf C^*})|_{A(P)}$. But
this is immediate since $A(P)$ is a ${\bf C}$-falsifier for $h_{\bf tp_1}$.

$\blacksquare$




\begin{corollary} \label{transfals}
Let $\mathcal{ORD}=(a_1, \dots, a_m)$
Let $g_1, \dots, q_m$ be $\varphi$-based functions
and let $L_1, \dots, L_m$ be graphs such that 
such that $g_i=h_{\bf tp}$ and $L_i=L_{\bf tp}$ if $a_i={\bf tp}$ and
$g_i=h_{\bf tp}|_A$ $L_i=L_{({\bf tp},A)}$ in case $a_i=({\bf tp},A)$.
Then $L_1, \dots, L_m$ is a transitional falsifying sequence.
\end{corollary}

{\bf Proof.}
Immediate from the combination 
Lemma \ref{validtype}, and Lemma \ref{validspair}.
$\blacksquare$
 
\begin{theorem} \label{validtypes}
With the data as in Corollary \ref{transfals}, let $R=L_1 \cup \dots \cup L_m$. 
Then the following statements hold.
\begin{itemize}
\item For each ${\bf tp} \in \mathcal{ORD}$, let $u_{\bf tp}$ be the source of $L_{\bf tp}$.
Then $R_{u_{\bf tp}}$ is a transitional resolution for $h_{\bf tp}$ 
and $Var(R_{u_{\bf tp}}) \subseteq Var(T_t) \setminus Var(p(t))$ 
where $t=t({\bf tp})$ and $p(t)$ is the parent of $t$ in $T$ (if $t$ is the root
of $t$, consider $Var(p(t))=\emptyset$). 
\item For each $({\bf tp},A) \in \mathcal{ORD}$, 
let $u_{({\bf tp},A)}$ be the source of $L_{({\bf tp},A)}$. 
Then $R_{u_{({\bf tp},A)}}$ is a transitional resolution for $h_{\bf tp}|_A$
and $Var(R_{u_{({\bf tp},A)}}) \subseteq Var(T_t) \setminus Var(t)$.
\end{itemize}
\end{theorem}

{\bf Proof.}
First of all, let us introduce one additional notation: for $1 \leq j \leq m$,
$R^j=L_1 \cup \dots \cup L_j$

Corollary \ref{transfals} gives us the possibility to apply Theorem \ref{assemble2}
without the read-onceness assumption.  
From this application, we obtain the following two statements.
\begin{enumerate}
\item For each $1 \leq i \leq m$, $R_{u_i}$ is a DAG with all the sinks labelled by either $C \rightarrow ()$
or $trans_{\bf C}$. where $u_i$ is the source of $L_i$
\item For each $1 \leq i \leq m$ and each $i \leq j \leq m$, $R_{u_i}^j=R_{u_i}^i$.
\end{enumerate}

\begin{claim} \label{clm:ronce}
\begin{enumerate}
\item For each ${\bf tp} \in \mathcal{ORD}$, 
$Var(R_{u_{\bf tp}} \subseteq Var(T_t) \setminus Var(p(t))$.
\item For each $({\bf tp},A) \in \mathcal{ORD}$, 
$Var(R_{u_{({\bf tp},A)}}) \subseteq Var(T_t) \setminus Var(t)$.
\item Each $R_{u_i}$ is read-once.
\end{enumerate}
\end{claim}

{\bf Proof.}
The proof is by induction on $1 \leq i \leq m$.
The above reasoning allows us to prove the claim for graphs $R^i_{u_i}$ only.
Assume first that $i=1$. Then $a_1={\bf tp}$ and $u_1=u_{\bf tp}$.
Moreover, $R^1_{u_{\bf tp}}=L_{\bf tp}$ and the claim follows by construction.

Assume now that $i>1$.
Assume first that $a_i={\bf tp}$.
Let $P$ be a source-sink path of $R^i_{u_{\bf tp}}$. 
We need to prove that $P$ is read-once and that $Var(P) \subseteq Var(T_t) \setminus Var(p(t))$
where $t=t({\bf tp})$. 
By construction, $P$ has a prefix $P'$ which is a source-sink path of $L_{\bf tp}$.
If $P=P'$ then we are done by definition of $L_{\bf tp}$.
Otherwise, the final node of $P'$ is $u_j$ for some $j<i$.
Let $P''$ be the suffix of $P$ starting from $u_j$.
We will prove that $P''$ is read-once, $Var(P'') \subseteq Var(T_t) \setminus Var(t)$
and hence does not intersect with $Var(P') \subseteq Var(t) \setminus Var(p(t))$
As a minor subtlety, note that this will imply 
that $Var(P'') \subseteq Var(T_t) \setminus Var(p(t))$ since, by the treewidth
connectivity condition, $Var(p(t)) \cap Var(T_t) \subseteq Var(t)$.

Path $P''$ is a source-sink path of $R^i_{u_j}$ hence it is a source-sink
path of $R_{u_j}$ and hence, in turn, it is read-once by the induction assumption. 
For the constraint on the set of variables, assume first that $a_j={\bf tp'}$.
Then ${\bf tp'}$ is the successor of ${\bf tp}$ through $A$ and $t'=t({\bf tp})$
is a child of $t$.
Hence $Var(P'') \subseteq Var(T_{t'}) \setminus Var(t) \subseteq Var(T_t) \setminus Var(t)$,
the first subset relation follows by the induction assumption.
 Otherwise, $u_j=({\bf tp},A)$ but then
$Var(P'') \subseteq Var(T_t) \setminus Var(t)$ is immediate by the induction assumption. 

It remains to assume that $a_i=({\bf tp},A)$.
Let ${\bf tp_1}$ and ${\bf tp_2}$ be the successors
of ${\bf tp}$ through $A$, $t=t({\bf tp})$, $t_1=t({\bf tp_1})$
and $t_2=t({\bf tp_2})$ are the children of $t$.  
We assume w.l.o.g. that ${\bf tp_1}$ is the main successor
of ${\bf tp}$ through $A$. 

As in the previous case, let $P$ be a source-sink path of $R^i_{({\bf tp},A)}$. 
We need to prove that $P$ is read-once and that $Var(P) \subseteq Var(T_t) \setminus Var(t)$. 
By construction, $P$ has a prefix $P'$  which is a source-sink path of $L_{({\bf tp},A)}$.
By construction, $P'$ is read-once and $Var(P') \subseteq Var(T_{t_1}) \setminus Var(t)$.
Therefore, if $P=P'$, we are done. Otherwise, the final node of $P'$ is $u_j$ for some $j<i$.
Let $P''$ be the suffix of $P$ starting at $u_j$. Then $P''$ is a source-sink path of $R^i_{u_j}$
and hence a source-sink path of $R_{u_j}$.
By construction, $a_j={\bf tp^*}$ such that $t({\bf tp^*})=t_2$.
It follows from the induction assumption that $P''$ is read-once 
and $Var(P'') \subseteq Var(T_{t_2}) \setminus Var(t) \subseteq Var(T_t) \setminus Var(t)$.
To conclude that the whole $P$ is read-once, we observe that the connecedness condition
implies that $Var(T_{t_1}) \cap Var(T_{t_2}) \subseteq Var(t)$ implyingthat $Var(P') \cap Var(P'')=\emptyset$
thus confirming the read-onceness of $P$.
$\square$

\begin{claim} \label{clm:rangedisj}
Let $a_i \in \mathcal{ORD}$.
Let $P$ be a source-sink path of $L_i$, let $u$ 
be the final vertex of $P$ and assume that $u$ is the source
of $L_j$ for $j<i$.
Then $Var(R_{u_j})$ is disjoint from 
$Var(range(g_i|_{A(P)} \setminus g_j))$. 
\end{claim}

{\bf Proof.}
Clearly, we may assume that $g_j$ is a proper subfunction of $g_i|_{A(P)}$.
Let ${\bf tp}={\bf tp}(a_i)$. 
Then, by the assumption and Lemma \ref{lem:ltp1},
${\bf tp}$ has two successors, ${\bf tp_1}$ and ${\bf tp_2}$ 
through $A$. Let $t=t({\bf tp})$, $t_1=t({bf tp_1})$, $t_2=t({\bf tp_2})$.

Assume that $a_i={\bf tp}$. Then, by assumption and 
Lemma \ref{lem:ltp1}, $({\bf tp},A)$ is not a special pair.
Let ${\bf tp_1}$ be the preferred successor of ${\bf tp}$ 
through $A$, that is $u_j=u_{\bf tp_1}$.
By Claim \ref{clm:ronce}, $Var(R_{u_j}) \subseteq Var(T_{t_1}) \setminus Var(t)$.
On the other hand, by Lemma \ref{lem:ltp1},
$h_{\bf tp} \setminus h_{\bf tp_1}=f_{\bf tp_2}$ and, by definition,
$Var(range(f_{\bf tp_2})) \subseteq Var(T_{t_2}) \setminus Var(t)$.
By the properties of tree decompositions, 
$Var(T_{t_1}) \setminus Var(t)$ is disjoint with $Var(T_{t_2}) \setminus Var(t)$. 

It remains to assume that $a_i=({\bf tp},A)$.
Then, by definition of $L_{({\bf tp},A)}$ $u_j={\bf tp^*}$ 
where $t({\bf tp^*})=t_2$. By Claim \ref{clm:ronce},
$Var(R_{u_{\bf tp^*}}) \subseteq Var(T_{t_2}) \setminus Var(t)$.
We are going to show that
$Var(range(h_{tp}|_{A \cup A(P)} \setminus h_{\bf tp^*})) \subseteq Var(T_{t_1}) \setminus Var(t)$
thus implying the same conclusion as in the previous paragraph. 
Clearly, $Var(range(h_{tp}|_{A \cup A(P)} \setminus h_{\bf tp^*})) \subseteq
Var(range(h_{tp}|_{A \cup A(P)} \setminus f_{\bf tp^*}))$. In the proof of 
Lemma \ref{validspair}, we established that $f_{\bf tp^*}=f_{\bf tp_2}$.
Furthermore, it is not hard to see that
$Var(range(h_{tp}|_{A \cup A(P)} \setminus f_{\bf tp_2}))
  \subseteq  Var(range(h_{tp}|_{A} \setminus f_{\bf tp_2}))$.
By Lemma \ref{lem:ltp1},
$h_{tp}|_{A} \setminus f_{\bf tp_2}=h_{\bf tp_1}$ and, by definition,

$Var(range(h_{\bf tp_1})) \subseteq Var(T_{t_1}) \setminus Var(t)$,
as required.
$\square$

It follows from Claims \ref{clm:ronce} and \ref{clm:rangedisj} that
the additional conditions of Theorem \ref{assemble2} are satisfied.
We can now fully apply Theorem \ref{assemble2} and hence to conclude that each $R_{u_i}$
is a transitional resolution for the functionas specified in the statement of
this theorem.
$\blacksquare$

We are now moving on to consider upper
bounds on the size of $R$. For this, we need to redefine the notion
of a descendant of a transitional falsifying sequence in terms of $\mathcal{ORD}$
and state a simple but important fact.

\begin{definition}
In light of Corollary \ref{transfals},
a descendant relation is defined on functions $g_1, \dots, g_m$.
We naturally extend it to elements $a_1, \dots a_m$ of $\mathcal{ORD}$:
$a_i$ is a descendant of $a_j$ if and only if $g_i$ is a descendant of $g_j$.
We denote the descendancy relation by $<_d$. That is $a_i<_d a_j$ means that
$a_i$ is a descendant of $a_j$. 
\end{definition}

\begin{proposition}
If $a_i<_d a_j$ then one of the following two statements hold.
\begin{enumerate}
\item $a_i={\bf tp}$ and $a_j=({\bf tp},A)$.
\item Let ${\bf tp}={\bf tp}(a_i)$ and ${\bf tp'}={\bf tp}(a_j)$,
$t=t({\bf tp})$ and $t'=t({\bf tp'})$. Then $t'$ is a descendant of $t$ in $T$. 
\end{enumerate}
\end{proposition}

\begin{definition} 
A node $t \in V(T)$ is \emph{splitting}
if $t$ has two children $t_1$ and $t_2$
such that $CL(T_{t_i}) \cap {\bf TR} \neq \emptyset$ for
each $i \in \{1,2\}$.  We denote by $split(t)$
the number of splitting nodes of $T_t$

The \emph{transitional number} $trans(t)$ of $T$
is $|CL(T_t \setminus t) \cap {\bf TR}|$.
\end{definition}

\begin{lemma} \label{splitnumber}
For each $t \in V(T)$,
$split(t) \leq max(0,trans(t)-1)$
\end{lemma}

{\bf Proof.}
The proof is by induction  
from the leaves to the root of $t$. 
If $t$ is a leaf then both $split(t)$
and $trans(t)$ are zeroes, so the inequality 
clearly holds. 

Suppose that $t$ is not a leaf.
If $split(t)=0$ then the inequality again clearly holds,
so we assume that $split(t)>0$. 
Assume first that $t$ is not a splitting node.
Then $t$ has a child  $t_1$ such that 
$CL(T_t \setminus t) \cap {\bf TR}=CL(T_{t_1}) \cap {\bf TR}$. 
It follows that all the splitting nodes of $T_t$ are in $T_{t_1}$.
Consequently, $split(t)=split(t_1)>0$ and hence, by the induction assumption, 
$split(t_1) \leq trans(t_1)-1$.
As $trans(t_1) \leq |CL(T_{t_1}) \cap {\bf TR}|=trans(t)$, the statement follows.

It remains to assume that $t$ is a splitting node.
Let $t_1$ and $t_2$ be the children of $t$. 
For $i \in \{1,2\}$ let $C_i=CL(T_{t_i}) \cap {\bf TR}$. 
Due to one-sidedness $C_1 \cap C_2=\emptyset$. 
In particular, it follows that
\begin{enumerate}
\item $trans(t) \geq 2$, thus immediately implying the statement for the case where
both $split(t_1)$ and $split(t_2)$ are zeroes. 
\item For each $i \in \{1,2\}$, $trans(t) \geq trans(t_i)+1$. 
This means that, if say $split(t_1)>0$ while $split(t_2)=0$,
then taking into account the induction assumption,
$split(t)=split(t_1)+1 \leq trans(t_1)-1+1 \leq trans(t)-1$.
\item $trans(t) \geq trans(t_1)+trans(t_2)$.
In particular, if both $split(t_1)>0$ and $split(t_2)>0$ hold
then $split(t)=split(t_1)+split(t_2)+1 \leq (trans(t_1)-1)+(trans(t_2)-1)+1 \leq transt(t)-1$,
the first equality accounts $1$ for $t$, the first ineuqality follows from the induction
assumption.
\end{enumerate}
$\blacksquare$

\begin{definition}
In this definition we introduce three notions of a rank. 
\begin{itemize}
\item {\bf Ranks for types and special pairs}
We define the rank $rank(x)$ of each element $x$ of $\mathcal{ORD}$
under assumption that the ranks have been defined for all
the smaller elements. Let $x={\bf tp}$. If $t({\bf tp})$ is a leaf 
then $rank(x)=0$. Otherwise, $rank(x)=min_{y<_dx} rank(y)$.
If $x=({\bf tp},A)$, let ${\bf tp_1}$ be the main successor of
${\bf tp}$ through $A$. Then $rank(x)=rank({\bf tp_1})+1$
\item {\bf Ranks for the nodes of $T$.}
For a node $t$, the main rank $mrank(t)$, and the splitting rank $srank(t)$
are defined recursively from the leaves to the root as follows.
If $t$ is a leaf then both $marnks(t)$ and $srank(t)$ are zeroes.
If $t$ is not a splitting node then $srank(t)=0$ and $mrank(t)$ is the maximum
over all $mrank(t')$ such that $t'<t$ (descendant of $t$ in $T$).
If $t$ is a splitting node, let $t_1$ be the main child of $t$.
Then $srank(t)=mrank(t_1)+1$ and $mrank(t)$ is the maximum between
$srank(t)$ and all $mrank(t')$ such that $t'<t$. 
\end{itemize}
\end{definition}

\begin{lemma} \label{redranks}
For each $x \in \mathcal{ORD}$, the following two statements hold. 
\begin{enumerate}
\item If $x={\bf tp}$ then $rank(x) \leq mrank(t({\bf tp}))$.
\item If $x=({\bf tp},A)$  then $rank(x) \leq srank(t({\bf tp}))$. 
\end{enumerate}
\end{lemma}

{\bf Proof.}
By induction 
from the leaves to the root of $T$.
Let $x \in \mathcal{ORD}$ be such that $t=t({\bf tp}(x))$ is a leaf. 
Since $t$ is not a splitting node, $x$ is not a special pair. 
By definition, $rank(x)=0$ and hence the statement holds because 
the $mrank$ function is non-negative. 

Assume now that $x={\bf tp}$ such that $t=t({\bf tp})$ is not
a splitting node. By definition, there is 
$y<_d x$ such that $rank(x)=rank(y)$.
As $t$ is not a splitting node, $y$ cannot be a special pair with type equal ${\bf tp}$.
Let ${\bf tp'}={\bf tp}(y)$ Then 
$t'=t({\bf tp})$ is a descendant of of $t$ in $T$.

This means that either $y={\bf tp'}$ or $y$ is a special
pair with the type equal ${\bf tp'}$.
It is not hard to observe that in the latter case ${\bf tp'}$ 
is also a descendant of ${\bf tp}$  (as the only parent of a
special pair is its type). Furthermore, in the latter case $rank({\bf tp'}) \geq rank(y)$.
Therefore, we may assume that $y={\bf tp'}$. 
Then, by the induction assumption
and definition of $mrank$,
$rank({\bf tp})=rank({\bf tp'})  \leq mrank(t')  \leq mrank(t)$.

Assume now that $x=({\bf tp},A)$. 
Let $t=t({\bf tp})$ and let $t_1$ be the main child of $t$. 
Let ${\bf tp_1}$ be the $(t_1,A)$-successor. 
Then, by definition of $rank$, and the induction assumption
$rank(x)\leq rank({\bf tp_1})+1 \leq mrank(t_1)+1=srank(t)$.

Finally, we assume that $x={\bf tp}$ such that
$t=t({\bf tp})$ is a splitting node.
Let $y<_d x$ be an element of $\mathcal{ORD}$ such that
$rank(x)=rank(y)$. 
If $y$ is a special pair $({\bf tp},A)$ then,
by the induction assumption, $rank(x) \leq srank(t) \leq mrank(t)$.
Otherwise, we 
apply the same argumentation as for the case where 
$t$ is not a splitting node. 
$\blacksquare$

\begin{lemma} \label{mrankbound}
For each $t \in V(T)$, 
$mrank(t) \leq max(0, \lceil \log(trans(t)) \rceil)$. 
\end{lemma}

{\bf Proof.}
By induction from the leaves to the root of $t$.
If $t$ is a leaf then $mrank(t)=0$ so the statement clearly holds.

Suppose that $t$ is not a leaf.
If $mrank(t)=0$ then the inequality again clearly holds,
so we assume that $mrank(t)>0$. 
Assume first that $t$ is not a splitting node.
Then $t$ has a child  $t_1$ such that 
$CL(T_t \setminus t) \cap {\bf TR}=CL(T_{t_1}) \cap {\bf TR}$. 
Consequently, $mrank(t)=mrank(t_1)>0$ and hence, by the induction assumption, 
$mrank(t_1) \leq \lceil \log(trans(t_1)) \rceil$.
As $trans(t_1) \leq |CL(T_{t_1}) \cap {\bf TR}|=trans(t)$, the statement follows.

It remains to assume that $t$ is a splitting node.
Let $t_1$ and $t_2$ be the children of $t$ with $t_1$ being the main child.
As $mrank$ is monotone non-decreasing on the direction from the leaves
to the root, $mrank(t)=max(mrank(t_1)+1,mrank(t_2))$ 
For $i \in \{1,2\}$ let $C_i=CL(T_{t_i}) \cap {\bf TR}$. 
Due to one sidedness $C_1 \cap C_2=\emptyset$. 
In particular, it follows that $trans(t) \geq 2$ thus immediately 
implying the statement in case both $mrank(t_1)$ and $mrank(t_2)$
are zeroes. Otherwise, assume first that $mrank(t)=mrank(t_2)$.
Then $mrank(t_2) \geq mrank(t_1)+1 \geq 0$ and the argumentation in the previous
paragraph applies. 

It remains to assume that $mrank(t)=mrank(t_1)+1$.
If $mrank(t_1)=0$, the statement is immediate since $trans(t) \geq 2$,
so we assume that $mrank(t_1)>0$
By definition of the main child, $|C_1| \leq trans(t)/2$.
Further on, $trans(t_1) \leq |C_1|$.
It follows that $\lceil \log(trans(t_1)) \rceil \leq \lceil \log(trans(t)) \rceil-1$.
By the induction assumption,
$mrank(t_1) \leq \lceil \log(trans(t)) \rceil-1$.
Hence, the statement holds.
$\blacksquare$

Recall, that by definition, $\mathcal{ORD}$ is the disjoint union
of the set $\mathcal{NS}$ of types and the set $\mathcal{SP}$
of special pairs. Let $ns=|\mathcal{NS}|$ and $sp=|\mathcal{SP}|$.
We observe that the definitions of local graphs $L_{\bf tp}$
and especially $L_{({\bf tp},A)}$ are not completely deterministic
in the sense that there may be several possible graphs meeting
the requirements of the respective definitions.
In the next lemma, we claim that these graphs can be such that a good 
upper bound can be claimed on the size of the respective $R$-graphs.

\begin{lemma} \label{sizesoftypes}
There are local graphs for the elements of $\mathcal{ORD}$
matching their respective definitions 
so that the following size upper bounds are observed. 
\begin{enumerate}
\item Let ${\bf tp} \in \mathcal{NS}$.
Then $|R_{u_{\bf tp}}| \leq 2^{k+1} \cdot ns \cdot \sum_{i=0}^r sp^i$
where $r$ is the rank of ${\bf tp}$.
\item Let $({\bf tp},A) \in \mathcal{SP}$. 
Then $|L_{({\bf tp},A}| \leq 2^{k+1} \cdot ns \cdot \sum_{i=0}^{r-1} sp^i$, 
where $r$ is the rank of $({\bf tp},A)$. 
\end{enumerate}
\end{lemma}

{\bf Proof.}
It is not hard to infer from the combination of Corollary \ref{transfals}
and Lemma \ref{assemble2} that for each $a_i \in \mathcal{ORD}$,
$R_{u_i}$ is the union of $L_i$ and all the $L_j$ such that $a_j$ is a descendant
of $a_i$. Therefore, for $a_i={\bf tp}$,
instead of $|R_{u_{\bf tp}}|$, we can use
$\sum_{a_j \leq_d a_i} |L_i|$.

Note also that for each ${\bf tp} \in \mathcal{NS}$ 
we can assume that $|L_{\bf tp}| \leq 2^{k+1}$ simply because
$L_{\bf tp}$ is a decision tree over at most $k$ variables.  

The proof is by induction along $\mathcal{ORD}$. 
Consider first the smallest element of $\mathcal{ORD}$.
This element is some ${\bf tp}$. In fact, we make a more general assumption
that $rank({\bf tp})=0$.
In this case, all the successors of ${\bf tp}$ are elements of $\mathcal{NS}$.
Hence, taking into account the reasoning in the first two paragraphs, $|R_{\bf tp}| \leq ns \cdot 2^{k+1}$.
In particular, the statement of the lemma holds for this case.

Consider now an element $x \in \mathcal{ORD}$ under assumption 
that the statement of the lemma holds for all the smaller elements of $\mathcal{ORD}$.
Assume first that $x=({\bf tp},A)$.
Let ${\bf tp_1}$ be the main successor of ${\bf tp}$,
$t=t({\bf tp})$ and $t_1=t({\bf tp_1})$  
By construction, $L_{({\bf tp},A)}$ is a transitional resolution for
$h_{\bf tp_1}$ under additional constraint that its set of variables
must be a subset of $Var(T_{t_1}) \setminus Var(t)$. 
It turns out that a good upper bound on such a transitional resolution
follows from the induction assumption. 
Indeed, ${\bf tp_1} \in \mathcal{ORD}$ and precedes $({\bf tp},A)$. 
By Theorem \ref{validtypes}, $R_{\bf tp_1}$ is a transitional resolution for $h_{\bf tp_1}$
with the required constraint on its set of variables, hence any upper bound
on its size can be used to upper-bound the size of $L_{({\bf tp},A)}$.
By definition, of the rank of $({\bf tp},A)$ the rank of ${\bf tp_1}$
is $r-1$. Therefore, by the induction assumption, 
$|R_{u_{\bf tp_1}}| \leq 2^{k+1} \cdot ns \cdot \sum_{i=0}^{r-1} sp^i$
thus confirming the lemma for the considered case. 

It remains to assume that $x \in \mathcal{NS}$, that is $x={\bf tp}$. 
Then we can upper bound the size of $R_{u_{\bf tp}}$ by the total size
of $ns$ local graphs of size at most $2^{k+1}$ each and at most
$sp$ special pairs of rank at most $r$ each (since these special pairs must be descendants of ${\bf tp}$).
Applying the induction assumption to the special pairs, observe that the sizes
of their respective local graphs are at most $2^{k+1} \cdot ns \cdot (\sum_{i=1}^{r-1} sp^i+1)$.
Thus $|R_{u_{\bf tp}}| \leq 2^{k+1} \cdot ns+ sp \cdot (2^{k+1} \cdot ns \cdot \sum_{i=0}^{r-1} sp^i=
2^{k+1} \cdot ns \cdot \sum_{i=0}^r sp^i$.
$\blacksquare$

{\bf Proof of Theorem \ref{transres}.}
Let $rt$ be the root of $T$.
Consider the type ${\bf tp}=(rt,\emptyset,\emptyset,\emptyset)$.
It is not hard to see that $h_{\bf tp}={\bf 1}_{\varphi}$. It follows that
${\bf tp} \in \mathcal{ORD}$ and hence, by Theorem \ref{validtypes}
$R_{u_{\bf tp}}$ is a transitional resolution for $\varphi$ and ${\bf TR}$.
Thus we need to demonstrate that the required upper bound holds
for $|R_{\bf tp}|$. For this purpose we employ Lemma \ref{sizesoftypes}
and conclude that 
$|R_{u_{\bf tp}}| \leq 2^{k+1} \cdot ns \cdot \sum_{i=1}^r sp^i$, 
where $r$ is the rank of ${\bf tp}$.
So, we need to upper-bound, $ns,sp$, and $r$. 

In order to upper-bound $ns$, let us compute for the given fixed $t$
the number of types ${\bf tp}$ such that $t=t({\bf tp})$. 
It is not hard to see that the number of such types is at most
the product of the possible elements of the second, third and fourth components,
that is $2^{|Var(t)|} \cdot 2^{|CL(t)} \cdot 2^{|{\bf TR}|}$. 
Taking into account that $|Var(t)|+|Cl(t)| \leq k$, we conclue that 
$2^{|Var(t)|} \cdot 2^{|CL(t)} \cdot 2^{|{\bf TR}|} \leq 2^k \cdot 2^{|{\bf TR}|}$.
Because of our assumption about $(T,{\bf B})$,
$|V(t)| \leq O(n)$ where $n=|Var(\varphi)|+|\varphi|$.
Thus $ns=O(n \cdot 2^k \cdot 2^{|{\bf TR}|})$.

In order to upper bound $sp$, we again fix $t \in V(t)$
and upper-bound the number of $({\bf tp},A)$ such that $t=t({\bf tp})$.
Taking into account that $Var(A) \subseteq Var(t)$, 
we obtain an upper bound of $4^k \cdot 2^{|{\bf TR}|}$.
Next, we observe that if $t=t({\bf tp})$ and $({\bf tp},A)$ is a special pair
then $t$ is a splitting node. 
This means that $sp$ is at most the number of splitting nodes of $T$
multiplied by $4^k \cdot 2^{|{\bf TR}|}$.
Applying Lemma \ref{splitnumber} and taking into account that
$trans(t) \leq |{\bf TR}|$ for each $t \in V(T)$, we conclude that
$sp \leq |{\bf TR}| \cdot 4^k \cdot 2^{|{\bf TR}|}$.

In order to upper bound $r$, we combine the inequality $trans(t) \leq |{\bf TR}|$
obtained in the previous paragraph and Lemmas
\ref{redranks} and \ref{mrankbound} to observe that 
$r \leq max(0,\lceil \log(|{\bf TR}|) \rceil)$. 
Substituting the data in the formula in the end of the first paragraph of this proof
and performing elementary transformations, 
we conclude that assuming that $|{\bf TR} \neq \emptyset$,
a transitional resolution for $\varphi$ can be upper bounded by
$n\cdot 2^{O(k \cdot \log(|{\bf TR}|)+ \log^2(|{\bf TR}|))}$.
If ${\bf TR}=\emptyset$ then the upper bound becomes $O(4^k)$.
$\blacksquare$

\section{Proof of Theorem \ref{mainres}}
We first prove two propositions concerning postorder traversals. 

\begin{proposition} \label{expcontr}
Let $\pi$ be a proper prefix of $\pi_T$ and let $t$ be its 
immediate successor. Then the following statements hold.  
\begin{enumerate}
\item If $t$ is an expanding node then 
$Trees_{\pi+t}=Trees_{\pi} \cup \{T_t\}$
and $T_t$ is the \emph{last} element of $Trees_{\pi+t}$.
The order between the remaining elements is the same as in $Trees_{\pi}$. 
\item Assume that $t$ is a contracting node.
and let $Trees'_{\pi}$ be the set of trees rooted by the children of $t$. 
Then $Trees'_{\pi}$ are the \emph{last elements} of $Trees_(\pi)$, 
$Trees_{\pi+t}=(Trees_{\pi} \setminus Trees'_{\pi}) \cup \{T_t\}$,
$T_t$ is the last node of  $Trees_{\pi+t}$,  and the order of the rest of
the elements is preserved as in $Trees_{\pi}$. 
\end{enumerate}
\end{proposition} 

{\bf Proof.}
By induction on $|V(T)|$
The statement is clearly true for $|V(T)|=1$
so assume that $|V(T)|>1$. 

Let $rt$ be the root of $T$.
Assume that $t=rt$.
Then clearly $t$ is a contracting node and $Trees_{\pi}=Trees'_{\pi}$.
It is not hard to see that $Trees(\pi+t)=\{T_{rt}\}$, hence the statement
holds for this case.

Let $t_1$ be the left child of $rt$ and let $\pi_1$ be the 
postorder traversal of $T_{t_1}$. 
Assume that $t \in T_{t_1}$. Then, by the induction assumption, the
statement is correct regarding $\pi_1$. Clearly, it remains correct if 
$\pi$ is used instead of $\pi_1$. 
It remains to assume that $rt$ has two children. 
Let $t_2$ be the right child of $rt$. and let $\pi_2$ 
be the postorder traversal of $T_{t_2}$. 
Since the previous cases do not hold,
$\pi=\pi_1+\pi'$. Morveor, $t$ is the immediate
successor of $\pi'$ in $\pi_2$. 
By the induction assumption, the statement holds regarding $\pi'$ and $t$
w.r.t. $\pi_2$. As $Trees_{\pi'+t}$ (and their order) are the same w.r.t. $\pi_2$ and $\pi_T$
and $T_{t_1}$ is located before all the elements of $Trees_(\pi'+t)$, the statement holds
regarding $\pi$ and $t$.
$\blacksquare$

\begin{proposition} \label{minpresexp}
Let $\pi$ be a proper prefix of $\pi_T$ and let $t$ be the immediate 
successor of $\pi$. Let $x \in Var(\pi)$.
\begin{enumerate}
\item Assume that $t$ is an expanding node.
 Then $MT_{\pi}(x)=MT_{\pi+t}(x)$.
\item Assume that $t$ is a contracting node. 
If $MT_{\pi}(x) \notin Trees'_{\pi}$ then $MT_{\pi}(x)=MT_{\pi+t}(x)$. 
Otherwise, $MT_{\pi+t}(x)=T_t$. 
\end{enumerate}
\end{proposition}

{\bf Proof.}
Assume that $t$ is an expanding node.
Let $T'=MT_{\pi}(x)$.
By Proposition \ref{expcontr}, $T' \in Trees_{\pi+t}$.
Suppose that $T'' \in Trees_{\pi+t}$ such that $x \in Var(T'')$
and $T''<T'$. Again by Proposition \ref{expcontr}, $T'' \in Trees_{\pi}$
and $T''<T'$ in contradiction to the minimality of $T''$.

Assume now that $t$ is a contracting node with $T'$
being as above.  If $T' \notin Trees'_{\pi}$  then 
$MT_{\pi}(x)=MT_{\pi+t}(x)$ by the same argument as in the 
previous paragraph. 
Otherwise,  by Proposition \ref{expcontr}, $x \in Var(T_t)$.
Suppose there is $T'' \in Trees_{\pi+t}$ such that $x \in Var(T'')$
and $T'' <T_t$. By Proposition \ref{expcontr}, 
$T'' \in Trees_{\pi}$ and $T''$ is smaller than each of $Trees'_{\pi}$,
a contradiction to the minimality of $T'$.
$\blacksquare$


When we consider the types of $\mathcal{ORD}$, we in fact concentrate 
on specific types that are called \emph{well-formed}. We define them below.

\begin{definition} \label{detwit}
Let $x \in Var_{fix}({\bf tp})$.
We denote by $W_{\bf tp}(x)$ the set of witnessing clauses of $x$
as per Definition \ref{innfixout}. 
$W_{\bf tp}(x)$ is the disjoint union of two sets $W^0_{\bf tp}(x)$ 
and $W^1_{\bf tp}(x)$ where $W^0_{\bf tp}(x)$ consists of the clauses
satisfying part (i) of definition of a witnessing clause as per Definition \ref{innfixout}
and $W^1_{\bf tp}(x)$ are those satisfying part (ii). 
\end{definition}
 
\begin{definition}
Continuing on Definition \ref{detwit},
we say that $x$ is \emph{consistent}  if it has \emph{the same} 
occurrence in all the clauses of $W_{\bf tp}(x)$.
Put it differently, $x$ is consistent if it either occurs positively in all
the clauses of $W_{\bf tp}(x)$ or occurs negatively in all of them. 
We say that ${\bf tp}$ is \emph{well-formed} if all the fixed variables
are consistent. 
\end{definition}

The rest of the proof is divided into five subsections.
In the first four subsections we prove that sinks of local
graphs of non-basic non-final types ${\bf tp}$  satisfy the validity conditions
in terms of Theorem \ref{assembly1}.
Each subsection is devoted to a specific condition concerning the kind
of the immediate successor $t$ of $\pi({\bf tp})$ and the assignment $A(u)$
of the considered sink. The actual proof, gathering all facts together,
is provided in the fifth subsection.  

\subsection{Expanding $t$, $A(u)$ does not satisfy new long clauses}
\begin{lemma} \label{lem:extnonsat}
Let ${\bf tp}=(\pi,map,CN,RA)$ be a well-formed non-basic non-final unsatisfiable type and let
$t$ be the immediate successor of $\pi$. assume that $t$ is an expanding
node.
Let $u$ be a sink of $L_{\bf tp}$ such that $A(u)$ does not satisfy any 
clauses of ${\bf LONG} \setminus {\bf LS}$.
Let ${\bf tp'}=(\pi({\bf tp})+t,map,CN',RA)$
where $CN'$ is obtained from $CN$ by removal of clauses that are satisfied by $A(u)$.
Then the following statements hold.
\begin{enumerate}
\item For each $x \in Var_{fix}({\bf tp}) \cap Var_{fix}({\bf tp'})$,
$W_{\bf tp}(x)=W_{\bf tp'}(x)$. 
\item  $Var_{fix}({\bf tp'})$ is  the disjoint union of $Var_{fix}({\bf tp})$ 
and $Det({\bf tp})$.
\item For each $x \in Det({\bf tp})$, $W_{\bf tp'}(x)=W^0_{\bf tp}(x)$.
\item $Var(A(u))=Det({\bf tp})$.
\item $FA({\bf tp'})=FA({\bf tp}) \cup A(u)$.
\item $Var_{out}({\bf tp})$ is the disjoint union of $Var_{out}({\bf tp'})$
and $Det({\bf tp})$.
\end{enumerate}
\end{lemma}

{\bf Proof.}
\begin{claim} \label{clm201}
Let $x \in Var_{fix}({\bf tp})$. Then $x \in Var_{fix}({\bf tp'})$
and $W_{\bf tp}(x) \subseteq W{\bf tp'}(x)$.
\end{claim}

{\bf Proof.}
First of all, it follows directly from definition of ${\bf tp'}$
that $Var_{in}({\bf tp})=Var_{in}({\bf tp'})$.
Then, $x \in Var(\pi+t) \setminus Var_{in}({\bf tp'})$. 

Let $C \in W_{\bf tp}(x)$.
Assume first that $C \in W^0_{\bf tp}(x)$.
By definition of ${\bf tp'}$, $[map({\bf tp'})](C)=none$ 
It follows that $x \in Var_{fix}({\bf tp'})$ and that $C \in W_{\bf tp'}(x)$.
Assume now that $C \in W^1_{\bf tp}(x)$. 
By definition of ${\bf tp'}$,  $[map({\bf tp}](C)=[map({\bf tp'}](C)$.
By Proposition \ref{minpresexp}, $MT_{\pi+t}(x)=MT_{\pi}(x)$.
As $MT_{\pi}(x)<[map({\bf tp})](C)$ (due to $C$ being witnessing for $x$
in ${\bf tp}$), we conclude that 
$MT_{\pi+t}(x)<[map({\bf tp'})](C)$ thus confirming that
$x \in Var_{fix}({\bf tp'})$ and that $C \in W_{\bf tp'}(x)$.
$\square$

The proof of the following claim is analogous 
to the proof of Claim \ref{clm201} with ${\bf tp}$
and ${\bf tp'}$ exchanging their roles. 

\begin{claim} \label{clm202}
Let $x \in Var_{fix}({\bf tp'}) \cap Var(\pi)$.
Then $x \in Var_{fix}({\bf tp})$ and 
$W_{\bf tp'}(x) \subseteq W_{\bf tp}(x)$.
\end{claim}

Taking into account that $Var_{fix}({\bf tp}) \subseteq Var(\pi)$
by definition, Claim \ref{clm201} in fact shows that
$Var_{fix}({\bf tp}) \subseteq Var_{fix}({\bf tp'}) \cap Var(\pi)$. 
Hence, combination of Claim \ref{clm201} and \ref{clm202}
imply that these two sets are equal and the witnessing sets
of each variable are the same w.r.t ${\bf tp}$ and ${\bf tp'}$.
This implies the first statement of the lemma. 

Let now $x \in Var_{fix}({\bf tp'}) \setminus Var(\pi)$. 
Then $MT_{\pi+t}(x)=T_t$, the largest tree of $Trees(\pi+t)$.
Hence $W^1_{\bf tp'}(x)=\emptyset$, implying that $W_{\bf tp'}(x)=W^0_{\bf tp'}(x)$
and, in particular (since $x \in Var_{fix}({\bf tp'})$ must be witnessed by at least 
one clause), that $W^0_{\bf tp'}(x) \neq \emptyset$.

Let $C \in W^0_{\bf tp'}(x)$. Then $C \in {\bf LONG} \setminus {\bf LS}({\bf tp'})=
{\bf LONG} \setminus {\bf LS}({\bf tp})$. As all the variables outside of $Var(\pi)$ belong 
to $Var_{out}({\bf tp})$, $x \in Var(t) \cap Var_{out}({\bf tp})$. Hence $x \in Det({\bf tp})$.

Conversely, let $x \in Det({\bf tp})$. This means that there is 
$C \in {\bf LONG} \setminus {\bf LS}({\bf tp})={\bf LONG}  \setminus {\bf LS}({\bf tp'})$.
such that $x \in Var(C)$. This already implies that $x \in Var_{fix}({\bf tp'})$. 
It remains to show that $x \notin Var(\pi)$. But if it were so then $x$ would belong
to $Var_{fix}({\bf tp})$ causing a contradiction since $x \in Var_{out}({\bf tp})$
by definition of $Det({\bf tp})$.  Thus the second and the third statements are settled.

For the fourth statement, note that, by construction, $Det({\bf tp}) \subseteq Var(A(u))$.
If $A(u)$ assigned any extra variables, it would have satisfied a 
clause of ${\bf LONG} \setminus {\bf LS}({\bf tp})$ which is not the
case. 

For the fifth statement, remains to show that for each $x \in Var_{fix}({\bf tp})$,
the occurrences of $x$ in $FA({\bf tp})$ and $FA({\bf tp'})$ are the same and 
that for each $x \in Det({\bf tp})$, the occurrences of $x$ in $A(u)$ 
and in $FA({\bf tp'})$ are the same. In the former case, the desired statement
follows from $W_{\bf tp}(x)=W_{\bf tp'}(x)$. In the latter case, the assignment
of $x$ in $A(u)$ is precisely the one to not satisfy any witnessing clause of $x$,
hence precisely the assignment of $x$ in $FA({\bf tp'})$. 

For the last statement, note that
$Var_{out}({\bf tp'})=
 Var(\varphi) \setminus (Var_{in}({\bf tp'}) \cup Var_{fix}({\bf tp'})=
Var(\varphi) \setminus (Var_{in}({\bf tp}) \cup Var_{fix}({\bf tp}) \cup Det({\bf tp}))=
(Var(\varphi) \setminus (Var_{in}({\bf tp}) \cup Var_{fx}({\bf tp})) \setminus Det({\bf tp})=
Var_{out}({\bf tp}) \setminus Det({\bf tp})$.
$\blacksquare$

\begin{theorem} \label{succextnonsat}
Let ${\bf tp}=(\pi,map,CN,RA)$ be a well-formed non-basic non-final unsatisfiable type and let
$t$ be the immediate successor of $\pi$. assume that $t$ is an expanding
node.
Let $u$ be a sink of $L_{\bf tp}$ such that $A(u)$ does not satisfy any 
clauses of ${\bf LONG} \setminus {\bf LS}$.
Then the following statements hold.
\begin{enumerate}
\item If $u$ is a labelled with $C \rightarrow ()$ 
then $C \in dom(F_{\bf tp})$, $C$ is not satisfied by $A(u)$
and $F(C)|_{A(u)}=()$.
\item If $u$ is identified with the source of $L_{\bf tp'}$
then ${\bf tp'}$ is a well-formed non-basic unsatisfiable type  
and $F_{\bf tp'}=F_{\bf tp}|_{A(u)}$. 
\end{enumerate}
\end{theorem}

{\bf Proof.}
The first statement is immediate by construction, so we prove the second statement. 
Assume the premise of the statement regarding $u$. 
By construction, ${\bf tp'}=(\pi({\bf tp})+t,map,CN',RA)$
where $CN'$ is obtained from $CN$ by removal of clauses that are satisfied by $A(u)$.
In order to show that ${\bf tp'}$ is a well-formed type, we demonstrate that 
each variable $x \in Var_{fix}({\bf tp'})$ is consistent.
If $x \in Var_{fix}({\bf tp'}) \cap Var_{fix}({\bf tp})$ then the consistency of $x$ follows from the
well-formity of ${\bf tp}$ by the first statement of Lemma \ref{lem:extnonsat}.
Otherwise, by the second statement of Lemma \ref{lem:extnonsat}, $x \in Det_{\bf tp}$. 
For the sake of contradiction, assume that there are two clauses of $W_{\bf tp'}(x)$
such that $x$ occurs positively, say in $C_1$ and negatively in $C_2$. Then the assignment
of $x$ in $A(u)$ must satisfy one of the clauses. However, by the third statement of
Lemma \ref{lem:extnonsat} $C_1,C_2 \in {\bf LONG} \setminus {\bf LS}({\bf tp})$ and $A(u)$
does not satisfy any of these clauses by assumption.

Let $C \in dom(F_{\bf tp}|_{A(u)}) \cap dom(F_{\bf tp'})$. Then $F_{\bf tp}(C)|_{A(u)}=Proj(C,Var_{out}({\bf tp}))|_{A(u)}=
Proj(C,Var_{out}({\bf tp}) \setminus Var(A(u)))=Proj(C,Var_{out}({\bf tp'}))=F_{\bf tp'}(C)$.
The penultimate equality follows form the combination of statements 4 and 6 of 
Lemma \ref{lem:extnonsat}.
It follows that it is enough to show that $dom(F_{\bf tp'})=dom(F_{\bf tp}|_A)$. 
Since ${\bf LS}({\bf tp})={\bf LS}({\bf tp'})$,
$dom(F_{\bf tp}) \cap {\bf LONG}={\bf LONG} \setminus {\bf LS}({\bf tp})={\bf LONG} \setminus {\bf LS}({\bf tp'})=dom(F_{\bf tp'}) \cap {\bf LONG}$.
Since, by assumption, $A(u)$ does not satisfy any clause of ${\bf LONG} \setminus {\bf LS}({\bf tp})$,
we conclude that $dom(F_{\bf tp}|_{A(u)}) \cap {\bf LONG}=dom(F_{\bf tp}) \cap {\bf LONG}$. 

Next, recall that ${\bf MT}({\bf tp})={\bf MT}({\bf tp'})$ and let $CL^*=CL(Roots({\bf MT}({\bf tp})))$. 
By definition,
$dom(FT_{\bf tp'}) \cap CL^*=CN'$ while $dom(F_{\bf tp}) \cap CL^*=CN$.
This means that $dom(F_{\bf tp}|_{A(u)}) \cap CL^*$ is the subset of $CN$ consisting of the 
clauses that are not satisfied by $A(u)$. But, by definition, this is exactly $CN'$!

Now, let $C \in dom(F_{\bf tp'}) \setminus ({\bf LONG} \cup CL^*)$. 
This means that $Var(C) \cap Var_{out}({\bf tp'}) \neq \emptyset$
and $C$ is not satisfied by $RA({\bf tp'}) \cup FA({\bf tp'})$.
By construction and Lemma \ref{lem:extnonsat},
$Var(C) \cap Var_{out}({\bf tp}) \neq \emptyset$,
$C$ is not satisfied by $RA({\bf tp}) \cup FA({\bf tp})$
and also not satisfied by $A(u)$. We conclude that $C \in dom(F_{\bf tp}|_{A(u)})$. 
Conversely, suppose that $C \in dom(F_{\bf tp}|_{A(u)})$.  
This means that $C$ is not satisfied by 
$RA({\bf tp}) \cup FA({\bf tp}) \cup A(u)=RA({\bf tp'}) \cup FA({\bf tp'})$
It remains to see that $Var(C) \cap Var_{out}({\bf tp'}) \neq \emptyset$. 
By Lemma \ref{lem:extnonsat}, the opposite can happen only if $F_{\bf tp}(C)|_{A(u)}=()$. 
But, in this case, the premise of the first statement is satisfied that is $u$
is a falsifying sink of $L_{\bf tp}$, which is not the case by assumption. 

It remains to be added that, based on the previous paragraph, the unsatisfiability of 
${\bf tp''}$ follows from the unsatisfiability of ${\bf tp}$ since this implies that
$F_{\bf tp}|_{A(u)}$ is unsatisfiable. Also ${\bf tp''}$ because $A(u)$ does not falsify
any $F_{\bf tp}(C)$ for any $C \in dom(F_{\bf tp})$ simply by assumption.  

$\blacksquare$

\subsection{Expanding $t$, $A(u)$ satisfies new long clauses}
Now, we are going to consider the case where $t$ is an expanding node
and $A(u)$ satisfies some clauses mapped to $none$ by $map$.
Throughout the consideration,   
${\bf tp'}=(\pi({\bf tp})+t,map',CN^*,RA')$
where $map'$ is obtained from $map$ by replacing $C \rightarrow none$ by $C \rightarrow last(Trees_{\pi({\bf tp})+t})$
for each $C \in {\bf LONG} \setminus {\bf LS}$ that is satisfied by $A(u)$.
$CN^*$  s obtained from $CN$ by removal of all clauses that are satisfied by $A(u)$
and adding those clauses of $dom(F_{\bf tp}) \cap CL(t)$ that are not satisfied by $A(u)$.
Finally $RA'=RA \cup A(u) \cup Proj(FA({\bf tp}), Var(t))$.

\begin{lemma} \label{validextsat}
The following statements hold.
\begin{enumerate}
\item $Var_{fix}({\bf tp'})=Var_{fix}({\bf tp}) \setminus Var(t)$.
\item For each $x \in Var_{fix}({\bf tp'})$, $W_{\bf tp}(x)=W_{\bf tp'}(x)$. 
\end{enumerate}
\end{lemma}

{\bf Proof.}
Let $x \in Var_{fix}({\bf tp}) \setminus Var(t)$.
Let $C \in W_{\bf tp}(x)$.
If $C \in W^1_{\bf tp}(x)$ then 
$map(C) \in Trees_{\pi}$.
By construction, $map'(C)=map(C)$. 
By Proposition \ref{minpresexp},
$MT_{\pi+t}(x)=MT_{\pi}(x)<map(C)=map'(C)$. 

Assume now that $C \in W_{\bf tp}^0(x)$.
If $C$ is not satisfied by $A(u)$ then $map'(C)=none$
by construction and we are done since 
$C \in W_{\bf tp'}^0(x)$. 
Finally, assume that $C$ is satisfied by $A(u)$. 
Let $y \in Var(A(u))$ be a variable whose occurrence
in $A(u)$ satisfies $C$. Then $MT_{\pi+t}(y)=T_t$.
Indeed, otherwise, $y \in Var(\pi)$ and hence,
by construction, $y$ cannot belong to $Var_{out}({\bf tp})$,
a contradiction. 
As $x \in Var(\pi)$. $MT_{\pi+t}(x)<map'(x)$
meaning that $x \in W_{\bf tp'}(x)$. 

Conversely, let $x \in Var_{fix}({\bf tp'})$.
By construction, $x \in Var(\pi) \setminus Var(t)$.
Let $C \in W_{\bf tp'}(x)$. If $C \in W^0_{\bf tp'}(x)$
meaning that $map'(C)=none$ then, clearly
$C \in W^0({\bf tp})(x)$ and hence $x \in Var_{fix}({\bf tp})$. 
If $C \in W^1_{\bf tp'}(x)$ then we consider two subcases. 
The first is if $map(C) \neq none$. In this case 
$MT_{\pi+t}(x)=MT_{\pi}(x)<map(C)=map'(C)$,
the first statement follows from Proposition \ref{minpresexp}.
Hence $C \in W_{\bf tp}(x)$ and hence $x \in V_{\bf fix}({\bf tp})$.
In the second case, $map(C)=none$ and hence $C \in W_{\bf tp}^0(x)$
and again $x \in Var_{fix}({\bf tp})$. 
$\blacksquare$

\begin{lemma} \label{newparamextsat}
The following two statements hold.
\begin{itemize}
\item $Var_{out}({\bf tp})=Var_{out}({\bf tp'}) \cup Var(A(u))$
\item $FA({\bf tp'}) \cup RA({\bf tp'})=FA({\bf tp}) \cup RA({\bf tp}) \cup A(u)$
\end{itemize}
\end{lemma}

{\bf Proof.}
By the first statement of Lemma \ref{validextsat},
$Var_{fix}({\bf tp'})=Var_{fix}({\bf tp}) \setminus Var(t)$.
On the other hand, by construction, 
$Var_{in}({\bf tp'})$ is the disjoint union of 
$Var_{in}({\bf tp}) \setminus Var(t)$, $Var_{\bf fix}({\bf tp}) \cap Var(t)$,
$Var_{in}({\bf tp}) \cap Var(t)$ and $Var(A(u))$.
By regrouping the items, we obtain
$Var_{fix}({\bf tp'}) \cup Var_{in}({\bf tp'})=Var_{fix}({\bf tp}) \cup Var_{in}({\bf tp}) \cup Var(A(u))$
The first statement now follows by taking the complement of both sides.

It follows from Lemma \ref{validextsat}
that $FA({\bf tp'})=Proj(FA({\bf tp}, Var_{fix}({\bf tp}) \setminus Var(t))$.
Next, by construction, $RA({\bf tp'})$ is the disjoint union of
$RA({\bf tp})$, $Proj(FA({\bf tp}),Var(t))$ and $A(u)$. 
Clearly, their union is the right-hand part of the second statement.
$\blacksquare$

\begin{theorem} \label{succextsat}
Let ${\bf tp}=(\pi,map,CN,RA)$ be a well-formed non-basic non-final unsatisfiable type and let
$t$ be the immediate successor of $\pi$. assume that $t$ is an expanding
node.
Let $u$ be a sink of $L_{\bf tp}$ such that $A(u)$ satisfies
clauses of ${\bf LONG} \setminus {\bf LS}$.
Then the following statements hold.
\begin{enumerate}
\item If $u$ is a labelled with $C \rightarrow ()$ 
then $C \in dom(F_{\bf tp})$, $C$ is not satisfied by $A(u)$
and $F(C)|_{A(u)}=()$.
\item If $u$ is identified with the source of $L_{\bf tp'}$
then ${\bf tp'}$ is a well-formed non-basic unsatisfiable type  
and $F_{\bf tp'}=F_{\bf tp}|_{A(u)}$. 
\end{enumerate}
\end{theorem}

{\bf Proof.}
The first statement is immediate by construction, so we prove the second statement. 
Assume the premise of the statement regarding $u$. 
By construction, ${\bf tp'}=(\pi({\bf tp})+t,map,CN^*,RA)$ is as defined
in the paragraph preceding Lemma \ref{validextsat}.

By the first statement of Lemma \ref{validextsat},
each $x \in Var_{fix}({\bf tp'})$ is a variable of $Var_{fix}({\bf tp})$
and hence the consistency of $x$ follows from the second statement
of Lemma \ref{validextsat} and the consistency of $x$ for ${\bf tp}$. 
We conclude that ${\bf tp'}$ is well-formed. 

Let $C \in dom(F_{\bf tp}|_{A(u)}) \cap dom(F_{\bf tp'})$
Then $F_{\bf tp}|_{A(u)}(C)=Proj(C,Var_{out})|_{A(u)}=
Proj(C,Var_{out}({\bf tp}) \setminus A(u))=Proj(C,Var_{out}({\bf tp'}))=F_{\bf tp'}(C)$,
the penultimate inequality follows from Lemma \ref{validextsat}. 
It follows that it is sufficient to establish that 
$dom(F_{\bf}|_{A(u)})=dom(F_{\bf tp'})$.

Let $C \in dom(F_{\bf tp}|_{A(u)})$.
If $C \in {\bf LS}({\bf tp})$ then, taking into account that $C$
is not satisfied by $A(u)$, $C \in {\bf LS}({\bf tp'})$ by construction
(and hence, by definition, belongs to $dom(F_{\bf tp'})$).

Assume that $C \in CN$.
Then as $C$ is not satisfied by $A(u)$, $C \in CN'$ 
(and, again, belongs to $dom(F_{\bf tp'})$.

Finally, let $C \notin {\bf LS}({\bf tp}) \cup CN$.
If $C \in CL(t)$ then $C \in CN'$ since it is not satisfied by $A(u)$.
Otherwise,
by definition of $F_{\bf tp}$, $C$ is not satisfied by
$FA({\bf tp}) \cup RA({\bf tp})$. Hence, by the second statement
of Lemma \ref{newparamextsat}, $C$ is not satisfied by
$FA({\bf tp'}) \cup RA({\bf tp'})$ and hence 
belongs to $dom(F_{\bf tp'})$ in the role of a clause outside
of ${\bf LS}({\bf tp'}) \cup CN'$. 

Conversely, assume that $C \in dom(F_{\bf tp'})$.
If $C \in {\bf LS}({\bf tp'})$ then, by definition 
$C \in {\bf LS}({\bf tp})$ and not satisfied by $A(u)$
and hence belongs to $dom(F_{\bf tp}|_{A(u)})$. 
Assume that $C \in CN'$. If $C \in CN$ then, taking into account
that $C$ is not satisfied by $A(u)$, we conclude that $C \in  dom(F_{\bf tp}|_{A(u)})$.
Otherwise the only way for $C$ to get into $CN'$ is if $C \in dom(F_{\bf tp})$ and
not satisfied by $A(u)$. 

Finally, if $C \notin {\bf LS}({\bf tp}) \cup CN'$ then, 
,by definition of ${\bf tp'}$,
$C$ is not satisfied by $FA({\bf tp'}) \cup RA({\bf tp'})=FA({\bf tp}) \cup RA({\bf tp}) \cup A(u)$
by Lemma \ref{validextsat}. 
By definition of ${\bf tp}$, $C \notin {\bf LS}({\bf tp}) \cup CN$.
Therefore $C \in dom(F_{\bf tp}|_{A(u)})$.
Finally, we note that ${\bf tp'}$ being non-basic and unsatisfiable follows from
the same argument as in the proof of Theorem \ref{succextnonsat}.
$\blacksquare$

\subsection{Contracting $t$, $A(u)$ not potentially complex}
\begin{lemma} \label{lem:contrnonsat}
Let ${\bf tp}=(\pi,map,CN,RA)$ be a well-formed non-basic non-final unsatisfiable type and let
$t$ be the immediate successor of $\pi$. assume that $t$ is a contracting
node.
Let $u$ be a sink of $D_{\bf tp}$ such that $A(u)$ 
is not potentially complex. 
Let ${\bf tp'}=(\pi({\bf tp})+t,map,CN',RA)$
where $CN'$ is obtained from $CN$ by removal of clauses that are satisfied by $A(u)$.
Then the following statements hold.
\begin{enumerate}
\item For each $x \in Var_{fix}({\bf tp}) \cap Var_{fix}({\bf tp'})$,
$W_{\bf tp}(x)=W_{\bf tp'}(x)$. 
\item  $Var_{fix}({\bf tp'})$ is  the disjoint union of $Var_{fix}({\bf tp})$ 
and $Det({\bf tp})$.
\item For each $x \in Det({\bf tp})$, $W_{\bf tp'}(x)=W^0_{\bf tp}(x)$.
\item $Var(A(u))=Det({\bf tp})$.
\item $FA({\bf tp'})=FA({\bf tp}) \cup A(u)$.
\item $Var_{out}({\bf tp})$ is the disjoint union of $Var_{out}({\bf tp'})$
and $Det({\bf tp})$.
\end{enumerate}
\end{lemma}

{\bf Proof.}
Unlike in the case of $t$ being an expanding note,
in the considered case the trees of $Trees'_{\pi}$ disappear
and we need to verify that ${\bf tp'}$ is valid in the sense that
none of the trees in the range of $map$ belongs to $Trees'_{\pi}$.
However, if it was so then $A(u)$ would be potentially complex.
So, we conclude that

\begin{claim} \label{claim300}
${\bf MT}(\pi) \subseteq Trees_{\pi} \cap Trees_{\pi+t}$. 
\end{claim}

\begin{claim} \label{clm301}
Let $x \in Var_{fix}({\bf tp})$. Then $x \in Var_{fix}({\bf tp'})$
and $W_{\bf tp}(x) \subseteq W_{\bf tp'}(x)$.
\end{claim}

{\bf Proof.}
First of all, it follows directly from definition of ${\bf tp'}$
that $Var_{in}({\bf tp})=Var_{in}({\bf tp'})$.
Then, $x \in Var(\pi+t) \setminus Var_{in}({\bf tp'})$. 

Let $C \in W_{\bf tp}(x)$.
Assume first that $C \in W^0_{\bf tp}(x)$.
By definition of ${\bf tp'}$, $[map({\bf tp'})](C)=none$.  
It follows that $x \in Var_{fix}({\bf tp'})$ and that $C \in W_{\bf tp'}(x)$.

Assume now that $C \in W^1_{\bf tp}(x)$. 
As $MT_{\pi}(x)<[map({\bf tp})](C)$ and $[map({\bf tp})](C) \in Trees_{\pi} \setminus Trees'_{\pi'}$
by Claim \ref{claim300}, $MT_{\pi}(x) \in Trees_{\pi} \setminus Trees'_{\pi}$ by Proposition \ref{expcontr}.
By Proposition \ref{minpresexp}, $MT_{\pi+t}(x)=MT_{\pi}(x)$.
As $[map({\bf tp'})](C)=[map({\bf tp})](C)$, $MT_{\pi+t}(x)<[map({\bf tp'}](C)$
by another application of Proposition \ref{expcontr}
thus implying that $x \in W^1_{\bf tp'}(x)$. $\square$

\begin{claim} \label{clm302}
Let $x \in Var_{fix}({\bf tp'}) \cap Var(\pi)$.
Then $x \in Var_{fix}({\bf tp})$ and 
$W_{\bf tp'}(x) \subseteq W_{\bf tp}(x)$
(note that the second part implies the first one). 
\end{claim}

{\bf Proof.}
Let $C \in W_{\bf tp'}(x)$.
Suppose that $C \in W_{\bf tp'}^0(x)$.
This means that $[map{\bf tp}](C)=none$ and, in particular,
that $C \in W_{\bf tp}^0(x)$. 

Assume that $C \in W^1_{\bf tp'}(x)$.
This means that $[map({\bf tp'})](C) \neq none$.
By Claim \ref{claim300}, $]map({\bf tp'})](C) \in Trees_{\pi} \cap Trees_{\pi+t}$. 
By Proposition \ref{expcontr}, $MT_{\pi+t}(x) \in Trees_{\pi} \cap Trees_{\pi+t}$.
By Proposition \ref{minpresexp}
$MT_{\pi+t}(x)=MT_{\pi}(x)$ and one more application of Proposition \ref{expcontr}
implies that $MT_{\pi}(x)<map(C)$.
We conclude that $C \in W^1_{\bf tp}(x)$.  $\square$

Taking into account that $Var_{fix}({\bf tp}) \subseteq Var(\pi)$
by definition, Claim \ref{clm301} in fact shows that
$Var_{fix}({\bf tp}) \subseteq Var_{fix}({\bf tp'}) \cap Var(\pi)$. 
Hence, combination of Claim \ref{clm301} and \ref{clm302}
imply that these two sets are equal and the witnessing sets
of each variable are the same w.r.t ${\bf tp}$ and ${\bf tp'}$.
This implies the first statement of the lemma. 

Let now $x \in Var_{fix}({\bf tp'}) \setminus Var(\pi)$. 
Then $MT_{\pi+t}(x)=T_t$, the largest tree of $Trees_{\pi+t}$.
Hence $W^1_{\bf tp'}(x)=\emptyset$, implying that $W_{\bf tp'}(x)=W^0_{\bf tp'}(x)$
and, in particular (since $x \in Var_{fix}({\bf tp'})$ must be witnessed by at least 
one clause), that $W^0_{\bf tp'}(x) \neq \emptyset$.

Let $C \in W^0_{\bf tp'}(x)$. Then $C \in {\bf LONG} \setminus {\bf LS}({\bf tp'})=
{\bf LONG} \setminus {\bf LS}({\bf tp})$. As all the variables outside of $Var(\pi)$ belong 
to $Var_{out}({\bf tp})$, $x \in Var(t) \cap Var_{out}({\bf tp})$. Hence $x \in Det({\bf tp})$.

Conversely, let $x \in Det({\bf tp})$. This means that there is 
$C \in {\bf LONG} \setminus {\bf LS}({\bf tp})={\bf LONG}  \setminus {\bf LS}({\bf tp'})$.
such that $x \in Var(C)$. This already implies that $x \in Var_{fix}({\bf tp'})$. 
It remains to show that $x \notin Var(\pi)$. But if it were so then $x$ would belong
to $Var_{fix}({\bf tp})$ causing a contradiction since $x \in Var_{out}({\bf tp})$
by definition of $Det({\bf tp})$.  Thus the second and the third statements are settled.

For the fourth statement, note that, by construction, $Det({\bf tp}) \subseteq Var(A(u))$.
If $A(u)$ assigned any extra variables, it would have satisfied a 
clause of ${\bf LONG} \setminus {\bf LS}({\bf tp})$ which is not the
case. 

For the fifth statement, remains to show that for each $x \in Var_{fix}({\bf tp})$,
the occurrences of $x$ in $FA({\bf tp})$ and $FA({\bf tp'})$ are the same and 
that for each $x \in Det({\bf tp})$, the occurrences of $x$ in $A(u)$ 
and in $FA({\bf tp'})$ are the same. In the former case, the desired statement
follows from $W_{\bf tp}(x)=W_{\bf tp'}(x)$. In the latter case, the assignment
of $x$ in $A(u)$ is precisely the one to not satisfy any witnessing clause of $x$,
hence precisely the assignment of $x$ in $FA({\bf tp'})$. 

For the last statement, note that
$Var_{out}({\bf tp'})=
 Var(\varphi) \setminus (Var_{in}({\bf tp'}) \cup Var_{fix}({\bf tp'})=
Var(\varphi) \setminus (Var_{in}({\bf tp}) \cup Var_{fx}({\bf tp}) \cup Det({\bf tp}))=
(Var(\varphi) \setminus (Var_{in}({\bf tp}) \cup Var_{fx}({\bf tp})) \setminus Det({\bf tp})=
Var_{out}({\bf tp}) \setminus Det({\bf tp})$.
$\blacksquare$

\begin{theorem} \label{succcontrnonsat}
Let ${\bf tp}=(\pi,map,CN,RA)$ be a well-formed non-basic non-final unsatisfiable type and let
$t$ be the immediate successor of $\pi$. assume that $t$ is a contracting
node.
Let $u$ be a sink of $D_{\bf tp}$ such that $A(u)$ 
is not potentially complex. 
Then the following statements hold.
\begin{enumerate}
\item If $u$ is a labelled with $C \rightarrow ()$ 
then $C \in dom(F_{\bf tp})$, $C$ is not satisfied by $A(u)$
and $F(C)|_{A(u)}=()$.
\item If $u$ is identified with the source of $L_{\bf tp'}$
then ${\bf tp'}$ is a well-formed non-basic unsatisfiable type  
and $F_{\bf tp'}=F_{\bf tp}|_{A(u)}$. 
\end{enumerate}
\end{theorem}

{\bf Proof.}
Analogous to Theorem \ref{succextnonsat}
with Lemma \ref{lem:contrnonsat} used instead of Lemma \ref{lem:extnonsat}.
$\blacksquare$

\subsection{Contracting $t$, $A(u)$ potentially complex}
In this subsection 
we assume that ${\bf tp}=(\pi,map,CN,RA)$ is a well-formed non-basic non-final unsatisfiable type,
that $t$, the immediate successor of $\pi$ is a contracting node, and that
$A(u)$ is potentially complex.

Let $CN_1$ be the subset of $CN$ that are not satisfied by $A(u)$.
Let $Trees^*=(MT({\bf tp}) \setminus Trees'_{\pi}) \cup \{T_t\}$.

\begin{lemma} \label{validcn1}
$CN_1 \subseteq CL(Roots(Trees^*))$.
\end{lemma}

{\bf Proof.}
Let $C \in CN_1$.
If there is $T' \in MT({\bf tp}) \setminus Trees'_{\pi}$
such that $C \in CL(Root(T'))$ then we are done.
Otherwise, $C \in CL(Roots(Trees'_{\pi}))$.
We show that in this case, $C \in CL(t)$.
First note that, by assumption, $C \in CL(t_0)$
for some child $t_0$  of $t$ such that $T_{t_0} \in MT({\bf tp})$. 
Next,  as ${\bf tp}$ is non-basic, $F_{\bf tp}(C) \neq ()$.
This means that there is $x \in Var_{out}({\bf tp})$ such
that $x \in Var(C)$. That is, by the edge containement property of tree decompositions, 
there is a node $t_1$ of $T$ such that $x \in Var(t_1)$ and $C \in CL(t_1)$.
Now $Var(T_{t_0}) \subseteq Var_{in}({\bf tp})$ and hence $t_1 \notin V(T_{t_0})$.
By the connectivity property, $C$ is contained in the bag of the parent of $t_0$ which is $t$.
$\blacksquare$

Let $CN_2$ be the subset of $dom(F_{\bf tp}) \setminus ({\bf LONG} \cup CL(Roots(MT)))$
consisting of all clauses $C$ such that $C$ is not satisfied by $A(u)$ and 
$C \in CL(t)$ and $Var(F_{\bf tp}(C)) \cap Var_{free}({\bf tp})=\emptyset$.

Let $map'$ be a function from ${\bf LONG}$ to $Trees_{\pi+t}$ obtained from $map$
as follows.
\begin{itemize}
\item For each $C \in {\bf LONG}$ such that $map(C)=none$ and $C$ is satisfied by $A(u)$,
$map'(C)=T_t$.
\item For each $C \in {\bf LONG}$ such that $map(C) \in Trees'_{\pi}$,
           $map'(C)=T_t$.
\end{itemize}

 Let $RA'=Proj(RA,Var(Roots(Trees^*))) \cup Proj(FA,Var(t)) \cup A(u)$.
Let ${\bf tp'}=(\pi+t,map',CN_1 \cup CN_2,RA')$
It follows from definitions of $map'$ and a potentially complex assignment 
that $dom(map')=Trees^*$.
Moreover, Lemma \ref{validcn1} implies that  all the elements of $CN_1 \cup CN_2$
occur in the bags of the roots of ${\bf MT}({\bf tp'})$ and hence the type is valid in this 
sense. Note that in the previous cases this validity was obvious by construction. 

\begin{lemma} \label{contrfix}
$Var_{fix}({\bf tp'})=Var_{fix}({\bf tp}) \setminus Var(T_t)$
and for each $x \in Var_{fix}({\bf tp'})$,
$W_{\bf tp}(x)=W_{\bf tp'}(x)$.
\end{lemma}

{\bf Proof.}

Let $x \in Var_{fix}({\bf tp}) \setminus Var(T_t)$.
We are going to show that for $W_{\bf tp}(x) \subseteq W_{\bf tp'}(x)$.
Clearly, this will imply that $x \in Var_{fix}({\bf tp'})$. 

It follows from  Proposition \ref{expcontr} that
$MT_{\pi}(x)  \in Trees_{\pi} \cap Trees_{\pi+t}$
and hence, by Proposition \ref{minpresexp}  $MT_{\pi}(x)=MT_{\pi+t}(x)$.

Let $C \in W_{\bf tp}(x)$.
Assume first that $C \in W^0_{\bf tp}$
If $map'(C)=none$ then $C \in W^0_{\bf tp'}(x)$ and we are done.
Otherwise, $map'(C)=T_t$.
In light of the previous paragraph plus another application
of Proposition \ref{minpresexp},
$MT_{\pi+t}(x)<map'(C)$, hence $C \in W^1_{\bf tp'}(x)$.

Assume now that $C \in W^1_{\bf tp}(x)$.
If $map(C) \notin Trees'_{\pi}$ then $map'(C)=map(C)$ through another 
application of Proposition \ref{expcontr},
$MT_{\pi}(x)<map(C)$ implies that $MT_{\pi+t}(x)<map'(C)$
and hence $C \in W^1_{\bf tp}(x)$. 
If $map(C) \in Trees'_{\pi}$ then 
$map'(C)=T_t$. Hence, by the second paragraph and
another application of Proposition \ref{expcontr},
$MT_{\pi+t}(x)<map'(C)$. Again, it follows that
$C \in W^1_{\bf tp'}(x)$.

Conversely, assume that $x \in Var_{fix}({\bf tp'})$. 
As $Var(T_t) \subseteq Var_{in}({\bf tp})$,
$x \notin Var(T_t)$. It thus remains to show that 
$W_{\bf tp'}(x) \subseteq W_{\bf tp}(x)$ thus 
implying that $x \in Var_{fix}({\bf tp})$. 

Let $C \in W_{\bf tp'}(x)$.
If $C \in W_{\bf tp'}^0(x)$ then $map'(C)=none$
and hence $map(C)=none$ thus impying that
$C \in W_{\bf tp}^0(x)$. 
Assume that $C \in W_{\bf tp'}^1(x)$.
If $map(C)=none$ then $C \in W^0_{\bf tp}(x)$. 
Otherwise, note that $MT_{\pi+t}(x)<map'(C)$ implies by Proposition \ref{expcontr}
that $MT_{\pi+t}<T_t$ and hence $MT_{\pi}(x)=MT_{\pi+t}(x)$ by Proposition \ref{minpresexp}. 
If $map'(C)=map(C)$ then $MT_{\pi}(x)<map(C)$ by Proposition \ref{expcontr} and
we are done. Otherwise, $map(C) \in Trees'_{\pi}$. 
It follows from $MT_{\pi}(x)=MT_{\pi+x}$ that $MT_{\pi}(x) \notin Trees'_{\pi}$
Hence, by Proposition \ref{expcontr}, $MT_{\pi}(x)<map(C)$ and hence
$C \in W^1_{\bf tp}(x)$ and we are done.
$\blacksquare$

\begin{lemma} \label{controut}
$Var_{out}({\bf tp})=Var_{out}({\bf tp'}) \cup Var_{free}({\bf tp}) \cup Var(A(u))$
\end{lemma}

{\bf Proof.}
By construction,
$Var_{in}({\bf tp'})=Var_{in}({\bf tp}) \cup Var(T_t)$.
Combining with Lemma \ref{contrfix}, we observe that
$Var_{fix}({\bf tp'}) \cup Var_{in}({\bf tp'})=Var_{in}({\bf tp}) \cup Var_{fix}({\bf tp}) \cup Var(T_t)$.
Consequently, 
$Var_{out}({\bf tp'})=Var_{out}({\bf tp}) \setminus Var(T_t)$.
This is the same as to say that
$Var_{out}({\bf tp'})=Var_{out}({\bf tp}) \setminus (Var_{out}({\bf tp}) \cap Var(T_t))$.
However, by construction, $Var_{out}({\bf tp}) \cap Var(T_t)=Var(A(u)) \cup Var_{free}({\bf tp})$. 
$\blacksquare$

\begin{lemma} \label{contrnofree}
Then for each $C \in dom(F_{\bf tp'})$, $Var(F_{\bf tp}(C)) \cap Var_{free}({\bf tp})=\emptyset$.
\end{lemma}

{\bf Proof.}
Let $C \in dom(F_{\bf tp'})$.
Assume fist that $C \in {\bf LONG}$.
Then $map'(C)=none$ and hence $map(C)=none$.
Let $x \in Var_{free}({\bf tp})$. Then (as $x \notin Var(t)$),
$x \in Var(\pi)$. If $x \in Var(C)$ then $x \in Var_{fix}({\bf tp})$
by definition in contradiction to the definition of $Var_{free}({\bf tp})$
being a subset of $Var_{out}({\bf tp})$. 

Assume next that $C \in CN({\bf tp'})$ which is, by definition $CN_1 \cup CN_2$.
If $C \in CN_2$ then $F_{\bf tp}(C)$ does not have joint variables with $Var_{free}({\bf tp})$
simply by definition. If $C \in CN_1$ then we need to use the definition of one-sided decomposition. 
In particular, let $x \in Var_{free}({\bf tp})$.
This means that there is a child $t_0$ of $t$ such that $x \in Var(T_{t_0})$.
Since $x \notin Var(t)$ by definition, $x$ is not present in any bag outside of $T_{t_0}$,
$x \in Var(C)$  implies that $C$ is present in a bag of $T_{t_0}$ by the containement property.
To this end note that $T_{t_0} \in Trees_{\pi} \setminus {\bf MT}({\bf tp})$ simply because
$Var(T_{t_0})$ contains a variable of $Var_{out}({\bf tp})$. On the other hand, $C \in CL(t_1)$
where $t_1$ is a root of a tree of ${\bf MT}({\bf tp})$. Now $V(T_{t_0})$ does not contain ancestors 
nor descendants of $t_1$ hence $C \notin CL(T_{t_0})$ by definition of one-sided tree decomposition.

Finally,  let $C \in dom(F_{\bf tp'}) \setminus ({\bf LONG} \cup CL(Roots({\bf MT}({\bf tp'}))))$.
Reusing the previous paragraph, we note that if $x \in Var(C)$ then $C \in CL(T_{t_0})$. 
On the other hand, by definition of $F_{\bf tp'}$, there is $y \in Var_{out}({\bf tp'})$
such that $y \in Var(C)$.  So, by the containement condition $C$ must be present in a bag 
containing $y$. On the other hand, $Var(T_{t_0}) \subseteq Var_{in}({\bf tp'})$ so 
$y$ is not present in a bags of $T_{t_0}$. It follows that $C$ must be contained in a bag outside
of $T_{t_0}$. By the connectivity condition, this means that $C$ must be present in the bag of the parent
of $t_0$ which is $t$, in contradiction to the definition of $C$.
$\blacksquare$

\begin{lemma}  \label{contrsubfun}
$F_{\bf tp'}$ is a subfunction of $F_{\bf tp}|_{A(u)}$.
\end{lemma}

{\bf Proof.}
First of all,  let  $C \in dom(F_{\bf tp'}) \cap dom(F_{\bf tp}|_{A(u)})$.
By combination of Lemmas \ref{controut} and \ref{contrnofree},
$Proj(C,Var_{out}({\bf tp})=Proj(C,Var_{out}({\bf tp'}) \cup Var(A(u)))$.
This means that $F_{\bf tp}|_{A(u)}(C)=Proj(C,Var_{out}({\bf tp'}))=F_{\bf tp'}(C)$.
It thus remains to verify that $dom(F_{\bf tp'}) \subseteq dom(F_{\bf tp}|_{A(u)})$.

So, let $C \in dom(F_{\bf tp'})$.
If $C \in {\bf LONG}$ then $map'(C)=none$.
By definition of $map'$ this means $map(C)=none$
(and hence $C \in dom(F_{\bf tp})$)
and $C$ is not satisfied by $A(u)$.
If $C \in CN_1 \cup CN_2$ then it is immediate from the 
definition of these sets that $C \in dom(F_{\bf tp})$ and
that is not satisfied by $A(u)$.

It thus remains to assume that 
$C \in dom(F_{\bf tp'}) \setminus ({\bf LONG} \cup CL(Roots({\bf MT}({\bf tp'}))))$.
One necessary condition for that is that 
$C$ is not satisfied by $FA({\bf tp'}) \cup RA({\bf tp'})$.
It follows from Lemma \ref{contrfix} and the definition of $RA'$ that 
$FA({\bf tp'})=Proj(FA,Var_{fix}({\bf tp}) \setminus Var(T_t))$ 
and $RA({\bf tp'})=Proj(RA({\bf tp}),Var(Roots(Trees^*))) \cup Proj(FA({\bf tp}),Var(t)) \cup A(u)$.
From this we already conclude that $C$ is not satisfied by $A(u)$,
so we only need to show that $C \in dom(F_{\bf tp})$. 

The other necessary condition for $C \in dom(F_{\bf tp'})$ is
that there is $y \in Var_{out}({\bf tp'})$ such that $y \in Var(C)$.
Since $Var_{out}({\bf tp'}) \subseteq Var_{out}({\bf tp})$ by 
Lemma \ref{contrfix}, we know now that $Var(C) \cap Var_{out}({\bf tp}) \neq \emptyset$.
It thus remains to verify that $C$ is not satisfied by $FA({\bf tp}) \cup RA({\bf tp})$.
More precisely, in light of the previous paragraph we need to prevify
that $C$ is not satisfied by $(FA({\bf tp}) \cup RA({\bf tp})) \setminus (FA({\bf tp'}) \cup RA({\bf tp'}))$.
Taking into account the reasoning in the previous paragraph, this amounts to showing 
that $C$ is not satisfied by $Proj(FA({\bf tp}) \cup RA({\bf tp}), Var(T_t) \setminus Var(t))$.
Assume the opposite. This means that there is $x \in Var(T_t) \setminus Var(t)$
such that $x \in Var(C)$. As $x \notin Var(t)$, by the connectivity condition, $x$ cannot appear in
any bag outside of $T_t$. This means that $C$ itself must appear in a bag inside $T_t$ 
by the containement condition of tree decompositions.
On the other hand, as $Var(T_t) \subseteq Var_{in}({\bf tp'})$, $y$ cannot appear in any
bag inside $T_t$. This means that by the containement condition, $C$ must appear in a bag 
outside $T_t$. By the connnectivity condition this means that $C \in CL(t)$ contradicting
the definition of $C$.
$\blacksquare$

\begin{lemma} \label{contrnonsat}
Assume that $F^*$
(the filling function) is satisfiable. 
Then $F_{\bf tp'}$ is not satisfiable. 
\end{lemma}

{\bf Proof.}
Assume the opposite and let $S_1$ be a satisfying assignment of
$range(F_{\bf tp'})$. Also, let $S_2$ be a satisfying assignment of
$range(F^*)$. Clearly, we may assume that
$Var(S_1) \subseteq Var(range(F_{\bf tp'}))$
and $Var(S_2) \subseteq Var(range(F^*))$.
According to Lemma \ref{contrnofree},
$Var(range(F_{\bf tp'})) \cap Var(range(F^*))=\emptyset$.
Therefore, $S=S_1 \cup S_2$ is a well formed set of literals. 

By definition of $F^*$ for each $C \in dom(F^*)$,
$Var(C) \cap Var_{free}({\bf tp}) \neq \emptyset$.
Therefore, by Lemma \ref{contrnofree},
$dom(F_{\bf tp'}) \cup dom(F^*)=\emptyset$. 
Hence, we can consider the function 
$F'=F_{\bf tp'} \cup F^*$. The set $S$ as in the previous paragraph satisfies
$range(F')$. We will establish a contradiction by showing that $F'$
is in fact unsatisfiable.

In order to do this, we need the following claim.
\begin{claim}
$dom(F_{\bf tp}|_{A(u)})=dom(F')$.
\end{claim}

Assume that the claim holds. 
Then for each $C \in dom(F')$, $F'(C) \subseteq F_{\bf tp}|_{A(u)}(C)$. 
Indeed, for $C \in dom(F_{\bf tp'})$ this follows from
Lemma \ref{contrsubfun}.
For $C \in dom(F^*)$,
$F^*(C)=Proj(C,Var_{free}({\bf tp})) \subseteq Proj(C,Var_{out}({\bf tp}) \setminus Var(A(u)))=F_{\bf tp}|_{A(u)}(C)$, 
the containement relation follows from Theorem \ref{controut}.
It thus follows that $S$ is a satisfiying assignment for $range(F_{\bf tp}|_{A(u)})$.
However, this is a contradiction: $F_{\bf tp}|_{A(u)}$ simply because $F_{\bf tp}$ is unsatisfiable
by assumption. It thus remains to prove the claim. 

In light of Lemma \ref{contrsubfun}, it is enough to show that 
each $C \in dom(F_{\bf tp}|_{A(u)})$ such that $Var(C) \cap Var_{free}({\bf tp})=\emptyset$
is contained in $dom(F_{\bf tp'})$.
Assume first that $C \in {\bf LONG}$. Then from $map(C)=none$ and $C$ not being
satisfied by $A(u)$, it follows that $map'(C)=none$ and hence $C \in dom(F_{\bf tp'})$. 
If $C \in CN$  then, since $C$ is not satisfied by $A(u)$, $C \in CN_1$.

It remains to assume that $C \in dom(F_{\bf tp}) \setminus ({\bf LONG} \cup CN)$.
If $C \in Var(t)$ then $C \in CN_2$. Otherwise, applying the same argument 
as in the first paragraph of Lemma \ref{contrsubfun},
we observe that $Var(F_{\bf tp}|_{A(u)}) \subseteq Var_{out}({\bf tp'})$.
As we assumed that $A(u)$ does not falsify $F_{\bf tp}(C)$,
$Var(F_{\bf tp}|_{A(u)}) \cap Var_{out}({\bf tp'}) \neq \emptyset$.
In ths remains to show that $C$ is not satisfied by $FA({\bf tp'}) \cup RA({\bf tp'})$. 
We note that $FA({\bf tp'}) \cup RA{\bf tp'}) \subseteq FA({\bf tp}) \cup RA({\bf tp}) \cup A(u)$
(see the third paragraph of the proof of Lemma \ref{contrsubfun}) for a detailed 
reasoning). Now $C$ is clearly not satisfied by $A(u)$ and since, 
$C \in dom(F_{\bf tp}) \setminus ({\bf LONG} \cup CN)$, $C$ is not satisfied by $FA({\bf tp}) \cup RA({\bf tp})$
simply by definition of $F_{\bf tp}$.
$\blacksquare$

Now we are considering the case where $F^*$ is not satisfiable. 
The reasoning is similar to the case where $F^*$ is satisfiable.
So, rather than reproving everything, we will refer to relevant
proofs or their fragments wherever appropriate. 

Let ${\bf TR^*}$ be the subset of $dom(F^*)$ consisting of all
the clauses $C$ such that $Var(F_{\bf tp}(C)) \setminus Var(T_t) \neq \emptyset$.
Let $R^*$ be a transitional resolution for $F^*$ with ${\bf TR^*}$ being 
the set of transitional clauses. 


\begin{lemma} \label{cint}
${\bf TR^*} \subseteq CL(t)$. 
\end{lemma}

{\bf Proof.}
Let $C \in {\bf TR^*}$.
By definition of $F^*$, there is a variable
$x \in Var(C)$ such that $x \in Var(T_t)$ but does not
occur in a bag outside $T_t$.
This means that $C \in CL(T_t)$ by the edge connectivity 
property. 
On the other hand, by definition of ${\bf TR^*}$,
there is a variable $y \in Var(C) \setminus Var(T_t)$.
Because of the containement condition, $C$ must occur 
in a bag outside $T_t$. We conclude that, by the
connectivity condition, $C \in CL(t)$. 
$\blacksquare$

\begin{lemma} \label{nontranscorr}
Let $u$ be a non-transitional sink of $R^*$ labelled with $C \rightarrow ()$.
Let $A$ be an assignment carried by a source-$u$ path of $R^*$.
Then $F_{\bf tp}(C)|_{A(u) \cup A}=()$. 
\end{lemma}

{\bf Proof.}
In other words, we need to show 
that $Var(F_{\bf tp}(C)) \subseteq Var(A(u)) \cup Var(A)$. 
By definition of $F_{\bf tp}$ and a non-transitional clause of $F^*$,
this is the same as to show that 

$Var(C) \cap Var_{out}({\bf tp}) \cap Var(T_t) \subseteq Var(A(u)) \cup Var(A)$.
Now, $Var_{out}({\bf tp}) \cap Var(T_t)=Var(A(u)) \cup Var_{free}({\bf tp})$.
That is, 
$Var(C) \cap Var_{out}({\bf tp}) \cap Var(T_t)=(Var(C) \cap Var(A(u))) \cup (Var(C) \cap Var_{free}({\bf tp}))$.
As $A$ falsifies $F^*(C)=Var(C) \cap Var_{free}({\bf tp})$,
we are done.
$\blacksquare$

Let ${\bf C} \subseteq {\bf TR^*}$ and $A$ be a ${\bf C}$-falsifier for $F^*$
with $Var(A) \subseteq Var_{free}({\bf tp})$. 
Let ${\bf tp''}=(\pi+t,map',CN_1 \cup CN_2 \cup {\bf C},RA')$.
It follows from the combination of Lemmas \ref{validcn1}
and \ref{cint} that $CN_1 \cup CN_2 \cup {\bf C}$ all occur
in bags of $Roots(MT({\bf tp''}))$, hence ${\bf tp''}$ is valid in
this sense. 

The following theorem is proved analogously 
to Lemma \ref{contrfix}.

\begin{lemma} \label{contrfix2}
$Var_{fix}({\bf tp''})=Var_{fix}({\bf tp}) \setminus Var(T_t)$
and for each $x \in Var_{fix}({\bf tp'})$,
$W_{\bf tp}(x)=W_{\bf tp''}(x)$.
\end{lemma}

The following lemma is proved analogously to
Lemma \ref{controut} with 
Lemma \ref{contrfix2} used instead of Lemma \ref{contrfix}. 

\begin{lemma} \label{controut2}
$Var_{out}({\bf tp})=Var_{out}({\bf tp''}) \cup Var_{free}({\bf tp}) \cup Var(A(u))$
\end{lemma}

\begin{lemma} \label{contrnofree2}
Then for each $C \in dom(F_{\bf tp''}) \setminus {\bf C}$, $Var(C) \cap Var_{free}({\bf tp})=\emptyset$.
\end{lemma}

{\bf Proof.}
Let $C \in dom(F_{\bf tp''}) $.
The subsequent reasoning is analogous to that of Lemma \ref{contrnofree}
with Lemma \ref{contrfix2} and Lemma \ref{controut2} used
instead of Lemma \ref{contrfix} and Lemma \ref{controut}, respectively.
$\blacksquare$

\begin{lemma}  \label{contrsubfun2}
$F_{\bf tp''}$ is a subfunction of $F_{\bf tp}|_{A(u) \cup A}$.
\end{lemma}

{\bf Proof.}
Let $F_1$ be the restriction of $F_{\bf tp''}$ 
to $dom(F_{\bf tp''}) \setminus {\bf C}$ and 
$F_2$ be the restriction of $F_{\bf tp''}$ to ${\bf C}$.

It follows from Lemma \ref{contrnofree} that 
$F_1$ being a subfunction of $F_{\bf tp}|_{A(u) \cup A}$
is equilvalent to $F_1$ being a subfunction of
$F_{\bf tp}|_{A(u)}$. This can be proved analogously
to Lemma \ref{contrsubfun} with $F_1$ used instead of $F_{\bf tp'}$
and Lemmas \ref{contrfix2}, \ref{controut2}, and \ref{contrnofree2}
being used instead of Lemmas \ref{contrfix}, \ref{controut}, 
and \ref{contrnofree}, respectively. 

So, it remains to show that $F_2$ is a subfunction of 
$F_{\bf tp}|_{A(u) \cup A}$.
By definition of ${\bf C}$, $C \in dom(F_{\bf tp})$ and not
satisfied by $A(u) \cup A$. So, it remains to show that
$F_2(C)=F_{\bf tp}|_{A(u) \cup A}(C)$. 
This is the same as to show that
$Var(F_2(C))=Var(F_{\bf tp}|_{A(u) \cup A(C)}$. 
Now, $Var(F_2(C))=Var(C) \cap Var_{out}({\bf tp''})$,
which is by Lemma \ref{contrfix2},
$(Var(C) \cap Var_{out}({\bf tp})) \setminus ((Var(C) \cap Var(A(u)))\cup(Var(C) \cap Var_{free}({\bf tp})))$. 
On the other hand, the only difference in the representation of $Var(F_{\bf tp}|_{A(u) \cup A}(C))$  
is that $Var(C) \cap Var(A)$ is used instead of $Var(C) \cap Var_{free}({\bf tp})$. 
The former is a superset of the latter since $A$ falsifies $F^*(C)$ by assumption.
On the other hand $Var(A) \subseteq Var_{free}({\bf tp})$,
therefore, $Var(C) \cap Var(A)=Var(C) \cap Var(A) \cap Var_{free}({\bf tp})$ meaning that the containement
in the opposite direction holds as well.
$\blacksquare$

\begin{lemma} \label{contrnonsat2}
Then $F_{\bf tp''}$ is not satisfiable. 
\end{lemma}

{\bf Proof.}
Assume the opposite. Let $S_1$ be a satisfying assignment
for $range(F_{\bf tp''})$. Also, let $F_0$ be the restriction of
$F^*$ to $dom(F^*) \setminus {\bf C}$ and let $S_2$ be a satisfying
assignment for $range(F_0)$ existing by definition of $A$.
Clearly, we assume that $Var(S_1) \subseteq Var_{out}({\bf tp''})$
and that $Var(S_2) \subseteq Var_{free}({\bf tp'})$.
That is, $Var(S_1) \cap Var(S_2)=\emptyset$ and hence 
$S=S_1 \cup S_2$ is a well formed set of literals. 
 
It follows from Lemma \ref{contrnofree2}
that $dom(F_{\bf tp''}) \cap dom(F_0)=\emptyset$,
therefore, we can consider a function 
$F'=F_{\bf tp''} \cup F_0$. Clearly, $S$ satisfies
$range(F')$.  We are going to demonstrate that
$dom(F')=dom(F_{\bf tp}|_{A(u)})$ and for each
$C \in dom(F')$, $F'(C) \subseteq F_{\bf tp}|_{A(u)}$.
This will imply that $S$ satisfies $range(F_{\bf tp}|_{A(u)})$,
a contradiction to the assumption that $F_{\bf tp}|_{A(u)}$.

In light of Lemma \ref{contrsubfun2}, it is enough
to show the following. 
\begin{itemize}
\item $dom(F_0) \subseteq dom(F_{\bf tp}|_{A(u)})$.
Just follows from the definition of $F^*$. 
\item For each $C \in dom(F_0)$,
$F_0(C) \subseteq F_{\bf tp}|_{A(u)}(C)$.
In fact this is proved for the whole $F^*$ in the proof of 
Lemma \ref{contrnonsat}.
\item $dom(F_{\bf tp''}) \cup dom(F_0)=dom(F_{\bf tp}|_{A(u)})$.
This is the claim in the proof of Lemma \ref{contrnonsat}
(wih Lemma \ref{contrsubfun2} used instead of Lemma \ref{contrsubfun}).
\end{itemize}
$\blacksquare$

\begin{theorem} \label{succcontrsat}
Let ${\bf tp}=(\pi,map,CN,RA)$ be a well-formed non-basic non-final unsatisfiable type and let
$t$ be the immediate successor of $\pi$. assume that $t$ is a contracting
node.
Let $u$ be a sink of $D_{\bf tp}$ such that $A(u)$ 
is potentially complex.
Let $F^*$ be the filling function. 

If $F^*$ is satisfiable then, 
$u$ is identified with the source of $L_{\bf tp'}$
where ${\bf tp'}$ is non-basic well-formed unsatisfiable type
Moreover, $F_{\bf tp'}$ is a subfunction of
$F_{\bf tp}|_{A(u)}$.

If $F^*$ is unsatisfiable then, by construction, $u$ is
the source of a transitional resolution of $F^*$ with ${\bf TR^*}$
being the set of transitional clauses. 
Let $v$ be a sink of $R^*$. 
Let $P$ be an arbitrary path from $u$ to $v$.
Suppose that $v$
is associated with $C \rightarrow ()$ for some non-transitional
clause $C$. 
Then $F_{\bf tp}(C)|_{A(u) \cap A(P)}=()$.
Otherwise, $v$ is a transitional link associated with $trans_{\bf C}$.
In this case $v$ is identified with the source of $L_{\bf tp''}$
where ${\bf tp''}$ is a well-formed non-basic  unsatisfiable type. 
Moreover, $F_{\bf tp''}$ is a subfunction of $F_{\bf tp}|_{A(u) \cup A(P)}$. 
\end{theorem}

{\bf Proof.}
If $F^*$ is satisfiable then the identification of $u$ with the source
of $L_{\bf tp'}$ follows by construction.
It follows from Lemma \ref{contrfix} that ${\bf tp'}$ is well-formed. 
Further on, it follows from Lemma \ref{contrsubfun} that $F_{\bf tp'}$
is a subfunction of $F_{\bf tp}|_{A(u)}$. 
As $A(u)$ does not falsify any clause of $range(F_{\bf tp})$, we conclude that
${\bf tp'}$ is a non-basic type. Finally, it follows from Lemma \ref{contrnonsat}
that ${\bf tp'}$ is a non-satisfiable type.  

Assume now that $F^*$ is not satisfiable.
If $v$ is a non-transitional sink then the related statement follows
from Lemma \ref{nontranscorr}.
If $v$ is transitional sink then it is identified with the source of $L_{\bf tp''}$
by construction. It follows from Lemma \ref{contrfix2} 
that ${\bf tp''}$ is well formed. Next, it follows from Lemma \ref{contrsubfun2}
that $F_{\bf tp''}$ is a sufunction of $F_{\bf tp}|_{A(u) \cup A}$.  
The argument that ${\bf tp''}$ is non-basic is now a little bit more
complicated. As $A(u)$ is potentially complex, $A(u)$ does not
fasify any clause of $range(F_{\bf tp})$ but there is also extra effect by $A$.
For each $C \in dom(F_{\bf tp''}) \setminus {\bf C}$  this does not matter
because by Lemma \ref{contrnofree2}, $Var(C) \cap Var(A)=\emptyset$
(recall that $Var(A) \subseteq Var_{free}({\bf tp})$ by assumption).
On the other hand, for each $C \in {\bf C}$,
$Var(C) \cap Var_{out}({\bf tp''}) \neq \emptyset$ and hence 
$F_{\bf tp}(C)$ is not falsified by $A(u) \cup A$ either. 
Finally, it follows from Lemma \ref{contrnonsat2} ${\bf tp''}$ 
is unsatisfiable. 
$\blacksquare$

\subsection{Combining things together}


\begin{theorem} \label{globalfrr}
Let $\mathcal{ORD}^*$ be a subsequence of $\mathcal{ORD}$
consisting of all the non-basic well-formed unsatisfiable types. 
Let $R=\bigcup_{{\bf tp} \in \mathcal{ORD}^*} L_{\bf tp}$. 
For each ${\bf tp} \in \mathcal{ORD}^*$ let $u_{\bf tp}$ be the
source of $L_{\bf tp}$. Then for each ${\bf tp} \in \mathcal{ORD}^*$
$R_{u_{\bf tp}}$ is a falsifying FRR. 
\end{theorem}

{\bf Proof.}
Let $\mathcal{ORD}^*=({\bf tp_1}, \dots, {\bf tp_m})$.
First, we observe that $F_{\bf tp_1}, \dots F_{\bf tp_m}$ is a falsifying 
sequence of functions. In order to do this, we need to demonstrate
that each $L_{\bf tp_i}$ satisfies the conditions of Definition \ref{def:falsesequence}.
If ${\bf tp_i}$ is final, this follows from the definition.
Otherwise, this follows from
Theorems \ref{succextnonsat}, \ref{succextsat}, \ref{succcontrnonsat}, \ref{succcontrsat}, 
$L_{\bf tp_1}, \dots, L_{\bf tp_m}$.   
$f_{\bf tp_1}, \dots f_{\bf tp_m}$.
Let us prove that the sequence is read-once. 

Let us say that a type ${\bf tp'}$ is a \emph{descendant} of type ${\bf tp}$
(both types belong to $\mathcal{ORD}^*$) if $f_{\bf tp'}$ is a descendant of $f_{\bf tp}$
as per the definition before Theorem \ref{assemble1}. 
Let $Var^*({\bf tp})$ be the union of $Var(L_{\bf tp})$ and all the sets
$Var(L_{\bf tp'})$ such that ${\bf tp'}$ is a descendant of ${\bf tp}$.

\begin{claim}
$Var^*({\bf tp}) \subseteq Var_{out}({\bf tp})$. 
\end{claim}

{\bf Proof.}
By induction on $\mathcal{ORD}^*$. 
For ${\bf tp_1}$, $Var^*({\bf tp_1})=Var(L_{\bf tp_1})$
and $Var(L_{\bf tp_1}) \subseteq Var_{out}({\bf tp_1})$
simply by construction.

Now consider ${\bf tp_i}$ for $i>1$. 
Then $Var^*({\bf tp_i})$ is the union 
of $Var(L_{\bf tp_i})$, which is a subset
of $Var_{out}({\bf tp_i})$ by construction
and the union of $Var^*({\bf tp_j})$ for children
${\bf tp_j}$ of ${\bf tp_i}$. 
By the induction assumption,
$Var^*({\bf tp_j}) \subseteq Var_{out}({\bf tp_j}$
It follows that $Var_{out}({\bf tp_j}) \subseteq Var_{out}({\bf tp_i})$
from the combination of the following statements:
\begin{itemize}
\item the last statement of Lemma \ref{lem:extnonsat} ;
\item the first statement of Lemma \ref{newparamextsat} ;
\item the last statement of Lemma \ref{lem:contrnonsat};
\item Lemma \ref{controut} and Lemma \ref{controut2}.
\end{itemize}
$\square$

Now, let ${\bf tp'}$ be a descendant of ${\bf tp}$.
This means that there is a child ${\bf tp''}$
of ${\bf tp}$ such that either ${\bf tp'}={\bf tp''}$
or ${\bf tp'}$ is a descendant of ${\bf tp''}$. 
In any case, $Var(L_{\bf tp'}) \subseteq Var^*({\bf tp''})$
and hence, by the claim $Var(L_{\bf tp'}) \subseteq Var_{out}({\bf tp''})$.
The combination of statements provided inthe end of the proof of the claim
implies that $Var_{out}({\bf tp''}) \subseteq Var_{out}({\bf tp}) \setminus Var(L_{\bf tp})$.
It follows that $Var(L_{\bf tp}) \cap Var(L_{\bf tp'})=\emptyset$ as required.
$\blacksquare$


{\bf Proof of Theorem \ref{mainres}.}
Assume that $\varphi$ does not contain empty clauses for otherwise,
the statement is trivial. 
Let ${\bf st}=(\emptyset,nm,\emptyset,\emptyset)$
be a type where $\emptyset$ at the first position denotes the empty prefix of $\pi_T$
and $nm$ maps every clause to $none$. 
We call ${\bf st}$  the \emph{starting type}.
We claim that $F_{\bf st}={\bf 1}_{\varphi}$. 
Indeed, as $\pi({\bf st})$ is empty,
$Var_{fix}({\bf st})=Var_{in}({\bf st})=\emptyset$.
That is,  $Var_{out}({\bf tp})=Var(\varphi)$. 
Thus for each $C \in dom(f_{\bf  st})$, $f_{\bf st}(C)=C$.
It remains to show that $dom(F_{\bf st})=\varphi$.  
As $nm$ maps all the long clauses to $none$, they all belong 
to $dom(F_{\bf st})$. As for a clause $C \in \varphi \setminus {\bf LONG}$,
direct inspection of the last condition of the type function definition  
shows that $C \in dom(F_{\bf st})$.
It follows from Theorem \ref{globalfrr} that $R_{u_{\bf st}}$ is an FRR for $\varphi$.

Now, we need to upper-bound the size of $R_{u_{\bf st}}$.
Let us denote $|Var(\varphi)|+|\varphi|$ by $n$. 
Clearly, $R_{u_{\bf st}}$ is the union of $L_{\bf st}$ and the local
graphs of all the descendants of ${\bf st}$ in the sense defined in the proof
of Theorem \ref{globalfrr}. 
We are going to upper bound the size of each local graph and the number of local
subgraphs in the union.

Consider first the local graph for a final type ${\bf tp}$.
by definition, $L_{\bf tp}$ is a falsifying FRR for $F_{\bf tp}$,
so it might look like a recursion. However, the situation is much simpler.
In particular, observe that $dom(F_{\bf tp}) \cap {\bf LONG}=\emptyset$. 
Indeed, suppose a long clause $C$ belongs to $dom(F_{\bf tp})$. 
This means that $[map({\bf tp})](C)=none$.  Consequently, by definition, 
$Var(\pi({\bf tp})) \cap Var(C) \subseteq Var_{in}({\bf tp}) \cup Var_{fix}({\bf tp})$.
As ${\bf tp}$ is final, it follows that $Var(C) \cap Var_{out}({\bf tp})=\emptyset$
in contradiction to ${\bf tp}$ being non-basic. 
As $F_{\bf tp}$ is in fact $\varphi \setminus {\bf LONG}$,
Corollary \ref{col:transres} is applicable. As there are no transitional clauses,
the upper bound for $L_{\bf tp}$ is at most $n \cdot 4^k$.

Each local graph of a non-final type consists of a dome of size at most $2^k$ plus at most $2^k$
(at most one per leaf transitional resolutions). By Lemma \ref{cint}, the number of
transitional clauses for such a resolution is at most $k$.
Therefore, by Corollary \ref{col:transres}, the size of each resolution is $n \cdot 2^{O(k^2)}$.
We concldue that thesize of each local subgraph is $n \cdot 2^{O(k^2)}$. 

In order to calculate the number of local graphs in the union we calculate the
number of respective types. A type consists of four components.  
The number of possible first components is the number of bags of $T$ plus one.
As the number of bags is $O(n)$, this is an upper bound on the number of the first components as well.
For the number of second components, recall that by  Proposition \ref{logtrees}
(taking into account that the number of bags is $O(n)$),  $|Trees_{\pi}|$ is $O(\log n)$ for each prefix
$\pi$ of  $\pi_T$. Therefore, with the first component being fixed,  each clause of 
${\bf LONG}$ can be mapped to $O(\log n)$ different values. Thus the number of possible
maps is $O(\log n^{|{\bf LONG}|})$ which is well known to be upper-bounded by 
$O(n+|{\bf LONG}|^{| {\bf LONG} |})$. 

For the last two components, observe that for each type ${\bf tp}$, ${\bf MT}({\bf tp})$
partitions a subset of ${\bf LONG}$. Therefore $|{\bf MT}({\bf tp})| \leq |{\bf LONG}|$.
By definition, $CN({\bf tp})$ is a subset of bags of the roots of ${\bf MT}({\bf tp})$
and  $RA({\bf tp})$ assigns the variables in the bags of the roots of ${\bf MT}({\bf tp})$. 
As the ${\bf MT}$ set of trees is 
completely determined by the first two components, with these components being 
fixed, the number of possible third components as well as the number of possible 
fourth components is $2^{O(k \cdot |{\bf LONG}|}$ (recall that the size of each bag is at most $k$). 
We conclude that the total size of $R_{u_{\bf st}}$ is
$(n+|{\bf LONG}|^{|{\bf LONG}|}) \cdot n 
\cdot 2^{O(k^2+k \cdot |{\bf LONG}|)}$. 
$\blacksquare$

\end{document}